\documentclass[12pt,a4paper]{article}

\usepackage[a4paper,text={160mm,247mm},centering]{geometry} 
\usepackage{amsmath,amssymb} 
\usepackage[bookmarks=true,hyperfigures=true]{hyperref} 
\usepackage{graphicx, bm} 
\usepackage[font=small,labelfont=bf,width=0.85\textwidth]{caption} 
\usepackage{verbatim} 
\usepackage[nosort]{cite} 
\usepackage[bulletsep]{collref} 
\usepackage{fixmath}
\usepackage{subfigure}
\usepackage[normalem]{ulem}
\usepackage{mathrsfs}
\usepackage[usenames,dvipsnames]{color}

\makeatletter \let\@keywords\@empty \let\@subject\@empty
\providecommand{\keywords}[1]{\gdef\@keywords{#1}}
\providecommand{\subject}[1]{\gdef\@subject{#1}}
\def\thetitle{\@title} \def\theauthor{\@author}
\def\thesubject{\@subject} \def\thedate{\@date}
\def\thekeywords{\@keywords} \makeatother \AtBeginDocument{
\hypersetup{pdftitle={\thetitle}}%
\hypersetup{pdfauthor={\theauthor}}%
\hypersetup{pdfsubject={\thesubject}}%
\hypersetup{pdfkeywords={\thekeywords}}%
}

\providecommand{\hypersetup}[1]{}
\providecommand{\texorpdfstring}[2]{#1}

\hypersetup{plainpages=false}
\hypersetup{pdfpagemode=UseNone} 
\hypersetup{bookmarksnumbered=true}
\hypersetup{pdfstartview=FitH} 
\hypersetup{colorlinks=false} \hypersetup{citebordercolor={.5 1 .5}}
\hypersetup{urlbordercolor={.5 1 1}} \hypersetup{linkbordercolor={1 .7
.7}}

\numberwithin{equation}{section}

\let\oldbfseries=\bfseries
\let\oldmdseries=\mdseries
\let\oldnormalfont=\normalfont
\renewcommand{\bfseries}{\oldbfseries\boldmath}
\renewcommand{\mdseries}{\oldmdseries\unboldmath}
\renewcommand{\normalfont}{\oldnormalfont\unboldmath}

\makeatletter
\def\mr@ignsp#1 {\ifx\:#1\@empty\else #1\expandafter\mr@ignsp\fi}%
\newcommand{\multiref}[1]{\begingroup
\xdef\mr@no@sparg{\expandafter\mr@ignsp#1 \: }%
\def\mr@comma{}%
\@for\mr@refs:=\mr@no@sparg\do{\mr@comma\def\mr@comma{,}\ref{\mr@refs}}%
\endgroup}
\makeatother

\newcommand{\hypref}[2]{\ifx\href\asklfhas #2\else\href{#1}{#2}\fi}
\newcommand{\secref}[1]{Section~\multiref{#1}}
\newcommand{\appref}[1]{Appendix~\multiref{#1}}

\newcommand{\figref}[1]{Figure~\multiref{#1}}
\renewcommand{\eqref}[1]{(\multiref{#1})}

\RequirePackage{verbatim}

\makeatletter
\newwrite\bibinl@out
\newenvironment{bibtex}[1]{%
  \immediate\openout\bibinl@out #1.bib
  \immediate\write\bibinl@out{\@percentchar generated from `\jobname' starting line \the\inputlineno^^J}%
  \def\verbatim@processline{\immediate\write\bibinl@out{\the\verbatim@line}}%
  \@bsphack\let\do\@makeother\dospecials\catcode`\^^M\active\verbatim@start
}%
{\immediate\closeout\bibinl@out\@esphack}
\makeatother

\makeatletter
\newlength{\apb@width}
\newcommand{\autoparbox}[2][c]{\settowidth{\apb@width}{#2}\parbox[#1]{\apb@width}{#2}}
\newcommand{\includegraphicsbox}[2][]{\autoparbox{\includegraphics[#1]{#2}}}
\makeatother

\usepackage{color}

\def\bea{\begin{eqnarray}} \def\eea{\end{eqnarray}} 
\def\beq{\begin{equation}} \def\eeq{\end{equation}}
\def\ba{\beq\begin{array}{c}} \def\ea{\end{array}\eeq}

\newcommand{\bn}{\begin{enumerate}} \newcommand{\en}{\end{enumerate}}
\newcommand{\bi}{\begin{itemize}} \newcommand{\ei}{\end{itemize}}

\def\cech{${\rm C}^{\kern-6pt
\vbox{\hbox{$\scriptscriptstyle\vee$}\kern2.5pt}}${\rm ech}}
\def\Cech{${\sl C}^{\kern-6pt
\vbox{\hbox{$\scriptscriptstyle\vee$}\kern2.5pt}}${\sl ech}}

\def\a{{\alpha}}

\def\b{{\beta}}

\def\CC{{\cal C}}
\def\CD{{\cal D}}

\def\CG{{\cal G}}

\def\CN{{\cal N}}
\def\CO{{\cal O}}
\def\CP{{\cal P}}

\def\d{\partial}

\def\inv{^{-1}}

\def\inv{^{\raise.15ex\hbox{${\scriptscriptstyle -}$}\kern-.05em 1}}
 \def\hf{\frac{1}{2}}

\def\({\left(}
\def\){\right)}
\def\<{\left\langle\,}
\def\>{\, \right\rangle}
\def\[{\left[}
\def\]{\right]}

\newcommand{\no}{\nonumber}

\def\la{\label}
\def\Li{{\rm Li}_2}

\def\p{{\rm P}}
\def\T{{\rm T}}
\def\D{{\rm D}}

\def\BDS{\text{BDS}}

\newcommand{\su}{\alg{su}}
\newcommand{\so}{\alg{so}}

\newcommand{\sll}{\alg{sl}}

\newcommand{\R}{{\rm R}}
\newcommand{\A}{{\rm A}}
\newcommand{\B}{{\rm B}}
\newcommand{\C}{{\rm C}}
\newcommand{\QQ}{{\rm Q}}

\newcommand{\EE}{{\rm E}}
\newcommand{\I}{{\rm I}}

\newcommand{\uketz}{|\raps{u}\rangle}
\newcommand{\vbra}{\langle\{v\}|}

\newcommand{\uketin}{|\{u\}\rangle_{{\rm ih}}}
\newcommand{\vbrain}{_{{\rm ih}}\langle\{v\}|}

\newcommand{\wketin}{|\{w\}\rangle_{{\rm ih}}}
\newcommand{\h}{{\rm H}}

\def\>{\rangle}
\def\<{\langle}
\newcommand{\LR}{{\rm LR}}
\newcommand{\SD}{{\mathscr D}}
\newcommand{\GV}{{\rm GV}}
\newcommand{\ssum}{\sigma}
\newcommand{\tcon}{\tau}

\newcommand{\uketinLR}{|{\bf u};{{\bm \theta} }\rangle}

\newcommand{\ubrainLR}{\langle{\bf u};{\bm \theta}}
\newcommand{\vbrainLR}{\langle{\bf v};{\bm \theta}}
 \newcommand{\caA}{{\mathscr A}}
  \newcommand{\caS}{{\mathscr S}}
 \newcommand{\hammone}{\{{\rm H}\}}

\def\bu{  {\bf u}^{(1)} }
\def\bv{  {\bf u}^{(2)} }
\def\bw{  {\bf u}^{(3)} }
 \def\thu{ {\bm \theta}^{(1)}}
  \def\thv{ {\bm \theta}^{(2)}}
    \def\thw{ {\bm \theta}^{(3)}}
    \def\thuv{ {\bm \theta}^{(12)}}
    \def\thuw{ {\bm \theta}^{(13)}}
     \def\thvw{ {\bm \theta}^{(23)}}
     \def\uthi{| {\bf u}^{(a)} , {\bm \theta ^{(a)}}\rangle }
 \def\uketin{ | \bu; \thu\rangle}

\def\wketin{ | \bw;\thw\rangle}

\def\vbrain{ \langle \bv;\thv |}

\def\uket{ | \bu \rangle }
\def\vket{ | \bv  \rangle}
\def\wket{ | \bw \rangle }

\def\vbra{ \langle \bv |}

\def\Normw{ \langle \bw  \wket }
\def\Normv{ \langle \bv \vket }
\def\Normu{ \langle \bu  \uket }

\def\uuket{ | {\bf u}  \rangle }

\def\uubra{ \langle  {\bf u}  |}

\newcommand{\comm}[2]{[#1,#2]}
\newcommand{\loc}{\mathcal{L}}

\newcommand{\boost}[1]{\mathcal{B}[#1]}
\newcommand{\biloc}[2]{[#1|#2]}
\newcommand{\charge}{{\mathrm Q}}
\newcommand{\chargeBDS}{\bar {\mathrm Q}}
\newcommand{\qBDS}{\bar q}

\newcommand{\id}{{\mathrm I}}
\newcommand{\X}{{\mathrm X}}
\newcommand{\rmat}{{\mathrm R}}
\newcommand{\rmathat}{ \hat {\mathrm  R}}

\newcommand{\trans}{{\mathrm T}}
\newcommand{\mon}{{\mathrm M}}
\newcommand{\shift}{{\mathrm U}}

\newcommand{\G}{{\mathrm G}}
\newcommand{\ctm}{\mathcal{A}}
\newcommand{\Sop}{{\mathrm S}}

\newcommand{\boo}{\mathcal{B}}
\newcommand{\Bexp}{{\mathrm \Phi}}
\newcommand{\Iexp}{{\mathrm \Theta}}

\newcommand{\ham}{{\mathrm H}}
\newcommand{\hamone}{{\mathrm [\ham ]}}
\newcommand{\hamtwo}{{\mathrm [[\ham]]}}
\newcommand{\qev}{E}

\newcommand{\Yop}{{\mathrm Y}}

\newcommand{\sfrac}[2]{{\textstyle\frac{#1}{#2}}}
\newcommand{\half}{\sfrac{1}{2}}
\newcommand{\ihalf}{\sfrac{i}{2}}
\newcommand{\alg}[1]{\mathfrak{#1}}
\newcommand{\sutwo}{\alg{su}(2)}
\DeclareMathOperator{\tr}{Tr}
\newcommand{\order}[1]{\mathcal{O}(#1)}

\newcommand{\inhom}{\theta}

\newcommand{\sone}{\nu}
\newcommand{\stwo}{\rho}
\newcommand{\sthree}{\mu}

\newcommand{\ups}[1]{{[#1]}}
\newcommand{\SR}{\text{SR}}

\newcommand{\raps}[1]{{\mathbf #1}}

\title{Fixing Loops for Three-Point Functions}

\author{%
Yunfeng Jiang\texorpdfstring{$^{1}$}{},
Ivan Kostov\texorpdfstring{$^{1}$}{}, 
Florian Loebbert\texorpdfstring{$^{2,3}$}{},
Didina Serban\texorpdfstring{$^{1}$}{}
}

\begin{document}

\pdfbookmark[1]{Title Page}{title}

\thispagestyle{empty}
\begin{flushright}\footnotesize
\texttt{IPhT-t14/002}%
\end{flushright}
\vspace*{1cm}

{\small
\vspace{1cm}
\begin{center}%

{\Large \bf Fixing the Quantum Three-Point Function}
\vspace{1cm}

\begingroup\scshape\theauthor\par\endgroup
\vspace{1cm}%

\begingroup\itshape

$^1$
Institut de Physique Th\'eorique,
DSM, CEA, URA2306 CNRS,\\
Saclay, F-91191 Gif-sur-Yvette,
France
\vspace{3mm}

$^2$
School of Natural Sciences, Institute for Advanced Study\\
Einstein Drive, Princeton, NJ 08540, USA
\vspace{3mm}

$^3$
Niels Bohr International Academy \& Discovery Center,
Niels Bohr Institute,\\
Blegdamsvej 17, 2100 Copenhagen, Denmark
\vspace{3mm}
\par\endgroup
\vspace{2mm}

\begingroup\ttfamily
\texttt{%
\{yunfeng.jiang,
ivan.kostov,
didina.serban\}@cea.fr,\\
loebbert@ias.edu
}
\par\endgroup

\vspace{1.5cm}

\textbf{Abstract}\vspace{7mm}

\begin{minipage}{11.7cm}

We propose a new method for the computation of quantum three-point
functions for operators in $\su(2)$ sectors of $\mathcal{N}=4$ super
Yang--Mills theory.  The method is based on the existence of a unitary
transformation relating inhomogeneous and long-range spin chains.
This transformation can be traced back to a combination of boost
operators and an inhomogeneous version of Baxter's corner transfer
matrix.  We reproduce the existing results for the one-loop structure
constants in a simplified form and indicate how to use the method at
higher loop orders.  Then we evaluate the one-loop structure constants
in the quasiclassical limit and compare them with the recent strong
coupling computation.
\end{minipage}

\end{center}
}
\vspace{.3cm}
\newpage
\setcounter{tocdepth}{2}
\hrule height 0.75pt
\pdfbookmark[1]{\contentsname}{contents}
\tableofcontents
\vspace{0.8cm}
\hrule height 0.75pt
\vspace{1cm}
\setcounter{tocdepth}{2}



\section{Introduction}

Integrability has already proven to be a powerful tool for finding a
solution to the spectral problem of supersymmetric gauge theories (see
{\it e.g.}\ \cite{Beisert:2010jr}), and to test their duality to
string theories \cite{Maldacena:1997re}.  In the last few years, the
applications of integrability methods were largely extended to other
fundamental objects in gauge theory, such as scattering amplitudes or
Wilson loops (see {\it e.g.}\ \cite{Drummond:2009fd,Bargheer:2009qu,Korchemsky:2010ut,Drummond:2010zv,Beisert:2010gn,Alday:2010vh,CaronHuot:2011kk,Correa:2012hh,Drukker:2012de,Sever:2012qp,Ferro:2013dga,Chicherin:2013ora,Muller:2013rta,Basso:2013vsa,Basso:2013aha,Elvang:2013cua}) as well as to
correlation functions.  The majority of these computations was
concerned with $\CN=4$ super Yang--Mills (SYM) theory which is also
subject to this work.

The first computations of correlation functions were performed in the
early days of the AdS/CFT correspondence for protected BPS operators
\cite{Lee:1998bxa,Freedman:1998tz}.  For non-protected operators at
weak coupling, progress was made using the map to spin chains
\cite{Roiban:2004va, Okuyama:2004bd, EGSV}.  The most advanced results
concerning ``heavy'' operators, {\it i.e.}\ operators with large
$R$-charge, were obtained at tree-level and in the $\su(2)$ sector
\cite{EGSV,GSV,Foda:2011rr,SL}, but results for the $\su(3)$
\cite{EGSV:Tailoring2,Wheeler:2013zja,Foda:su3} and $\sll(2)$
\cite{Georgiou:sl2,Kazakov:3pttwist2,Sobko:3ptsl2,Pedro:sl23pt}
sectors are also available.  To extend the computation of structure
constants to higher loops, one needs as a crucial input the
field-theoretical computation of loop corrections to the three-point
function \cite{Okuyama:2004bd, Alday:2005nd, Plefka:3pttwist2,
Plefka:3ptlength5}.  Results at loop order were obtained for the
$\su(2)$ sector in \cite{Gromov:2012vu,Serban:2012dr,GV} using the
spin chain technology, and in \cite{Bissi:holographic3pt} using the
coherent state representation and the Landau-Lifshitz model.  At
strong coupling, an important effort was invested in formulating the
problem and in computing special configurations of three-point
functions, both using integrability methods
\cite{Janik:3pt1,Janik:3pt2,Komatsu:3pt1,Komatsu:3pt2,Caetano:3ptstrong,KKnew}
and string techniques \cite{ Zarembo:3pt1,Costa:3pt,
Tseytlin:3ptvertex,Bissi:2011dc,Tseytlin:3pt2,Klose:2011rm,Minahan:2012fh,Bargheer:2013faa}.
Here the conformal bootstrap was also successfully applied
\cite{Costa:2011mg, Caetano:4ptweak, Alday:higherspin3pt}.

Each of these results covers a particular case of three- (or higher-)
point functions, and we do not yet have a comprehensive understanding
of the generic structure of correlation functions, as we do for the
spectrum.  In particular, we do not yet have a method which provides
an acceptable recipe for obtaining a particular three-point function.
Nevertheless, a coherent picture starts to emerge, and an important
step forward is the very recent calculation of $\su(2)$ correlation
functions by Kazama and Komatsu at strong coupling \cite{KKnew}.

\bigskip

In this work, we revisit the computation of quantum three-point
functions in \cite{Gromov:2012vu,Serban:2012dr,GV}, with the purpose
1) to get reliable expressions in the semi-classical limit, which can
be compared to the strong coupling results and 2) to set up a
systematic formalism for proceeding to higher loop orders.  In order
to extend the results from tree-level to loops, we need to have a good
description of the wave functions and scalar products of long-range
interacting spin chains.

  With this motivation in mind, we study a method of generating
  long-range deformations of nearest-neighbor spin chains.  Here we
  consider the case of the XXX spin chain with spin equal to 1/2,
  since it is directly applicable to the computation of correlation
  functions in the $\su(2)$ sector of $\mathcal{N}=4$ super
  Yang--Mills theory.  As a prototype of long-range deformation, we
  consider the BDS model proposed by Beisert, Dippel and Staudacher
  \cite{Beisert:2004hm}, which was shown \cite{Rej:2005qt} to be
  equivalent to a spin-sector reduction of the one-dimensional Hubbard
  model at half-filling.  The method we use here is very general and
  it encompasses a large class of deformations.

Several different methods were employed to describe and solve
long-range spin chains, at least partially.  Historically, one of the
first methods to completely solve a long-range system is based on
so-called Dunkl operators \cite{Bernard:1993va}, and it was used
successfully for the Haldane-Shastry model, and for some aspects of
the infinite length Inozemtsev model
\cite{Serban:2004jf,Serban:2012dr}.  The drawback of this method is
that an explicit representation of the Dunkl operators is known only
for a restricted class of models.  Another restriction is that, with
the exception of the Haldane-Shastry model, the Dunkl operators cannot
be rendered periodic on a finite lattice.  The price to pay for
rendering the lattice finite is to introduce a defect
\cite{Serban:2013jua}.  The advantage is that explicit exact
expressions for the monodromy matrix can be obtained, and the scalar
products are relatively straightforward to compute
\cite{Serban:2012dr}.  Another method to deform the XXX spin chain
uses so-called boost and bilocal charges and was proposed in
\cite{Bargheer:2008jt,Bargheer:2009xy}.  This method works again
fairly well for long spin chains, but does not include wrapping
interactions.

Here, we use yet another method, which is to map the inhomogeneous XXX
model to a long-range model.  The authors of \cite{Beisert:2004hm}
noticed that the spectral equations, ({\it i.e.}\ the Bethe ansatz
equations) of the long-range model they have proposed, can be obtained
from those of an inhomogeneous spin chain by carefully choosing the
values of the inhomogeneities.  This equivalence ceases to hold when
wrapping interactions, {\it i.e.}\ interactions of range equal or
greater than the length of the spin chain, are taken into account.
However, the Hamiltonian of the inhomogeneous spin chain is not a
homogeneous long-range spin chain, because it depends on
inhomogeneities, which are site-dependent.  The observation of BDS was
taken further in \cite{Bargheer:2009xy}, where it was noticed that if
the two spin chains have the same spectrum, then they should be
related by a unitary transformation, which was computed up to
two-impurity order (or two-loop order in $\CN=4$ SYM terms).  This
unitary operator was not explicitly used before to construct the
eigenfunctions of the long-range spin chain.  Instead, the wave
functions of long-range spin chains were constructed via another
relation to inhomogeneous spin chains \cite{Gromov:2012vu,GV} or by
the relation to Dunkl operators \cite{Serban:2012dr,Serban:2013jua}.

In this paper we elaborate on the observation by Bargheer, Beisert and
one of the authors \cite{Bargheer:2009xy} and give a systematic method
to construct eigenvectors and scalar products of the BDS model which
are exact up to wrapping order.  We emphasize that we consider the
periodic model.  The computation of scalar products is straightforward
if the existence of the unitary similarity transformation (the
S-operator) from the inhomogeneous to the long-range spin chain is
assumed.  The method is general and it applies to all spin chains that
can be obtained perturbatively with boost deformations from the XXX
model.  To compute explicitly the wave functions, one needs the
explicit expression of the unitary transformation $\Sop$, which we
derive here up to
quadratic order in the inhomogeneities.%
\footnote{For the specific BDS inhomogeneities this unitary
transformation was already given in \cite{Bargheer:2009xy}.} We find
that the unitary transformation can be constructed using the
long-range boost deformations and an inhomogeneous version of Baxter's
corner transfer matrix:
\begin{eqnarray}
\begin{array}{c}\text{Inhomogeneous spin chain} \\ \text{with
inhomogeneities} \ \theta_k \end{array} \quad\,\,
\begin{array}{c}\text{{\color{Fuchsia} \bf S-operator}} \\
{\color{Fuchsia}\longleftrightarrow }\end{array} \quad\,\,
\begin{array}{c}\text{Homogenous long-range\ spin\ chain} \\ \text{ \
with coupling constants } \sigma_n= \sum_k\theta_k^n \end{array}
 \no
\end{eqnarray}

For verification, we demonstrate that applying the following two
operations, reproduces the differential operator found in
\cite{Gromov:2012vu,GV} plus the required boundary terms: 1)~shifting
the inhomogeneities from zero to their non-zero ({\it e.g.}\ BDS-like)
values and 2)~applying the unitary transformation $\Sop$ which
transforms the chain to a homogeneous long-range chain.  The
procedure, although relatively tedious, is straightforward and can be
applied at higher orders.  It furthermore proves the conjectures and observations 
on the all-loop norms (without dressing phase) in \cite{Gromov:2012vu,GV}.  The results are relatively
simple and elegant, due to the manifest structure of the
transformations.

In the next step, we apply the method described above to the
computation of three-point correlation functions in the $\su(2)$
sector of $\CN=4$ SYM theory in the planar limit.  The key property
that we use is the freedom to choose the values of the
inhomogeneities, as long as their symmetric sums $\sigma_n$ are kept
at the model-specific values.  This can be done in perturbation theory
for sufficiently large chains.  The results for the three-point
function are summarized in \secref{TPFint}.
 
 Our result resembles the asymptotic solution of the spectral problem
 in the $\alg{su}(2)$ sector, where \emph{fixing} the inhomogeneities
 in the Bethe ansatz to the BDS values was enough to obtain the
 long-range Bethe ansatz encoding the higher loop spectrum.  Here we
 get the long-range three-point function in a similar way: We take the
 inhomogeneous three-point vertex, and after \emph{fixing} the
 inhomogeneities, we add a correction given by an operator that acts
 merely on the splitting points of the involved spin chains.

 \bigskip
 
 The result is a very concise expression for the structure constant in
 terms of the rapidities of the three states.  An attractive property
 of this expression is that it allows to obtain without pain the
 semiclassical limit of three heavy operators.  We computed the
 quasiclassical limit of the one-loop structure constant and compared
 it with the Frolov--Tseytlin limit of the result of \cite{KKnew}.
 Both expressions are given by contour integrals of dilogarithm
 functions, up to terms that vanish in the Frolov--Tseytlin limit.  In
 the Frolov--Tseytlin limit the insertions at the splitting points are
 of subleading order, and the gauge theory result is given by the
 inhomogeneous three-point vertex, after {\it fixing} the
 inhomogeneities.  We find that the integrands match, which is already
 a strong evidence that the correspondence with the string theory
 persists at one loop.  Moreover, we reveal through this comparison
 the reason for the asymmetric form of the gauge theory structure
 constant, while the string theory result is completely symmetric with
 respect to permutations of the three operators.  To complete the
 result one should also compare the integration contours.  This is a
 subtle issue which is still lacking complete understanding, both in
 the gauge and in the string theory.  At the present stage the
 contours of integration are chosen case by case by taking into
 account the analytic properties of the solution.

 \bigskip

The structure of the paper is the following: In Section 2 we remind of
the definition of the inhomogeneous XXX spin chain and we define its
conserved charges.  In section 2.1 we define the corner transfer
matrix (CTM) and its inhomogeneous version and we remind of the link
between the CTM and the (first) boost operator.  Section 3 is devoted
to long-range spin chains, including the BDS spin chain, and more
generally to the local boost deformations of the XXX Hamiltonian.  In
Section 4, we make explicit the map between the local boost
deformations and inhomogeneous spin chains, by defining the operator
$\Sop$ and determining it to order $g^2$.  We compare with the result
obtained from the CTM and then compute the scalar products up to
wrapping order.  We also explore the morphism of the Yangian algebra
defined by the operator $\Sop$ and we derive the action of this
morphism on the elements of the monodromy matrix and on the Bethe
vectors.  In Section 5 we show how to compute the three-point function
at one-loop order.  In Section 6 we take the semiclassical limit of
the one-loop expression and compare it with the Frolov--Tseytlin limit
of \cite{KKnew}.

\subsection{The Result for the Three-Point Function}
\la{TPFint}

In this section we summarize our results for the three-point function
of operators in different $\su(2)$ sectors taking the generic form

\begin{eqnarray}
\la{threepointfunction}
\langle\CO^{(1)}(x_1)\CO^{(2)}(x_2)\CO^{(3)}(x_3)\rangle
=\frac{N_c^{-1}\ \sqrt{L^{(1)} L^{(2)} L^{(3)}}\ \
C_{123}(g^2)}{|x_{12}|^{\Delta^{(1)}+\Delta^{(2)}
-\Delta^{(3)}}|x_{13}|^{\Delta^{(1)}+\Delta^{(3)}
-\Delta^{(2)}}|x_{23}|^{\Delta^{(2)}
-\Delta^{(3)}-\Delta^{(1)}}}.
\end{eqnarray}

The operators are chosen such that they have definite conformal
dimensions $\Delta^{(1)}$, $\Delta^{(2)}$ and $\Delta^{(3)}$, and
belong to two different $\sutwo$ sectors
\begin{eqnarray}
\la{3opersint}
\begin{aligned}
\CO^{(1)} \in\{Z,X\}, \quad \CO^{(2)} \in\{\bar Z,\bar X\}, \quad
\CO^{(3)} \in\{Z,\bar X\}\;.
\end{aligned}
\end{eqnarray}
In the language of spin chains, they are characterized by three Bethe
vectors $\uket,$ $\vket,$ and $ \wket$ with lengths $L^{(1)}$,
$L^{(2)}$ and $L^{(3)}$, respectively.  By ${\bf u}^{(a)}$ we denote
the set of the magnon rapidities $\{ u_1^{(a)} ,\dots , u^{(a)}_{
N^{(a)}}\}$.  The renormalization scheme invariant part $C_{123}(g^2)$
can be expressed in the spin-chain language as
\cite{Okuyama:2004bd,Roiban:2004va}
\begin{eqnarray}
\label{c123genint}
C_{123}(g^2)&=& \frac{\langle {\bf u}^{(1)}, {\bf u}^{(2)}, {\bf
u}^{(3)} \rangle }{ \(\Normu\Normv\Normw\)^{1/2}}\;.
\end{eqnarray}
In the expression above, $\langle {\bf u}^{(a)}|{\bf u}^{(a)}\rangle$
are the square norms of the Bethe vectors, which were evaluated in
\cite{EGSV,SL} using the Gaudin-Korepin formula \cite{PhysRevD.23.417,
korepin-DWBC}.  Our result for the three-point function concerns the
loop expression of the cubic vertex
\begin{equation}
\la{cuv}
\langle {\bf u}^{(1)}, {\bf u}^{(2)}, {\bf
u}^{(3)}\rangle=\(1+  
  g^2\hat\Delta_{21}+\CO(g^4)\)\caA_{{\bf
u}^{(1)}\cup {\bf u}^{(2)},\,{\bm
\theta}^{(12)}}
\(1+  
  g^2\hat\Delta_{03} +\CO(g^4)\)\caA_{{\bf
u}^{(3)},\,{\bm
\theta}^{(13)}}.
\end{equation}
Above, the functional $\caA_{{\bf u},{\bm \theta}}$ is expressible in
terms of a determinant, see \secref{sec:Sandscalar}, where the set of
inhomogeneities ${\bm \theta}^{(ab)}$ is given by the BDS-like values
\cite{Beisert:2004hm,Rej:2005qt}
 \begin{eqnarray}
 \la{thetaBDSint} \theta^{(ab)}_l=2g \sin \frac{2\pi l}{L^{(ab)}}\;, 
 \quad l=1,\ldots ,L^{(ab)}\;, \quad 2L^{(ab)}=L^{(a)}+L^{(b)}-L^{(c)}\;.
\end{eqnarray}
It gives the main contribution to the loop-order three-point function
and captures the main effect of the mixing of operators at loop order,
at least for heavy operators.  The operators $\hat\Delta_{21}$ and
$\hat\Delta_{03}$ in \eqref{cuv} compute the effect of the insertions
\cite{Okuyama:2004bd,Alday:2005nd} and of the mixing near the
splitting points.  The operators $\hat\Delta_{ab}$ act as follows (by
convention we take the vacuum to be the state $|{\bf
u}^{(0)}=\emptyset\rangle=|\Omega\rangle$)
\begin{eqnarray}
\hat\Delta_{ab}\,\caA_{{\bf u}^{(a)}\cup {\bf u}^{(b)},\,{\bm
\theta}^{(bc)}} =
\(\partial_1^{(b)}\partial_2^{(b)}-i\delta\EE_2\partial_1^{(b)}
+i\delta\EE_3-\frac{1}{2}\delta\EE_2^2\) \caA_{{\bf u}^{(a)}\cup {\bf
u}^{(b)},\,{\bm \theta}^{(bc)}} \Big|_{\theta=0}\;, \no
\end{eqnarray}
with $\partial^{(a)}_j\equiv \partial/\partial \theta^{(a)}_j$ and
$\delta\EE_r=\EE_r^{(b)}-\EE_r^{(a)}$ being the difference of the
conserved charges between the ket and bra states.

We obtained the quasiclassical limit of (\ref{cuv}) and compare it
with the Frolov--Tseytlin \cite{Frolov-Tseytlin} limit\footnote{If we
introduce a scale for the lengths, with $L^{(a)}\to\infty$,
$L^{(a)}/L$ and $N^{(a)}/L$ finite, then the Frolov--Tseytlin limit
means that $g /L\ll 1$.} of the strong coupling result of
\cite{KKnew}.  In the quasiclassical limit the roots from the set
${\bf u}^{(a)}$ condense into one or several cuts (describing
macroscopic Bethe strings) and the state $|{\bf u}^{(a)}\rangle$ is
characterised by its quasimomentum $p^{(a)}$, which has
discontinuities across the cuts.  Up to terms that can be neglected in
the Frolov--Tseytlin limit,\footnote{The subleading terms are the
contributions of the operators $\hat\Delta_{ab}$.} the logarithm of
the structure constant is given by the contour integral
\begin{align}
\label{clsF123a}
\log C_{123}( g) \simeq& \oint \limits _ {\CC ^{(12|3)} }
\frac{du}{2\pi } \ \text{Li}_2\big( e^{ i p ^{(1)}(u) + i p ^{(2)}(u)
- i q ^{(3)}(u) } \big)\no\\
&+
  \oint \limits _ { \CC^{(13|2)}} \frac{du}{2\pi } \ \text{Li}_2\big(
 e^{ i p ^{(3)}(u) + i q^{(1)}(u) - i q^{(2)}(u) } \big)
  -   \half \; \sum_{a=1}^3
     \; \int\limits_{\CC^{(a)}}{dz\over 2\pi} \ \text{Li}_2\big( e^{
     2i p ^{(a)}(z)} \big) \, .
\end{align}
The integration contours should be placed taking into account the
analytical properties of the integrand.  The three quasimomenta
$p^{(a)} $ depend on $g$ through the distribution of the
inhomogeneities.  The functions $q^{(a)} $ are obtained from $p^{(a)}
$ by subtracting the resolvent for the Bethe roots ${\bf u}^{(a)}$.
Assuming that the contours of integration are the same, the difference
between (\ref{clsF123a}) and the contour integral obtained in
\cite{KKnew} resides in the integrand.\footnote{Also, there are
certain terms that vanish by kinematical reasons in the gauge theory
computation and which do not seem to vanish in the string theory
computation.  We believe that this issue will be resolved soon.} In
this paper we show that the integrand of (\ref{clsF123a}) coincides
with the linear order in $g^2$ of the expansion of the integrand in
the string solution.  The comparison shows that the asymmetry of the
integrand in (\ref{clsF123a}) in the three quasimomenta is a
consequence of the specific choice of the three $\su(2)$ sectors in
$\so(4)$ used in the weak coupling computation.


\section{Inhomogeneous XXX Spin Chain}

In this section we gather some well-known facts about the (periodic)
inhomogeneous XXX spin chain.  It is is defined by the expression of
its monodromy matrix
\begin{eqnarray}
\label{monodtheta}
\mon_\a(u;\theta)=\prod_{k=1}^L \R_{\a 
k}(u-\theta_k-\sfrac{i}{2}),         
\end{eqnarray}
where the rational R-matrix takes the form\footnote{This normalization
for the R matrix is convenient for obtaining the good conserved
quantities, however, for constructing the eigenvectors we find it more
convenient to use the normalization $\R'=\I+i\p/u$.}
\begin{equation}
\R_{\a\b}(u)=\frac{u}{u+i}\,\I_{\a\b}+\frac{i}{u+i}\,\p_{\a\b}.
\end{equation}
and the operator $\p_{\a\b}$ represents a permutation of the spins in
the spaces $\a$ and $\b$.  The monodromy matrix $\mon_\a(u;\theta)$
obeys the Yang-Baxter equation\footnote{In the following we will skip
the variables $\theta=\{\theta_1,\ldots,\theta_L\}$ from the
notations, since the algebraic relations are generic.  To denote the
homogeneous (short-range) quantities we will use the index $\ups{0}$
or SR. }
\begin{eqnarray}
\R_{\a\a'}(u-v) \mon_\a(u) \mon_{\a'}(v)= \mon_{\a'}(v)\mon_a(u)\R_{\a\a'}(u-v).
\end{eqnarray}
When the inhomogeneities $\theta_k$ are set to zero, or they are all
equal to each other, this is the usual homogeneous XXX spin chain.
The inhomogeneities can be interpreted as some extra degrees of
freedom which have been frozen.  It will be convenient to write the
monodromy matrix in the auxiliary space denoted by the index $\a$:
\begin{eqnarray}
\mon_\a(u)=\left(\begin{array}{cc}\A(u) & \B(u) \\ \C(u) &
\D(u)\end{array}\right)_\a
\end{eqnarray}
As for the homogeneous spin chain, the transfer matrix
\begin{equation}
\trans(u)=\tr_\a \mon_\a(u)=\A(u)+\D(u)
\end{equation}
commutes with itself for any value of the spectral parameter ({\it i.e.}\
$[T(u),T(v)]=0$) and it therefore generates the integrals of motion.

Since $\mon(u)$ obeys the Yang-Baxter equation with the rational
R-matrix $\R(u)=(u+i\p)/(u+i) $ the algebra of the matrix elements is
the same as for the homogeneous XXX model:
\begin{eqnarray}
\label{eq:matelalg}
\begin{aligned}
\A(v)\B(u)&=\frac{u-v+i}{u-v} \B(u) \A(v)-\frac{i}{u-v} \B(v) \A(u)\;,\\
\D(v)\B(u)&=\frac{u-v-i}{u-v} \B(u) \D(v)+\frac{i}{u-v} \B(v) \D(u)\;.
\end{aligned}
\end{eqnarray}
The Hilbert space is spanned by states obtained from the pseudo vacuum
$|\Omega\rangle =|\uparrow \uparrow\ldots \uparrow\rangle$ by acting
with the ``raising operators" $\B(u)$:
\begin{eqnarray}
|\raps{u}\rangle=\B(u_1)\ldots \B(u_M)|\Omega\rangle\;.
\end{eqnarray}
If the rapidities $\raps{u}=\{u_1,\ldots,u_M\}$ are generic, the state
is called ``off-shell'', and the sate is called ``on-shell'' if the
rapidities obey the Bethe ansatz equations
\begin{eqnarray}
\label{eq:BAEin}
\frac{a(u_j)}{d(u_j)}=\mathop{\prod_{k=1}^M}_{k\neq j}
\frac{u_j-u_k+i}{u_j-u_k-i}\;,
\end{eqnarray}
where $a(u)$ and $d(u)$ are the eigenvalues of the diagonal
operators $\A(u)$ and $D(u)$:
\begin{eqnarray}
a(u)=1\; & \qquad&
d(u)=\prod_{l=1}^L\frac{\(u-\sfrac{i}{2}-\theta_l\)}{\(u+\sfrac{i}{2}-\theta_l\)}\;.
\end{eqnarray}
The ``on-shell'' states are eigenstates of the transfer matrix $\T(u)$
with the eigenvalue
\begin{eqnarray}
t(u)=\frac{Q(u-i)}{Q(u)}+\frac{d(u)}{a(u)}\frac{Q(u+i)}{Q(u)}\;, \quad
{\rm with} \quad Q(u)=\prod_{k=1}^M(u-u_k)\;.
\end{eqnarray}
We define the integrals of motion of the inhomogeneous spin chain
model conventionally as the logarithmic derivatives of the transfer
matrix $\trans(u)$ around the point $u=i/2$:
\begin{eqnarray}
\QQ_r^\inhom=\left.\frac{1}{{i(r-1)!}}\frac{d^{r-1}}{d u^{r-1}}\ln 
\trans(u)\right |_{u=i/2}.
\end{eqnarray}
Any combination of the above integrals of motion is an integral of
motion and we are going to use later this property in order to define
a more convenient basis of charges.  The definition given above is
convenient if the values of the inhomogeneities are small,
$\theta_k\sim g$, where $g$ is a perturbation parameter which will be
specified later.  It extends the definition of the homogeneous case
({\it i.e.}\ $\theta_k =0$), for which the first conserved quantity is
the shift operator:
\begin{eqnarray}
\shift_0\equiv \trans_0(\sfrac{i}{2})=\tr_\a\prod_{k=1}^L
\p_{\a k}=\p_{L-1,L}\p_{L-2,L-1}\ldots\p_{12}\;.
\end{eqnarray}
The homogeneous shift $\shift_0$ translates the chain by one lattice
spacing, that is we have
\begin{align}
 \shift_0 \p_{k,k+1}\shift_0^{-1}=\p_{k-1,k}.
 \end{align}
  Periodicity of the chain means that $\shift_0^L=1$.  The first
  homogeneous Hamiltonians take the form
\begin{align}
\QQ_{2}^\SR&=\sum_{k=1}^L\ham_k, & 2\QQ_{3}^\SR&=i \sum_{k=1}^L
\hamone_{k-1}, & 3\QQ_{4}^\SR&=\sum_{k=1}^L
\Big(\hamtwo_{k-1}+\ham_k\hamone_{k-1}-\hamone_{k-1}-\ham_k\Big),
\label{eq:SRhamiltonians}
\end{align}
where we have introduced the compact recursive notation
\begin{align}
\ham_k&=\id_{k,k+1}- \p_{k,k+1}, &
\hamone_k&=\comm{\ham_k}{\ham_{k+1}}, &
\hamtwo_k&=\comm{\hamone_k}{\ham_{k+2}}.
\end{align}
For completeness we note that in terms of the R-matrix the
nearest-neighbor Hamiltonian is given by
\begin{equation}
\left.\ham_k=-i\rmat_{k,k+1}^{-1}(u)\frac{d \rmat_{k,k+1}(u)}{du}\right|_{u=0},
\end{equation}
and the homogeneous transfer matrix can be expressed in the convenient
form
\begin{equation}\label{eq:Tconv}
\trans_0(u+\sfrac{i}{2})=\shift_0\exp\Big[i\sum_{r=2}^L u^{r-1}\charge_r^\SR\Big].
\end{equation}
In the inhomogeneous case the conserved quantities do not take the
simple form \eqref{eq:SRhamiltonians}, but it is useful for later
purposes to write them as an expansion in the value of the
inhomogeneities.  The momentum is no longer a conserved quantity,
since the inhomogeneous chain is not translationally invariant.
However, the periodicity condition $\shift_0^L=1$ still holds.  The
conserved quantity which replaces the shift $\shift_0$ is the operator
 \begin{eqnarray}\label{eq:shiftinhomallorders}
\shift_\inhom=\;\tr_\a \prod_{k=1}^L \rmat_{\a k}(-\theta_k)\;,
\end{eqnarray}
whose expansion in $\theta$ exponentiates to%
\footnote{We assume periodic boundary conditions, $k+L\equiv k$.}
\begin{equation}
\label{lntrtbis}
\shift_\inhom= \shift_0\exp \Big[-i\sum_k\theta_k \ham_{k}-\frac{1}{2}
\sum_k\theta_{k-1}\theta_k\hamone_{k-1}+\order{\theta^2}\Big]\;.
\end{equation}
Note that for $\theta_k=-u$ the inhomogeneous shift $\shift_\inhom$
gives back the homogeneous transfer matrix $\trans_0(u+i/2)$
\eqref{eq:Tconv}.  The expansion of the inhomogeneous Hamiltonian
takes the form
\begin{equation}
\charge_2^\inhom =\sum_{k=1}^L\bigg[ \ham_k -i\theta_k[\ham]_{k-1}
+\theta_k^2\,\big(\ham_k[\ham]_{k-1}-[\ham]_{k-1}-\ham_k \big)
+\theta_{k}\theta_{k+1}[[\ham]]_{k-1} \bigg]+\order{\theta^3}.
\end{equation}
The $M$-magnon eigenvalues $\qev_r$ of the conserved quantities
$\charge_r$ are the sum over one-magnon eigenvalues
\begin{eqnarray}
\label{ezerou}
\qev_r^\inhom=\sum_{j=1}^Mq_r(u_j)+\CO(\theta^L),
\end{eqnarray}
where $q_r(u)$ takes the standard form of the XXX one-magnon eigenvalues
\begin{equation}\label{eq:NNeigen}
q_r(u)=\frac{i}{r-1}\bigg(\frac{1}{(u+\frac{i}{2})^{r-1}}
-\frac{1}{(u-\frac{i}{2})^{r-1}}\bigg).
\end{equation}
Here $u_1,\ldots,u_M$ are solutions of the Bethe ansatz equations
(\ref{eq:BAEin}) and as such they depend on the values of the
inhomogeneities $\theta_k$.
\subsection{Corner Transfer Matrix}
\label{sec:ctm}

An interesting quantity with regard to the construction of integrals
of motion is Baxter's corner transfer matrix (CTM)
\cite{Baxter:1976uh}.  After a brief review of some aspects of the CTM
for homogeneous spin chains, we define an inhomogeneous CTM that will
be useful in the subsequent sections.

\begin{figure}
\centering
\includegraphicsbox{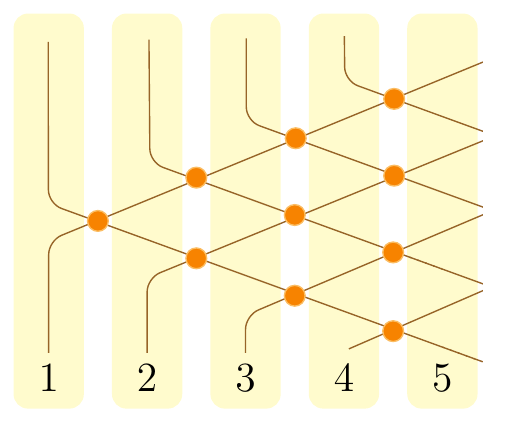} $\dots$ \caption{Corner transfer
matrix (CTM) acting on a spin chain.  We assume an infinite lattice on
the right hand side.  The small discs denote $\rmathat$-matrices
defined in (\ref{eq:rmathat}).}
\label{fig:ctm}
\end{figure}

\paragraph{Homogeneous chains.}
Let us briefly review the definition of a \emph{homogeneous} CTM and
its relation to the so-called nearest-neighbor boost operator, cf.\
\cite{Baxter:1976uh,Thacker:1985gz,Itoyama:1986ad}.
In the following we will assume to work on infinite chains ($L\to \infty$) 
or in the bulk of a periodic chain, respectively.%
\footnote{Note that typically some spins on the edge of the CTM are fixed.}
First we introduce a half-row matrix $\G_A$ ranging from site $A(<L)$ to site $L$:
\begin{equation}
\G_A(u)= \rmathat_{L-1,L}(u)\rmathat_{L-2,L-1}(u)
\dots\rmathat_{A+1,A+2}(u)\rmathat_{A,A+1}(u).
\end{equation}
Here we have defined the symbol $\rmathat(u)$ as the R-matrix times
the permutation operator:%
\footnote{Usually the CTM is defined in terms of ordinary R-matrices
or vertex weights and the ingoing site $k$ is identified with the
outgoing site $k+1$ when mapping the vertex model to a spin chain.
Here it seems convenient to circumvent the vertex model interpretation
to avoid confusion.  }
\begin{equation}\label{eq:rmathat}
\rmathat_{k,k+1}(u)=\p_{k,k+1}\rmat_{k,k+1}(u).
\end{equation}
Then we define the CTM as a stack of half-row matrices of different
lengths according to (cf.\ \figref{fig:ctm})
\begin{equation}
\ctm(u)=\G_1(u)\ldots \G_{L-2}(u)\G_{L-1}(u).
\end{equation}
Note that the triangular definition of the CTM originates in the
context of vertex models.  In fact, this matrix can be defined for
every quadrant of a square lattice of R-matrices (vertices).  In the
bulk the half-row matrix $\G_A$ has (up to the shift) the same
structure as the parity inverted row-to-row transfer matrix
$\trans^{-1}(-u+\sfrac{i}{2})$
and consequently a similar expansion%
\footnote{Note that $\rmat^{-1}(u)=\rmat(-u)$.}
\begin{equation}
\G_A(u)=1+iu\sum_{k=A}^L\ham_{k}+\order{u^2}.
\end{equation}
This form makes it clear that the CTM expands as
\begin{equation}
\ctm(u)=1+iu\boost{\charge_2^\SR}+\order{u}^2,
\end{equation}
where $\boost{\charge_2^\SR}$ denotes the so-called boost operator of
the nearest-neighbor Hamiltonian $\charge_2^\SR=\sum_k \ham_k$.  For a
generic local operator $\loc$ with local density $\loc_k$, the boost
is defined as
\begin{equation}\label{eq:defboostop}
\boost{\loc}=\sum_k k \loc_k.
\end{equation}
It is well-known that the boost of the nearest-neighbor Hamiltonian
allows to obtain higher integrable Hamiltonians of a short-range spin
chain model based on a rational (or trigonometric) R-matrix
\cite{Tetelman:1981xx}:
\begin{equation}\label{eq:srboost}
\charge_{r+1}^\text{SR}=-\frac{i}{r}\comm{\boost{\charge_2^\text{SR}}}{\charge_{r}^\text{SR}}.
\end{equation}
In fact, on infinite chains the homogeneous CTM can be expressed as 
the exponential of the nearest-neighbor boost operator as shown 
in \cite{Baxter:1976uh,Thacker:1985gz} for the XYZ model:
\begin{equation}
\ctm(u)=\exp\big(i u \boost{\charge_2^\SR}\big).
\end{equation}
Since the row-to-row transfer matrix $\trans(u)$ is the generating
function of the local integrals of motion, \eqref{eq:srboost} is
equivalent to the differential equation
\cite{Thacker:1985gz,Sklyanin:1991ss}
\begin{equation}
\frac{d}{du}\trans(u+\sfrac{i}{2})=i\comm{\boost{\charge_2^\SR}}{\trans(u+\sfrac{i}{2})},
\qquad \trans(\sfrac{i}{2})=\shift_0,
\end{equation}
where we have fixed the initial value of the transfer matrix to be the
homogeneous shift operator.  This implies that a finite boost
transformation corresponds to a shift of the rapidity parameter of the
row-to-row transfer matrix:
\begin{equation}
\ctm^{-1}(u)\trans(v)\ctm(u)=\trans(u+v).
\end{equation}
In particular, one can understand the row-to-row transfer matrix as
being generated by the CTM through a transformation of the shift
operator $\shift_0=\trans(i/2)$:
\begin{equation}\label{eq:transfromshift}
\trans(u+\sfrac{i}{2})=\ctm^{-1}(u)\shift_0\ctm(u).
\end{equation}

\paragraph{Inhomogeneous chains.}
Now we would like to extend the above considerations to \emph{inhomogeneous} spin chains. 
We define the inhomogeneous CTM as a stack of homogeneous half-row 
matrices with different rapidity shifts:%
\footnote{In \cite{Loebbert:2010thesis} it was speculated on the
connection of the long-range deformations discussed in the subsequent
sections to an inhomogeneous version of the CTM. We have not found any
discussion of the inhomogeneous CTM defined in \eqref{eq:ctminhom} in
the literature.}
\begin{equation}\label{eq:ctminhom}
\ctm_\inhom(u)=\G_1(u-\theta_1)\G_{2}(u-\theta_2)\dots\G_L(u-\theta_L).
\end{equation}
Expanding this inhomogeneous CTM evaluated at $u=0$ in terms of the inhomogeneities $\theta$ we find%
\footnote{For an expansion of the inhomogeneous CTM at order
$\theta^3$ see \appref{app:CTMthree}.}
\begin{equation}
\label{eq:ctminhomex}
\ctm_\inhom(0)
=\exp\Big[i\sum_{k}\sone_k \ham_{k}
-\frac{1}{2}\sum_{k}\hat\stwo_k\hamone_{k-1}+\order{\theta^3}\Big].
\end{equation}
where the coefficients $\sone_k$ and $\hat\stwo_k$ are given by
\begin{equation}\label{eq:sonestwohat}
\sone_k=-\sum_{x=1}^k\theta_x,\qquad  \hat\stwo_k=-\theta_k\sone_k-\sum_{x=1}^k\theta_x^2.
\end{equation}
In analogy to \eqref{eq:transfromshift} we may interpret the
inhomogeneous shift operator as being generated by the operator
$\ctm_\inhom$ on infinite chains:
\begin{equation}
\shift_\inhom=\ctm_\inhom^{-1}(0)\,\shift_0\,\ctm_\inhom(0).
\end{equation}
While we have no proof for this transformation property in general, we
have verified it up to order $g^2$.  Similarly one can check that the
inhomogeneous bulk Hamiltonian is generated according to
$\charge_2^\inhom=\ctm_\inhom^{-1}(0)\charge_2^\text{SR}\ctm_\inhom(0)$,
at least up to order $g^2$.  In \secref{sec:MapLRInh} we will
rediscover the inhomogeneous CTM in the context of a map between
inhomogeneous and long-range spin chains.  It would be very
interesting to investigate in greater detail how this CTM generates
the asymptotic inhomogeneous spin chain model from an ordinary
short-range chain.


\section{Long-Range Integrable Models}

By deforming the homogeneous short-range XXX model, one can obtain
long-range spin chain models.  One possibility is to define these
models exactly, for any value of the deformation parameter and for any
length of the chain.  For example this is the case for the Inozemtsev
model \cite{Inozemtsev:2002vb} whose Hamiltonian takes the form
\begin{eqnarray}
\ham_{\I}=\mathop{\prod_{k=1}^L}_{k\neq l}  \CP_{L,i\pi/\kappa}(k-l)\;\p_{kl}.
\end{eqnarray}
Here $ \CP_{L,i\pi/\kappa}$ is the Weierstrass function with periods
$L$ and $i\pi/\kappa$.  At $\kappa \to \infty$ this model gives back
the short-range Heisenberg model.  Another limiting case of this model
is the $\kappa\to 0$ limit, which yields the Haldane-Shastry model
\cite{Haldane:1987gg,Shastry:1987gh}, and which was widely studied in connection
with exclusion statistics.  Another possibility to define long-range
deformations is to define the model through a series expansion in the
deformation parameter.  This was done for example for the dilatation
operator of $\CN=4$ SYM theory \cite{Beisert:2003tq}, which corresponds
to an (asymptotically) integrable spin chain Hamiltonian.
Integrability can then be defined perturbatively; for example if the
deformed conserved charges are given by an expansion in the
deformation parameter $g$ of the form\footnote{Here we suppose that
only even powers of $g$ appear in the small $g$ expansion, as it is
the case for the ${\cal N}=4$ SYM dilatation operator in the $\alg{su}(2)$ sector.}
\begin{eqnarray}
\charge_r(g)=\sum_{k\geq 0} \charge_r^\ups{k}g^{2k},
\end{eqnarray}
then the terms in the expansion can be computed order by order and to
test integrability to order $\ell$, one checks that
\begin{eqnarray}
[\charge_r(g),\charge_s(g)]=\CO(g^{2(\ell+1)})\;.
\end{eqnarray}
We say then that the model is integrable up to $\ell$-loop order.
\vskip5pt \noindent {\bf The BDS Spin Chain.} An important example of
a long-range spin chain that we will use in this work is the BDS chain
\cite{Beisert:2004hm}.  It was defined in the perturbative sense as a
long-range spin chain whose first three orders coincide with the
dilatation operator of $\CN=4$ supersymmetric Yang--Mills theory in
the $\sutwo$ sector:
\begin{eqnarray}
\label{eq:dilop}
D(g)=L+2\sum_{k\geq 1}g^{2k}D^\ups{k-1}\;.
\end{eqnarray}
The coupling constant $g$ is related to the 't~Hooft coupling constant
of the gauge theory as $\lambda=16\pi^2 g^2$. 
The first three
non-trivial orders of the dilatation operator were computed by
Beisert, Kristjansen and Staudacher \cite{Beisert:2003tq} and they are
given by
\begin{eqnarray}
\label{pertdilat}
&&D^\ups{0}=\sum_{k=1}^L (1-\p_{k,k+1})\;,\\ \nonumber
&&D^\ups{1}=\sum_{k=1}^L (4\p_{k,k+1}-\p_{k,k+2}-3)\;,\\ \nonumber
&&D^\ups{2}=\sum_{k=1}^L \left(
-14\p_{k,k+1}+4\p_{k,k+2}+10-\p_{k,k+3}\p_{k+1,k+2}+\p_{k,k+2}\p_{k+1,k+3}\right)\;.
\end{eqnarray}
In the initial BDS paper \cite{Beisert:2004hm}, the model was defined
beyond three-loop order by the Bethe ansatz equations it was supposed
to obey:
\begin{align}
\label{eq:BAEBDS}
\(\frac{x(u_j+\frac{i}{2})}{x(u_j-\frac{i}{2})}\)^L&=\mathop{\prod_{k=1}^M}_{k\neq
j} \frac{u_j-u_k+i}{u_j-u_k-i}\;, &
e^{ip}&=\frac{x(u+\frac{i}{2})}{x(u-\frac{i}{2})}\,,
\end{align}
with the rapidity map $x(u)$ and its inverse given by the Zhukovsky relation
\begin{align}\label{eq:Zhuk}
x(u)&=\frac{u}{2}\(1+\sqrt{1-\frac{4g^2}{u^2}}\),
&
u(x)&=x+\frac{g^2}{x}.
\end{align}
In \cite{Rej:2005qt} it was shown that the Hamiltonian
(\ref{eq:dilop}) and the Bethe ansatz (\ref{eq:BAEBDS}) can be
obtained by reducing the one-dimensional half-filled Hubbard model to
the spin sector.  In principle, the higher order terms in
(\ref{eq:dilop}) can be computed from perturbation theory of the
Hubbard model, and at increasing perturbative order they involve
interactions connecting more and more spins.  The difference between
the Hubbard model prediction and the Bethe ansatz equations appears at
order $ g^{2L}$, which is the order at which \emph{wrapping
interactions} start to contribute.

Notably, the above Bethe equations for the BDS model equal the inhomogeneous 
Bethe equations \eqref{eq:BAEin} up to wrapping, if the inhomogeneities 
are fixed to \cite{Beisert:2004hm}%
\footnote{For odd values of the length $L$ one should add a twist to
the inhomogeneities that we neglect here for simplicity
\cite{Rej:2005qt}.}
\begin{equation}
\label{eq:valimp}
\theta_k^\text{BDS}=2g \sin \frac{2\pi k}{L}.
\end{equation}
In consequence, the spectra of the two models are the same up to
wrapping order and their Hamiltonians can be related by a similarity
transformation \cite{Bargheer:2009xy}.  In the subsequent sections we
will pursue the investigation of this relation between the two spin
chain models.

Beyond three loops neither the Hubbard model nor the inhomogeneous or
BDS Hamiltonian yield the correct asymptotic dilatation operator of
$\mathcal{N}=4$ super Yang--Mills theory.  The different physical
quantities obtained from these models have to be corrected by the
so-called \emph{dressing phase} contributions
\cite{Arutyunov:2004vx,Beisert:2006ib,BES}.


\subsection{Boost Operators}
\label{sec:boostops}
In this section we review a general method for the construction of
long-range spin chains using a deformation equation that preserves
integrability \cite{Bargheer:2008jt,Bargheer:2009xy}.  We then discuss
the BDS spin chain in this context.

The starting point for these long-range deformations is a given
short-range system with mutually commuting Hamiltonians
$\charge_r^\text{SR}$, $r=2,3,\dots$, ({\it e.g.}\ generated through
\eqref{eq:srboost}) that act locally and homogeneously on a spin
chain.  The long-range charges $\charge_r(g)$ are then defined by the
deformation equation
\begin{equation}\label{eq:defeq}
\frac{d}{dg}\charge_r(g)=i\comm{\X(g)}{\charge_r(g)},
\qquad\quad
\charge_r(0)\equiv\charge_r^\ups{0}=\charge_r^\text{SR},
\end{equation}
whose solutions $\charge_r(g)$ are mutually commuting by construction.
The generators of long-range deformations $\X(g)$ are constrained by
the requirement that the $\charge_r(g)$ are local and homogeneous
operators.  In \cite{Bargheer:2008jt,Bargheer:2009xy} two main classes
of generators
$\X$ were identified and their physical interpretation was studied:%
\footnote{Note that more types of generators can be specified
depending on the deformed short-range model (see for instance the
discussions of open boundary conditions \cite{Loebbert:2012yd} or the
XXZ model \cite{Beisert:2013voa}).}
\begin{align}
&\text{Boost charges:} &&\X=\boost{\charge_r}=\biloc{\id}{\charge_r}
&&\text{(rapidity map)}\label{eq:boostcharges}\\
 &\text{Bilocal charges:} &&\X=\biloc{\charge_r}{\charge_s}
 &&\text{(dressing phase)}\label{eq:bilocalcharges}
\end{align}
Here the bilocal composition of two local operators $\loc_1$ and $\loc_2$ is defined as
\begin{equation}
\biloc{\loc_1}{\loc_2}=\sum_{k<\ell}\loc_{1,k}\loc_{2,\ell}.
\end{equation}
Furthermore one may deform the charges by local operators $\X=\loc$
which amounts to a similarity transformation not changing the
spectrum; deformations with local conserved charges $\X=\charge_r$ are
trivial.  As mentioned before, the basis of local charges can be
transformed without spoiling integrability.  Typically the initial
basis of short-range Hamiltonians is chosen in such a way that the
charge $\charge_r(0)$ acts on at most $r$ neighboring spin chain sites
at the same time.

Let us note that the boost operator \eqref{eq:defboostop} transforms under translations as
\begin{equation}\label{eq:boostshift}
\shift_0\,\boost{\loc}\,\shift_0^{-1}=\boost{\loc}+\loc,
\end{equation}
and is therefore not well-defined globally, since it is not compatible
with the periodicity condition $\shift_0^L=1$.  However, if $\loc$ is
a conserved charge, the above boost recursions
$\eqref{eq:srboost,eq:defeq}$ are well-defined locally, since the
defining relations yield a local homogeneous operator.  The fact that
the boost is not well-defined globally insures that the deformation
\eqref{eq:defeq} is not just a similarity transformation and that the
spectrum of the deformed model is different from the spectrum of the
undeformed model.  Similar arguments apply to deformations with
bilocal charges.

The BDS spin chain introduced in the previous sections is obtained
from a specific combination of the above boost deformations and basis
transformations.  Therefore we will here focus on boost deformations
\eqref{eq:boostcharges} of the XXX chain and leave the study of
bilocal deformations in this context for future work.  In order to
obtain the full integrable model describing the asymptotic $\sutwo$
sector of $\mathcal{N}=4$ SYM theory (including the dressing phase
contributions), also the bilocal charges \eqref{eq:bilocalcharges}
have to be switched on.  The BDS and the full $\mathcal{N}=4$ SYM
theory chain in the $\sutwo$ sector correspond to a specific choice of
parameters in the large class of different long-range models that can
be generated by the above method.

It is important to note that generically the interaction range of the
solutions $\charge_r(g)$ to \eqref{eq:defeq} increases with each order
of the coupling parameter $g$.  This implies that for a given spin
chain, the range of the charge $\charge_r(g)$ exceeds the length of
the chain from a given perturbative order in $g$.  It is not known how
to define the action of the charges beyond this so-called
\emph{wrapping order}.  Hence, the validity of the considered
long-range model is limited to the \emph{asymptotic} regime of long
states.

For our purposes in the following sections it is useful to distinguish two
sets of commuting charges defined as deformations of the same
short-range spin chain model (below we will consider deformations of the homogenoeus XXX spin chain).
The charges $\charge_r(g)$ and $\chargeBDS_r(g)$ are defined 
by the following two deformation equations:%
\footnote{For a more detailed discussion of the relation between long-range 
deformations and the Bethe ansatz see
\cite{Bargheer:2008jt,Bargheer:2009xy}.  For comparison to the
notation used in \cite{Bargheer:2009xy} we note that
$\tcon_k=\Pi_k/dg$, where $\Pi_k$ is a one-form defined in that
paper.}
\begin{align}
\frac{d}{dg}\charge_r(g)
&=\sum_{k=3}^\infty\tcon_ki\comm{\boost{\chargeBDS_k(g)}}{\charge_r(g)},\label{eq:remarkdefeqone}
\\
\frac{d}{dg}\chargeBDS_r(g)
&=
\sum_{k=3}^\infty\tcon_k\big(i\comm{\boost{\chargeBDS_k(g)}}{\chargeBDS_r(g)}
+
\sfrac{r+k-2}{k-1}\,\chargeBDS_{r+k-1}\big).
\label{eq:remarkdefeqtwo}
\end{align}
Furthermore we set $\charge_r(0)=\chargeBDS_r(0)$ such that the two sets of charges differ only by the perturbative basis transformation on the r.h.s.\ of \eqref{eq:remarkdefeqtwo}.
 As indicated above, the charges from the two sets commute among themselves by construction. In addition, the charges from different sets commute among each other since we have
\begin{equation}
\frac{d}{dg}\comm{\charge_r(g)}{\chargeBDS_s(g)}
=\sum_{k=3}^\infty\tau_k\big(i\comm{\boost{\chargeBDS_k(g)}}{\comm{\charge_r(g)}{\chargeBDS_s(g)}}
+\sfrac{r+k-2}{k-1}\,\comm{\charge_r(g)}{\chargeBDS_{s+k-1}(g)}\big).
\end{equation}
Thus, if the charges commute at order zero ({\it e.g.}\ they are deformations of the same short-range model), they also commute at higher orders in $g$. 
We conclude that both sets of charges should have the same basis of eigenstates.%
\footnote{Below we will fix $\tau_k$ such that the charges $\chargeBDS_r(g)$ correspond to the BDS charges with minimized
interaction range at each order in $g$. Ultimately we will be interested in comparing the eigenstates of the BDS charges to the eigenstates of the inhomogeneous model; for this purpose the basis of charges $\charge_r(g)$ is more convenient.}
Following the lines of \cite{Bargheer:2009xy}, one finds that both sets of charges are diagonalized by the
same Bethe ansatz equations:
\begin{align}\label{eq:Bethe}
\left(\frac{f(u+\ihalf)}{f(u-\ihalf)}\right)^L&=\mathop{\prod_{i=1}^M}_{i\neq
k}\frac{u_j-u_k+i}{u_j-u_k-i}, &
e^{ip}&=\frac{f(u+\frac{i}{2})}{f(u-\frac{i}{2})},
\end{align}
where the rapidity map $f(u)$ is related to $\tcon_k$ by
\begin{equation}\label{eq:defPix}
\frac{df(u)}{dg}=-\sum_{k=3}^\infty\frac{\tcon_k}{(k-1)}\frac{1}{f^{k-2}}.
\end{equation}
When expressed as functions of the rapidity $u$, the one-magnon
eigenvalues of the charges $\charge_r$ take the ordinary
short-range form \eqref{eq:NNeigen} and the deformation enters only
via the Bethe equations. For the charges $\chargeBDS_r$, the one-magnon
eigenvalues are given by
\begin{equation}
\qBDS_r(u)=\frac{i}{r-1}\bigg(\frac{1}{f(u+\frac{i}{2})^{r-1}}
-\frac{1}{f(u-\frac{i}{2})^{r-1}}\bigg).
\end{equation}
The eigenvalue functions turn out to be related by the following relation \cite{Beisert:2013voa}:
\begin{equation}
\qBDS_r(u)=q_r(u)+\sum_{s=r+1}^\infty \gamma_{r,s}(g)q_s(u),
\qquad
\frac{1}{f(u)^{r-1}}=\sum_{s=r}^\infty \gamma_{r,s}(g)\frac{r-1}{s-1}\frac{1}{u^{s-1}},
\end{equation}
where $\gamma_{r,s}(g)$ is defined by expansion of the second equation.
Similarly, the corresponding charge operators should be related to each other.

Let us now compare the Bethe Ansatz equations \eqref{eq:Bethe} for the deformed spin
chain with those for the inhomogeneous spin chain
\eqref{eq:BAEin}.  We notice that they look similar, up to terms of
order $\theta^L$ at least, if we write
\begin{equation}\label{eq:defpnx}
\frac{d\ln f(u)}{du}=\frac{1}{L}\sum_{k=0}^\infty
\frac{\ssum_k}{u^{k+1}},\quad {\rm with} \quad \ssum_k=\sum_{i=1}^L
\theta_i^k,
\end{equation}
and we relate $\tcon_k$ to the symmetric sums $\ssum_k$ as prescribed
by the relations \eqref{eq:defPix} and \eqref{eq:defpnx}.  Since the
functional form of the charge eigenvalues $q_r(u)$ also coincides with \eqref{eq:NNeigen}, 
we conclude that the spectra of the inhomogeneous model and
the corresponding deformed model are the same.  Because the spectrum
depends only on the value of the symmetric sums, any permutation of
the values of the impurities gives a model with the same spectrum (but
not the same Hamiltonian).  One may therefore suspect that the two
types of models are mutually related by a unitary transformation, cf.\
\cite{Bargheer:2009xy}.  In the next sections, this transformation is
defined, and determined explicitly for the first two orders in
perturbation.  The values of the symmetric sums $\ssum_k$ in
\eqref{eq:defpnx} can be translated into values of the coupling
constants $\tcon_k$ for the long-range deformations.  These coupling
constants define a whole family of long-range integrable models, since
the values of the first $L$ symmetric sums can be tuned independently.
Among these models we are particularly interested in the BDS model.

\paragraph{The BDS spin chain.}
Let us consider the recursive definition of the BDS chain in some more
detail.  In this case the rapidity map and its inverse are given by
$f^\text{BDS}(u)\equiv x(u)$, see \eqref{eq:Zhuk}.  We may use this
explicit form to evaluate \eqref{eq:defPix} according to
\begin{equation}
-\frac{dx(u)}{dg}=\left.\frac{du(x)}{dg}\right/\frac{du(x)}{dx}
=\frac{2gx}{x^2-g^2}=\sum_{k=3}^\infty\frac{\tcon^\text{BDS}_k}{(k-1)x^{k-2}}\;,
\end{equation}
such that writing the left hand side as a series we find the BDS expressions%
\begin{align}\label{eq:BDSPi}
\tcon_{2k}^\text{BDS}&=0,
&
\tcon_{2k+1}^\text{BDS}&=4k g^{2k-1}.
\end{align}
Then the BDS charges are given by $\chargeBDS_r(g)$ as defined by \eqref{eq:remarkdefeqtwo}, where the $\chargeBDS_r(0)\equiv\chargeBDS_r^\ups{0}$ denote the integrable charges of the XXX Heisenberg spin chain. 
In \appref{app:BDScharges} we give explicit solutions for the BDS
charges in terms of boost deformations up to four-loop order.

In the following we will be interested in studying the effect of the
above deformation on eigenstates of the charge operators and compare
them to the inhomogeneous spin chain model. For this purpose it is more convenient to consider the charges $\charge_r(g)$.
As indicated above,
these charges have the same eigenstates as the BDS Hamiltonians $\chargeBDS_r(g)$,
and they have the same eigenvalues as those of the
inhomogeneous models with the values of the symmetric sums given by%
\footnote{Let us emphasize that two inhomogeneous models obtained from
one another by permutation of inhomogeneities have the same spectrum
but different conserved charges, so they can be considered as being
different.}
\begin{align}
\ssum_{2k+1}^\BDS&=0\;, & \ssum_{2k}^\BDS&=Lg^{2k} \frac{(2k)!}{(k!)^2}\;.
\end{align}
In particular this is true for the model with the inhomogeneities
specified in formula \eqref{eq:valimp}.  

We thus consider the deformation equation
\eqref{eq:remarkdefeqone} evaluated for the BDS connection
\eqref{eq:BDSPi}.  We may solve this equation in the form
\begin{equation}
\charge_r(g)=e^{i \Bexp(g)}\, \charge_r^\ups{0}\, e^{-i\Bexp(g)}\equiv
\Sop_\boo \,\charge_r^\ups{0}\, \Sop_\boo^{-1},
\end{equation}
where $\charge_r^\ups{0}=\charge_r^\SR$ is the nearest-neighbor
charge, and at the first perturbative orders we have
\begin{equation}\label{eq:solY}
\Bexp(g)=g^2\,
\Yop^\ups{0}+\frac{g^4}{2}\Yop^\ups{1}+\frac{g^3}{6}\big(\Yop^\ups{2}+\frac{i}{2}
\comm{\Yop^\ups{1}}{\Yop^\ups{0}}\big)+\order{g^8},
\end{equation}
with
\begin{equation}
\Yop^\ups{\ell}=2\sum_{k=1}^{\ell+1} k\,\boost{\chargeBDS_{2k+1}^\ups{\ell-k+1}}\;.
\end{equation}
This explicitly defines the unitary boost transformation
$\Sop_\boo=\exp[i\Bexp(g^2)]$ up to terms corresponding to the
four-loop order in $\mathcal{N}=4$ SYM theory and can easily be
written down to higher loops.  Note that the local density
$\Bexp_k(g)$ of the operator $\Bexp(g)$ is not periodic in the spin
chain site $k$ since it is defined in terms of boost charges.

\section{Map from Long-Range to Inhomogeneous Models}
\label{sec:MapLRInh}

In this section we elaborate on the relation between long-range and
inhomogeneous spin chains.  After studying the unitary transformation
$\Sop$ that relates the charge operators of the two models at leading
orders, we argue that the operator $\Sop$ originates from a
combination of boost operators and an inhomogeneous version of
Baxter's corner transfer matrix discussed above.  Finally we comment
on the morphism defined by the S-operator and the relation to the
theta-morphism introduced in \cite{GV}.

\subsection{S-Operator}

We would like to understand better how the inhomogeneous spin chain
model is related to the long-range system.  To this end we follow the
observation in \cite{Bargheer:2009xy} that the inhomogeneous charge
operators can be mapped to the BDS charges by a unitary similarity
transformation $\Sop$.  In \cite{Bargheer:2009xy} this transformation
was specified for the BDS model and up to terms of order $g^2$.  Here
we study the relation between the inhomogeneous charges
$\charge_r^\inhom$ and generic long-range charges $\charge_r(g)$
obtained by the method
of boost deformation:%
\footnote{Note that the $\charge_r$ differ from the charges with
minimized interaction range ({\it e.g.}\ the BDS charges) by a basis
transformation as explained above.  This does not make any difference
for the transformation of eigenstates.}
\begin{equation}
\charge_r(g)=\Sop \,\charge_r^\inhom\,\Sop^{-1}.
\end{equation}
The existence of the operator $\Sop$ is motivated by comparison of the
spectra of the BDS chain and the inhomogeneous model for the
BDS-values of the inhomogeneities \cite{Beisert:2004hm}.
\paragraph{Definition of the S-operator.}
In order to define the operator $\Sop$, it is simplest to evaluate 
the same transformation as for the local charges on the shift operator
\begin{equation}\label{eq:shiftS}
\shift(g)=\Sop \,\shift_\inhom\,\Sop^{-1},
\end{equation}
and to extract the form of $\Sop$ from this equation.  Since both
shift operators $\shift(g)$ and $\shift_\inhom$ are defined to all
orders, this yields an all-order definition of the S-operator in the
parameter $g$.  On the one hand, the inhomogeneous shift operator is
defined by equation \eqref{eq:shiftinhomallorders} from which we can
read off its expansion ($\theta\sim g$):
\begin{equation}\label{eq:shiftinhom}
\shift_\inhom=\shift_0\bigg[1-i\sum_{k=1}^L\theta_k\ham_{k}
-\frac{1}{2}\sum_{k,l=1}^L\theta_k\theta_l\ham_{k}\ham_{l}
-\frac{1}{2}\sum_{k=1}^L\theta_{k-1}\theta_k\hamone_{k-1}\bigg]+\order{g^3}.
\end{equation}
For the boost induced long-range models on the other hand, we may
apply the deformation equation \eqref{eq:remarkdefeqone} to the shift
operator in analogy to deforming the local charges:\begin{align}
\frac{d}{dg}\shift(g)&=i\sum_{k=3}^\infty
\tcon_k\comm{\boost{\chargeBDS_k(g)}}{\shift(g)}, & \shift(0)&=\shift_0.
\end{align}
Here $\tcon_k$ is defined by the rapidity map $f(u)$, which in turn is
defined by the spectrum of the shift operator of the underlying model,
cf.\ \eqref{eq:Bethe,eq:defPix} and
\cite{Bargheer:2008jt,Bargheer:2009xy}.

Let us assume that the expansion of $\tcon_{2k+1}$ 
starts at $g^{2k-1}$ and that $\tcon_{2k}=0$.%
\footnote{The former assumption is motivated by the interaction range
of the local charges being constrained by the gauge theory.  The
latter assumption corresponds to a parity conserving model.  Both
assumptions are satisfied for the BDS chain.} Then we can use the
shift property \eqref{eq:boostshift} of the boost charges
($\comm{\boost{\loc}}{\shift_0}=-\shift_0\loc$) to write down the
first two non-trivial orders of $\shift(g^2)$:
\begin{equation}\label{eq:shiftlongrange}
\shift(g^2)=\shift_0\bigg[1-i\bar\tcon_3\charge_3^\SR+
\bar\tcon_3^2\Big(\comm{\boost{\charge_3^\SR}}{\charge_3^\SR}
-\frac{1}{2}\big(\charge_3^\SR\big)^2\Big)
-i\big(\bar\tcon_5+\bar\tcon_3^2\big)\charge_5^\SR\bigg]+\order{g^6}.
\end{equation}
Here we have defined%
\footnote{For instance we have $\bar
\tcon_{2k+1}^\text{BDS}=2g^{2k}$.}
\begin{equation}
\bar\tcon_k=\int_0^g \tcon_k(g')dg'.
\end{equation}
Thus, both shift operators $\shift_\theta$ and $\shift(g)$ are defined
to all orders in $g$ and \eqref{eq:shiftS} furnishes an all-order
definition of the operator $\Sop$.  In the following \secref{sec:ctm}
we will elaborate more on the generic structure of the S-operator.

Let us now explicitly derive the perturbative expression for the
unitary transformation that relates the two models up to order $g^2$.
We make the same ansatz as in \cite{Bargheer:2009xy}, namely
\begin{equation}\label{eq:ansatzS}
\Sop=\exp i \sum_k
\Big[\sone_k\ham_{k}+\frac{i}{2}\stwo_k\hamone_{k-1}+\order{g^3}\Big],
\end{equation}
but do not fix the constants $\sone_k$ and $\stwo_k$ to their BDS
values.  Instead we will obtain generic expressions for $\sone_k$ and
$\stwo_k$ in terms of the periodic inhomogeneities $\theta_k$ that
obey certain constraints.  Here we assume that $\theta_k\sim g$,
$\sone_k\sim g$ and $\stwo_k\sim g^2$.  We can now compare the two
shift operators \eqref{eq:shiftinhom} and \eqref{eq:shiftlongrange}
and derive the constraints following from \eqref{eq:shiftS}.
\paragraph{First order.}
We apply the ansatz \eqref{eq:ansatzS} for the S-operator to the
inhomogeneous shift and evaluate the expression at order $g$:
\begin{equation}
\Sop \,\shift_\inhom\,\Sop^{-1}
=\shift_0\bigg[1-i\sum_{k=1}^L
\ham_{k}(\sone_k-\sone_{k-1}+\theta_k)+(\sone_L-\sone_0)\ham_{1}\bigg]+\order{g^2}.
\end{equation}
Here we have used that $\ham_{k}\shift_0 =\shift_0 \ham_{k+1}$.  Since
the long-range shift operator has no contribution at order $g^1$,
\eqref{eq:shiftS} yields the constraints
\begin{align}\label{eq:inhomone}
\sone_k-\sone_{k-1}+\theta_k&=0, & \sone_L-\sone_0&=0.
\end{align}
These equations are solved by the explicit expression
\begin{equation}\label{eq:sone}
\sone_k=\sone_0-\sum_{x=1}^k\theta_x,
\end{equation}
and the periodicity condition for the first-order parameters yields
\begin{equation}\label{eq:firstsymsum}
\sone_{k+L}=\sone_k \qquad \Rightarrow \qquad
\sum_{x=1}^{L}\theta_x=0.
\end{equation}
The latter condition guarantees that the operator S is periodic, {\it i.e.}\
it represents a well-defined transformation on a periodic spin chain
at the considered order.
\paragraph{Second order.}
Proceeding to terms at order $g^2$ in \eqref{eq:shiftS} we assume that
the above constraints \eqref{eq:inhomone} hold.  Also at this order we
require $\Sop$ to be periodic which amounts to $\stwo_{k+L}=\stwo_k$.
We may again evaluate the right hand side of \eqref{eq:shiftS} and
after some manipulations (cf.\ \appref{app:Satgsquare}) we arrive at
\begin{equation}\label{eq:finally}
\Sop \, \shift_\theta \,
\Sop^{-1}=\shift_0\bigg[1+\frac{1}{2}\sum_{k=1}^L
\big(\stwo_k-\stwo_{k-1}+\sone_{k-1}(\theta_k
-\theta_{k-1})\big)\hamone_{k-1}\bigg]+\order{g^3}.
\end{equation}
We may now compare this expression to the long-range shift operator
\eqref{eq:shiftlongrange} which gives the constraint equation for the
second order parameters $\stwo_k$:
\begin{equation}
\label{eq:sonestwo}
\stwo_k-\stwo_{k-1}=\bar\tcon_3-\sone_{k-1}(\theta_k-\theta_{k-1}).
\end{equation}
This equation is solved by (here we assume for simplicity $\sone_0=0$)
\begin{equation}\label{eq:rhosol}
\stwo_k =\stwo_0+\bar\tcon_3 k
-\sum_{x=1}^k\sone_{x-1}(\theta_x-\theta_{x-1}) =\stwo_0+\bar\tcon_3
k-\theta_k\sone_k-\sum_{x=1}^k\theta_x^2,
\end{equation}
and periodicity for the second order parameter yields
\begin{equation}\label{eq:secsymsum}
\stwo_{k+L}=\stwo_k \qquad \Rightarrow \qquad
\sum_{x=1}^L\theta_x^2=\ssum_2=\bar\tcon_3 L.
\end{equation}
\paragraph{Conclusion.}
In this section we have given an all-order definition of the operator
$\Sop$ by the transformation relating the inhomogeneous and long-range
shift operators \eqref{eq:shiftS}.  We have then computed the operator
$\Sop$ up to terms of order $g^2$.  The S-operator takes the form
\begin{align}\label{eq:Sopfinal}
\Sop= \exp i\Big[\bar\tcon_3 \boost{\charge_3^\SR}+\sum_{k=1}^L
\big(\sone_k\ham_k+\frac{i}{2}\hat\stwo_k\hamone_{k-1}\big)\Big]+\order{g^3},
\end{align}
and for the inhomogeneities set to the BDS values it gives the
expression already determined in \cite{Bargheer:2009xy}.  Notably the
S-operator can be split into two contributions
\begin{align}\label{eq:SboostSinh}
\Sop&=\Sop_\boo\times\Sop_\inhom^{-1}, & \Sop_\boo&=\exp i\Bexp, &
\Sop^{-1}_\inhom&= \exp i\Iexp.
\end{align}
Here the boost and inhomogeneous generator are given by
\begin{align}\label{eq:expboostexpinhom}
\Bexp&=\bar\tcon_3\boost{\charge_3^\SR}+\order{g^4}, &
\Iexp&=\sum_{k=1}^L
\big(\sone_k\ham_k+\frac{i}{2}\hat\stwo_k\hamone_{k-1}\big)+\order{g^3}.
\end{align}
and we have defined the inhomogeneous parameter $\hat\stwo_k$ to
separate the boost and inhomogeneous piece:
\begin{equation}\label{eq:stwohat}
\hat\stwo_k=\stwo_0-\theta_k\sone_k-\sum_{x=1}^k\theta_x^2.
\end{equation}
Note that the boost part agrees with the first order of
\eqref{eq:solY} as expected.  This splitting into a boost and an
inhomogeneous piece is natural knowing that the boost deformations
generate the long-range model from the short-range (here Heisenberg)
model (cf.\ \secref{sec:boostops}).  In particular this implies two
important features for the inhomogeneous part $\Sop_\inhom^{-1}$ of
the S-operator:
\begin{itemize}
\item In the bulk, $\Sop_\inhom^{-1}$ sets all inhomogeneities to zero
and hence represents the generator of the inhomogeneous rapidity shift
as indicated in \secref{sec:ctm}.  \item At the boundary,
$\Sop_\inhom^{-1}$ completes $\Sop_\boo$ to a periodic operator
$\Sop$.
\end{itemize}
Remarkably, the expression for $\Sop_\inhom^{-1}$ in
\eqref{eq:SboostSinh} agrees with the expansion of the inhomogeneous
corner transfer matrix $\ctm_\inhom$ \eqref{eq:ctminhomex}.  That is
to say that the parameters $\sone_k$ and $\hat\stwo_k$ are the same
functions of $\theta$ as defined in \eqref{eq:sonestwohat} (for
$\sone_0=0$ and $\stwo_0=0$).  We have thus found that the expansion
of $\Sop_\inhom^{-1}$ is identical with the expansion of the
inhomogeneous CTM at first orders:
\begin{equation}
\Sop_\inhom^{-1}=\ctm_\inhom(0)+\order{\theta^3}.
\end{equation}
Assuming that the map between $\Sop_\inhom$ and $\ctm_\inhom$ holds at
higher orders, it seems natural to use the CTM to define the operator
$\Sop_\inhom$.  In fact, the inhomogeneous CTM is defined to all
orders in $\theta$ according to \eqref{eq:ctminhom} in terms of
R-matrices.  Together with the boost deformations discussed in the
previous sections this could furnish an explicit definition of the
complete S-operator.  Note that at higher orders it remains to be
shown that an operator $\Sop_\inhom$ defined in this way has the
desired property to combine with the boost part into the
transformation translating between long-range and inhomogeneous spin
chains.

\subsection{Morphism Property and Scalar Products}
\label{sec:Sandscalar}

In the previous chapter we have shown how to obtain the integrals of
motion for the long-range (LR) model by transforming the inhomogeneous
integrals of motion with the unitary operator $\Sop$.  The same
unitary transformation can be applied to the monodromy matrix $\mon$
as well,
\begin{equation}
\la{morph}
\mon^\LR(u)=\Sop\, \mon(u;{\bm \theta})\,\Sop^{-1}\;,
\end{equation}
where the values of ${\bm \theta}$ are chosen as explained in
\eqref{eq:defpnx}.  It is straightforward to show that the monodromy
matrix of the long-range model $\mon^\LR(u)$ obeys the Yang-Baxter
equation, and that its matrix elements obey the same algebra as the
inhomogeneous (or homogeneous) ones \eqref{eq:matelalg}.  The unitary
transformation is therefore a {\it morphism} of the Yangian algebra,
\begin{equation}
\mon_\a^\LR(u)\,\mon_\b^\LR(v)=\Sop\, \mon_\a(u;{\bm \theta})\,
\mon_\b(v;{\bm \theta} )\,\Sop^{-1}\;,
\end{equation}
 for any spaces $\a$ and $\b$.  It is important to note that this
 morphism works for periodic chains of arbitrary length, up to
 wrapping order $g^{2L}$.  This is in contradistinction to the
 morphism considered in \cite{Serban:2012dr,Serban:2013jua}, based on
 the Dunkl operators, where a defect was added at the point where the
 chain closes.  Of course, the difference between the two is small for
 large chains.  Let us explore the consequences of this morphism.
 First, the Bethe vectors for the long-range model, on-shell or
 of-shell, can be written simply as
\begin{equation}
\uuket_\LR=\Sop\,\uketinLR\;, \qquad _\LR\uubra=\ubrainLR|\,\Sop^{-1}\;.
\end{equation}
This means that the scalar products, including the norms, are the same
for the long-range model and the corresponding inhomogeneous model,
\begin{equation}
_\LR \langle {\bf v}\uuket_\LR=\vbrainLR \uketinLR
\;.
\end{equation}
The evaluation of the scalar products in the long-range model, up to
wrapping order, is then straightforward.  According to Slavnov
\cite{NSlavnov1}, the scalar product of an on-shell and and off-shell
vector can be written in terms of a determinant
\begin{eqnarray}
\vbrainLR \uketinLR
= \prod_{j =1}^N
    d(u_j )a(v_j )
    \ 
   \caS_{{\bf u} , {\bf v}}\, ,  
    \la{defscpr} 
 \quad 
   \caS_{{\bf u} , {\bf v}} = {\det_{j k}\Omega(u_j , v_k ) \over \det_{j k}
 {1\over u_j -v_k +i} }\,  .
 \la{defSuv} 
\end{eqnarray}
The Slavnov kernel $\Omega(u,v)$ is\footnote{We are using a different
normalization than in the original paper \cite{NSlavnov1}.}
\begin{eqnarray}
 \la{defOhat} \Omega(u,v)= t(u-v) - e^{2 i p_{\bf u}(v)} \ t(v-u)\, ,
  \qquad t(u) = {1\over u } - {1\over u+i}\,  ,
     \end{eqnarray}
where $p_{\bf u}(u)$ is  the quasimomentum 
 defined modulo $\pi$ by
\begin{eqnarray}
\la{defquasimom}
e^{2 i p(u)} = {Q_{\bf u} (u+i) \over Q_{\bf u} (u-i)} \ { Q_ { \bm
{\theta } }(u-i/2)\over Q_ { \bm {\theta } }(u+i/2)}.
\end{eqnarray}
Here $Q_{\bm \theta}(u)=\prod_{j=1}^L(u-\theta_j)$ and $Q_{{\bf
u}}(u)=\prod_{j=1}^N(u-u_j)$ are Baxter polynomials.  Taking the limit
${\bf v}\to{\bf u}$, one obtains a determinant expression for the norm
(the Gaudin-Korepin determinant)
\begin{eqnarray}
\< {\bf u}; {\bm \theta}|{\bf u}; {\bm \theta}\> &=& {\prod_{j=1}^N
{2\, d^2(u_j) }\over \prod_{j<k} (u_j-u_k)^2}
\det_{jk}\(\frac{(1-\delta_{jk})}{u_{jk}^2+1}+\d_{u}p_{{\bf
u}}(u)\Big|_{u_j}\delta_{jk}\)\no \\
&=&
{\prod_{j=1}^N {2\, d^2(u_j)  }\over \prod_{j<k} (u_j-u_k)^2}
\det_{jk}\d_{u_j}p_{{\bf u}}(u_k) .
 \end{eqnarray}
The diagonal term in the above determinant should be understood as
$\d_{u}p_{{\bf u}}(u)\big|_{u=u_j}$.
     
In \cite{sz} the Slavnov determinant formula was simplified, based on
the results in \cite{SL}, to the symmetric-looking formula\footnote{In
proving \eqref{scalarpr} it is essential that the rapidities ${\bf u}$
are on-shell.  In \cite{sz} only the first identity is proved; the
second one is obtained by the same method.  With the chosen
normalization of the monodromy matrix we should use the second
representation.}
\begin{eqnarray}
\label{scalarpr}
\vbrainLR \uketinLR
= \prod_{j =1}^N
    d(u_j )a(v_j )\ \caA^+ _{{\bf u}\cup {\bf v}, { \bm {\theta } }}
=
    \prod_{j =1}^N
   d(u_j ) d(v_j )\ \caA^- _{{\bf u}\cup {\bf v}, { \bm {\theta } }}
  \end{eqnarray}
where the functional $\caA^\pm _{{\bf w},{\bm \theta}}$ is defined
as follows:
\begin{eqnarray}
\label{defcaA}
\caA^\pm _{{\bf w } , { \bm {\theta } }}\equiv {1\over \Psi_{{\bf w } , {
\bm {\theta }\pm  i/2 }} }\prod _{j=1}^{2N} (1- e^{\pm i \d/\d w_j})\Psi_{{\bf w }
, { \bm {\theta } \pm i/2}}, \qquad \Psi_{{\bf w } , { \bm {\theta } }} =
{\prod_{j<k}^{2N} (w_j - w_k)^2 \over \prod_{j=1}^{2N} \prod _{ l=1}^L (w_j -\theta _l )}.
\end{eqnarray}
%
%
The two representations are compatible 
due to the 
 identity below, which follows directly from the definition, 
  \begin{eqnarray}
  \la{APAM} \caA^+ _{{\bf w } , { \bm {\theta } }} =
  \prod_{j=1}^{2N}{Q_{\bm {\theta } }^- (w_j) \over Q_{\bm {\theta }
  }^+(w_j)} \ \caA^-_{{\bf w},\bm {\theta }} , \qquad {\bf w } = {\bf
  u}\cup{\bf v},
\end{eqnarray}
 where we use the consequence of the Bethe equations,
\begin{eqnarray}
 \prod_{j=1}^N{Q_{\bm {\theta } }^- (u_j) \over Q_{\bm {\theta } }^+(u_j)}
 \equiv
 \prod_{j=1}^N {d(u_j)\over a(u_j)} = 1.
\end{eqnarray}
 As a consequence, the functionals $\caA^+$ and $\caA^-$ differ by a
 phase factor, with the phase equal to the total momentum of the
 magnons with rapidities ${\bf v}$ of the off-shell state.
 Since the phase factor does not have physical meaning, sometimes we
 will omit the $\pm$.  The $\caA$-functional can be also written in
 the form of a Fredholm-like determinant \cite{EldadI}
\begin{eqnarray}
\la{cafr} \caA^\pm _{{\bf u} , { \bm {\theta } }} = \det_{jk} \(
\delta_{jk} - { i \, E^\pm _j\over u_j-u_k+ i } \) , \quad E^\pm _j \equiv {Q_ {
\bm {\theta } } (u_j \mp i/2) \over Q_ { \bm {\theta } } (u_j\pm  i/2 )}
\prod_{k(\ne j)} {u_j-u_k\pm  i \over u_j-u_k}\;.
\end{eqnarray}
 The inhomogeneous $\caA$-functional is a symmetric function of the
variables ${\bm \theta}$ and it can be straightforwardly transformed
into a long-range $\caA$-functional, $\caA_{{\bf u} , { \bm {\theta }
}}=\caA_{{\bf u} }^\LR+\CO(g^{2L})$, with
\begin{eqnarray}
\la{calr}
 \caA_{{\bf u} }^{\LR, \pm} = \det_{jk} \( \delta_{jk} - { i \,
E_j^{\LR, \pm}\over u_j-u_k+ i } \) , \quad E_j^{\LR, \pm} \equiv \({f (u_j \mp i/2)
\over f (u_j\pm   i/2 )}\)^L \prod_{k(\ne j)} {u_j-u_k\pm  i \over u_j-u_k}.
\end{eqnarray}
The function $f(u)$ was defined in \eqref{eq:defpnx} in terms of the
symmetric sums $\ssum_k=\sum_j \theta_j^k$ and it enters the
long-range Bethe ansatz \eqref{eq:Bethe}.  A formula equivalent to
\eqref{calr} is implicit in \cite{Serban:2012dr} and was subsequently
conjectured and checked for the norms in \cite{GV}.  Let us emphasize
that we do not need to know the explicit form of the operator $\Sop$
in order to compute the scalar products of the long-range spin chain.
The above formulas are readily adapted for going to the semiclassical
limit where $L$ and $N$ are large.  \paragraph{The dressing phase and
the inhomogeneities.} The inhomogeneities can also be used to emulate
the effect of the dressing phase, provided that we allow their value
to depend on the value of the rapidities.  This amounts to allowing
the symmetric sums to be symmetric functions of the rapidities ${\bf
u}$ ($\ssum_k=\ssum_k({\bf u})$) so that we have for the eigenstates
of the model with the dressing phase, for example with the BES phase
\cite{BES},
\begin{eqnarray}
\la{vecdp}
|{\bf u}\rangle_{\rm BES}=\Sop({\bf u})|{\bf u};{\bm \theta({\bf u}})\rangle.
\end{eqnarray}
Since the operator $\Sop({\bf u})$ depends now on the state on which
it acts, we cannot compute the scalar products in the same
straightforward manner, but at least we can compute the norms of the
Bethe ansatz vectors,
\begin{eqnarray}
\la{normdp} _{\rm BES}\langle {\bf u} |{\bf u}\rangle_{\rm
BES}=\langle{\bf u};{\bm \theta({\bf u}})|{\bf u};{\bm \theta({\bf
u}})\rangle= \lim_{{\bf v}\to {\bf {u}}}\langle{\bf v};{\bm
\theta({\bf u}})|{\bf u};{\bm \theta({\bf u}})\rangle
\end{eqnarray}
The above scalar product can be computed as a usual scalar product in
the inhomogeneous model. In particular, the last expression, before
taking the limit, is a usual scalar product with one vector
on-shell and the other off-shell, which can be computed using
\eqref{defSuv}.  After replacing the symmetric sums with their values
$\ssum_k({\bf u})$ we get the matrix \eqref{defOhat}, with $p_{\bf
u}(v)$ replaced by $p^{\rm BES}_{\bf u}(v)$.  After taking the limit
${\bf v}\to {\bf {u}}$, we obtain
\begin{eqnarray}
_{\rm BES}\< {\bf u}|{\bf u}\>_{\rm BES} &=& {\prod_{j=1}^N {2\,
d^2(u_j) }\over \prod_{j<k} (u_j-u_k)^2}
\det_{jk}\(\frac{(1-\delta_{jk})}{u_{jk}^2+1}+\d_{u}p^{\rm BES}_{{\bf
u}}(u)\Big|_{u_j}\delta_{jk}\)\;.
 \end{eqnarray}
This formula looks different from the conjecture for the norm with the
dressing phase given by Gromov and Vieira \cite{GV}.

\subsection{Morphism and Theta-Morphism}
\label{sec:Sandtheta}

In this section we analyze the relation between the unitary
S-transformation and the ``theta-morphism'' introduced by Gromov and
Vieira \cite{Gromov:2012vu,GV}.  The idea of Gromov and Vieira was to
construct the states of the BDS long-range model by acting with a
differential operator, which they called theta-morphism, on the states
of the inhomogeneous model.  We find that a purely differential
operator cannot realize the morphism property, see below.  The failure
to fulfill the morphism property results in the cross-terms of
\cite{GV}.  Instead, we introduce the morphism associated to the
S-operator via
\begin{align} \label{thetmodef}
\mon^\BDS(u) =\Sop\,\mon(u;{\bm \theta}^\BDS)\,\Sop^{-1}
\equiv \SD_{\theta}\mon(u;{\bm \theta})|_{\theta=0}\;.
\end{align}
This definition  has the following advantages:
\begin{itemize}
\item Unlike the Gromov-Vieira theta-morphism, the inhomogeneity
translation plus the unitary transformation amounts to {\it an exact
morphism} of the Yangian algebra.  It works not only on Bethe vectors,
but on arbitrary elements of the monodromy matrix.
%
%
\item The S-transformation produces the required boundary terms and
therefore it is free of the ``cross-term'' issue which complicates the
computation in \cite{GV}.
\item The relation to boost deformations and the inhomogeneous corner
transfer matrix indicates a natural extension to higher orders.
\end{itemize}
We find that, up to terms of order 
$\order{g^3}$, the action of
$\SD_\theta$ on (an arbitrary matrix element of) the monodromy matrix
$\mon$ amounts to
\begin{align} \label{thetmo}
&\SD_{\theta}\mon(u;{\bm \theta})|_{\theta=0}\\
&\quad\left.\equiv\mon-\frac{g^2}{2}\sum_{k=1}^L\D_k^2{\mon}
-g^2\sum_{k=1}^L\D_k{\mon}(\p_{k,k+1}+\delta_{k,L}
\QQ_2)+{g^2}\left[\frac{\QQ_2^2}{2}+i\QQ_3+\p_{1L}\QQ_2, \mon\right]
\right|_{\theta=0}\;.\no
\end{align}
To avoid cumbersome notations, in this section and below, we drop the
indices on $\charge_r^\ups{0}=\charge_r^\SR$ to
denote the leading order charges simply by $\charge_r$, and we have%
\footnote{Our definition of $\D_k$ differs from that of \cite{GV} by a
factor of $i$.}
\begin{equation}
\D_k\equiv i(\d_{k}-\d_{k+1})= i(\d_{\theta_k}-\d_{\theta_{k+1}})\,.
\no
\end{equation}
The operator $\SD_\theta$ differs from the theta-morphism of Gromov
and Vieira \cite{GV} given by
\begin{eqnarray}
\CD_\theta^\GV=1-\frac{g^2}{2}\sum_{k=1}^L\D_k^2+\CO(g^4),
\end{eqnarray}
by the last two terms in the second line of equation \eqref{thetmo}.
These two extra terms account for the cross-terms in \cite{GV} and
they insure that the morphism property is exact
\begin{eqnarray}
\SD_{\theta}\(\mon(u_1;{\bm \theta})\mon(u_2;{\bm
\theta})\)|_{\theta=0}=\SD_{\theta}\mon(u_1;{\bm
\theta})|_{\theta=0}\; \SD_{\theta} \mon(u_2;{\bm
\theta})|_{\theta=0}\;.
\end{eqnarray}
On the Bethe vectors, the action of the operator $\SD_\theta$
reduces to
 \begin{align}
\uketz_\BDS
=\Sop |\raps{u};{\bm\theta}^\BDS\>= \SD_{\theta} |\raps{u};{\bm
\theta}\>|_{\theta=0}
=\(\CD_\theta^\GV+\frac{g^2}{2}\({\QQ_2^2}+2i\QQ_3+2\p_{1L}\QQ_2\)\)
|\raps{u};{\bm \theta}\>|_{\theta=0}\;.
\end{align}
To obtain this expression, we use that $\sum_k\D_k=0$ and that the
vacuum eigenvalues of $\QQ_2$ and $\QQ_3$ are zero.  If the Bethe
vectors are on-shell, the charges $\QQ_2$ and $\QQ_3$ become numbers
and we obtain
 \begin{align}
\uketz_\BDS&=\big[\CD_\theta^\GV+{g^2}\(\half{E_2^2}+E_2+iE_3-\h_{1L}\charge_2\)\big]
|\raps{u};{\bm \theta}\>|_{\theta=0}\nonumber \\
&=\big[1+g^2\(\half E_2^2+E_2+iE_3\)\big]\uketz_\BDS^\GV\;.
\end{align}
We thus see that our eigenvectors differ from those of Gromov and
Vieira by a state-dependent factor.  The imaginary contribution does
not affect the norms, while the real part changes the normalization.
The scalar products of two arbitrary Bethe states, on-shell or
off-shell, is:
\begin{eqnarray}
\label{eq:morstate}
_\BDS\uubra \raps{v}\>_\BDS =\SD_\theta \<\raps{u};{\bm \theta}|
\raps{v};{\bm \theta}\>|_{\theta=0}=\CD_\theta^\GV \<\raps{u};{\bm
\theta}| \raps{v};{\bm \theta}\>|_{\theta=0}\;.
\end{eqnarray}
%

Let us now sketch the derivation of the expression \eqref{thetmo}.
More details are  given in \appref{sec:pdrel}.  First, we account
for the shift in the inhomogeneities by
 \begin{align}
 \label{eq:pdgv}
\mon(u;{\bm \theta}^\BDS)=\exp\(\sum_{j=1}^L\theta_l^\BDS\,
\partial_{\theta_l}\)\;\mon(u;{\bm \theta})|_{\theta=0}\;.
\end{align}
The second ingredient is to transform the action of the permutation
operators contained in $\Sop$ into derivatives.  The simplest ones
were given in \cite{Gromov:2012vu},
 \begin{align}
[\ham_k , \mon(u)]=(\charge_2\,\delta_{k,L}-\D_k)\,\mon(u;{\bm
\theta})|_{\theta=0} \;.
\end{align}
At higher orders in the expansion we have to deal with multiple
permutations.  The case of non-overlapping permutations is simple,
 \begin{align}
[\ham_l,[\ham_k ,
\mon]]=(\charge_2\,\delta_{l,L}-\D_l)(\charge_2\,\delta_{k,L}-\D_k)\,\mon
\;, \quad |l-k|>1\;.
\end{align}
For overlapping permutations in the bulk, $k\neq L-1,L$, we obtain,
 \begin{eqnarray}
 \label{eq:pdcomm}
[\hamone_k ,
\mon]&=&\half\(\D_k^2-\D_{k+1}^2\)\mon+\D_k\mon\p_{k,k+1}-\D_{k+1}\mon\p_{k+1,k+2}\;,\\
 \label{eq:pdacomm}
[\hammone_k ,
\mon]]&=&\half\(\D_k^2+\D_{k+1}^2\)\mon+\D_k\mon\p_{k,k+1}+\D_{k+1}\mon\p_{k+1,k+2}\\
\no &&+2\D_k\D_{k+1}\mon +
2\D_k\mon\p_{k+1,k+2}+\D_{k+1}\mon\p_{k,k+1},
\end{eqnarray}
where $\hamone_k =\[\ham_k,\ham_{k+1}\]$ and $\hammone_k
=\{\ham_k,\ham_{k+1}\}$.  When $k= L-1,L$ the action in
\eqref{eq:pdcomm} has to be supplemented with boundary terms,
 \begin{eqnarray}
 \label{eq:pdbound}
\left[\hamone_{L-1} , \mon\right]&=&[\hamone_{L-1} , \mon]_{\rm
bulk}+[\delta_{\rm bound}-2i\charge_3,\mon] \\
\left[\hamone_{L} , \mon \right]&=&[\hamone_{L} , \mon]_{\rm bulk} -[
\delta_{\rm bound},\mon]\;, \qquad \delta _{{\rm bound}}
=\(\half{\QQ^{2}_2}+i\QQ_3+\p_{1L}\QQ_2\)\;.  \no
\end{eqnarray}
These expressions, together with the action of overlapping $\D_k$ and
$\ham_l$ that are derived in \appref{sec:pdrel}, are all we need to
obtain \eqref{thetmo}, provided that we choose $\sone_0=\stwo_0=0$.
Let us notice that the expressions \eqref{eq:pdcomm, eq:pdacomm,
eq:pdbound} obey the Leibniz rule, {\it e.g.}
  \begin{eqnarray}
[\hamone_k , \mon_1\mon_2]=\mon_1 [\hamone_k , \mon_2]+[\hamone_k ,
\mon_1]\mon_2\;.
\end{eqnarray}
We can thus safely replace $\mon$ by any product of monodromy matrices
in all the commutators above.  This feature is at the origin of the
morphism property.  \vskip5pt

\section{Three-Point Function of $\sutwo$ Fields Beyond Tree Level
  }

\def\tG{\tilde G}
\newcommand{\sofour}{\alg{so}(4)}
\newcommand{\sosix}{\alg{so}(6)}

%

We consider operators which have definite conformal dimensions
$\Delta^{(1)}$, $\Delta^{(2)}$ and $\Delta^{(3)}$.  The three-point function of
three renormalized operators $\CO^{(1)},\CO^{(2)}$ and $\CO^{(3)}$ in
the ${\cal N}=4$ gauge theory is almost entirely fixed by conformal
symmetry,
\begin{eqnarray}
\langle\CO^{(1)}(x_1)\CO^{(2)}(x_2)\CO^{(3)}(x_3)\rangle
=\frac{N_c^{-1}\ \sqrt{L^{(1)} L^{(2)} L^{(3)}}\ \
C_{123}(g^2)}{|x_{12}|^{\Delta^{(1)}+\Delta^{(2)}
-\Delta^{(3)}}|x_{13}|^{\Delta^{(1)}+\Delta^{(3)}-\Delta^{(2)}}|x_{23}|^{\Delta^{(2)}
-\Delta^{(3)}-\Delta^{(1)}}}\;.
\end{eqnarray}
The only part which remains to be evaluated is the scheme-independent
structure constant
\begin{equation}
C_{123}(g^2) = \sum _{k\ge 0}C_{123}^{[k]} \ g^{2k}.
\end{equation}
\paragraph{A particular embedding of $\sutwo$ fields in the  $\sofour$ sector.}
The structure constant depends on the quantum numbers of the three
$\sutwo$ fields.  Each $\sutwo$ type field is characterized by an
on-shell Bethe state in the XXX chain, as well as by the embedding
of the $\sutwo$ sector in $\sosix$.\footnote{ The choice of the
$\sutwo$ sector is determined by a set of global coordinates (angles).
One can argue that the dependence of the three-point function on the
global angles factorizes \cite{Didina-globalangles}, but we will not
discuss this issue here.} The correlation function considered here, as
well as in \cite{EGSV:Tailoring2,EGSV,GSV,GV},
corresponds to a particular choice of the $\sutwo$ sectors to which
the three operators belong.  With this choice, the three $\sutwo$
operators are traces of two complex bosons
\begin{eqnarray}
\la{3opers}
\begin{aligned}
\CO^{(1)} \in\{Z,X\}, \quad \CO^{(2)} \in\{\bar Z,\bar X\}, \quad
\CO^{(3)} \in\{Z,\bar X\},
\end{aligned}
\end{eqnarray}
for example $\CO^{(1)}(x_1)\sim \tr ZZXXX\ldots XXZX(x_1)$.  Chosen in
this way, the three $\sutwo$ operators belong to the $\sofour \sim
\sutwo_{_R} \oplus \sutwo_{_L}$ subsector of $\sosix$.  The two
$\sutwo$ groups act by left and right multiplication of the complex
matrix
\begin{eqnarray}
 \left(\begin{array}{cc} \Phi_1+i \Phi_2 & \Phi_3 +i\Phi_4 \\ -\Phi_3
 +i\Phi_4 & \Phi_1-i\Phi_2\end{array}\right)= \left(\begin{array}{cc}
 Z & \, X \\ - \bar X & \bar Z\end{array}\right) .
\end{eqnarray}
The operators $\CO^{(1)}$ and $\CO^{(2)}$ belong to the $\sutwo_R$
sector, while the operator $\CO^{(3)}$ belongs to the $\sutwo_L$
sector.
%
Under the right multiplications, $\{Z,X\}$ and $\{-\bar X, \bar Z\}$
transform as $\sutwo$ doublets.  Under the left multiplications, the
pairs of fields $\{Z, -\bar X\}$ and $\{X, \bar Z\}$ transform as
$\sutwo_R$ doublets \cite{Kazakov:2004qf}.
%
%

The spin-chain lengths and the magnon numbers of the three states are
related to the two $\sofour$ charges by
\begin{eqnarray}
\begin{aligned}
L^{(1)}&= +J_1^{(1)} + J_2^{(1)}, \quad N^{(1)}= +J_2^{(1)},
\\
L^{(2)}&= -J_1^{(2)} - J_2^{(2)}, \quad N^{(2)}= - J_2^{(2)},
\\
L^{(3)}&= +J_1^{(3)} - J_2^{(3)}, \quad N^{(3)}= -J_2^{(3)}.
\end{aligned}
\end{eqnarray}
In order to have a non-zero three-point function, the sum of the
$R$-charges of the three operators must be zero:
\begin{eqnarray}
\sum_{a=1}^3 J_1^{(a)}= \sum_{a=1}^3 J_2^{(a)}=0.
\end{eqnarray}
The conservation of charges gives
\begin{eqnarray}
N^{(1)}=N^{(2)}+N^{(3)}, \qquad 2N^{(3)} =L^{(1)}+ L^{(3)}-L^{(2)}.
\no\end{eqnarray}
The tree-level structure constant for the case when $\CO^{(2)}$ is a
BPS field was computed in \cite{EGSV,GSV}, and in the general case in
\cite{SL}.  Here we will apply the formalism of this paper to compute
the one-loop result, previously obtained in \cite{Gromov:2012vu, GV}.
Our computation agrees with that of \cite{Gromov:2012vu, GV}, giving
the result in a concise and elegant form.  The classical limit of the
one loop result is taken in Section \ref{sec:classical limit}.

\subsection{Three-Point Functions at One Loop}
 \label{oneloop3pf}
 
\begin{figure}[t]
     \begin{center}
     		\includegraphics[scale=0.45]{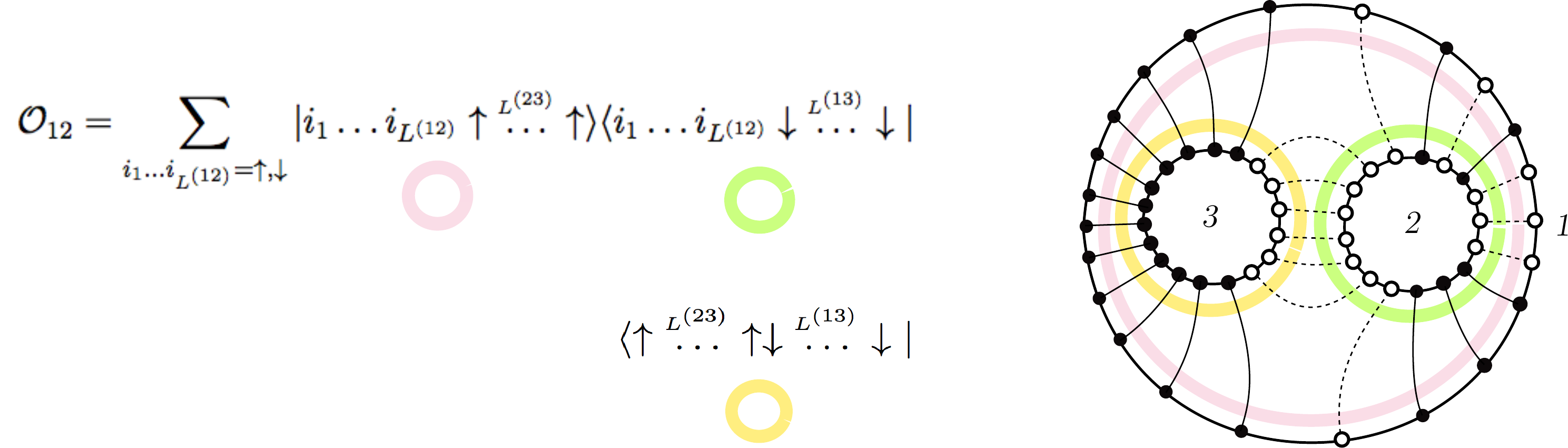}
	  \bigskip \caption{\small Structure of contractions captured
	  by the operator $\CO_{12}$ defined by (\ref{defO12}) and by
	  the vector $ \langle \uparrow\overset{^{L^{(23)}}}\ldots
	  \uparrow\downarrow\overset{^{L^{(13)}}}\ldots \downarrow|$ .
	  }
	\label{fig:ET}
	\end{center}
\end{figure}

In the language of integrable spin chains one identifies
$$ \tr ZZXXX\ldots
XXZX\;\quad\leftrightarrow\quad|\uparrow\uparrow\downarrow\downarrow\downarrow\ldots
\downarrow\downarrow\uparrow\downarrow\rangle,$$
 and the three operators are in correspondence with three eigenstates
 $\uket$, $\vket$ and $\wket$ of the dilatation operator.  The
 structure constant is expressed as \cite{EGSV}
\begin{eqnarray}
\label{c123gen}
C_{123}(g^2)&=& \frac{\langle {\bf u}^{(1)}, {\bf u}^{(2)}, {\bf
u}^{(3)} \rangle }{ \(\Normu\Normv\Normw\)^{1/2} },
\end{eqnarray}
with the cubic vertex in the numerator given by \cite{GV}\footnote{We
prefer to use the formulation from \cite{GV} for the three-point
functions, since it is free from the complications which arise when
considering ``flipping''.}
 \begin{eqnarray}
 \label{tpfho}
 \langle {\bf u}^{(1)}, {\bf u}^{(2)}, {\bf u}^{(3)} \rangle &\equiv &
 \vbra\;\mathbb{I}_{2}\; \CO_{12}\;\mathbb{I}_{1}\; \uket\ \langle
 \uparrow\overset{^{L^{(23)}}}\ldots
 \uparrow\downarrow\overset{^{L^{(13)}}}\ldots \downarrow|
 \;\mathbb{I}_{3} \; \wket \;.
\end{eqnarray}
The operator $\CO_{12}$ captures the particular structure of the
contractions of the elementary fields at the splitting point, cf.\
\figref{fig:ET}:
\begin{equation}
 \label{defO12}
\CO_{12}=\sum_{i_1\ldots i_{L^{(12)}}=\uparrow,\downarrow} |i_1\ldots
i_{L^{(12)}}\uparrow\overset{^{L^{(23)}}}\ldots \uparrow \rangle
\langle i_1\ldots i_{L^{(12)}} \downarrow\overset{^{L^{(13)}}}\ldots
\downarrow|\;.
\end{equation}

\bigskip

The insertions $\mathbb{I}_{1,2,3}$ represent the Hamiltonian
insertions and they have to be determined by perturbative gauge theory
computations.  Up to one-loop order they have been computed in
\cite{Okuyama:2004bd,Alday:2005nd}
 \begin{eqnarray}
 \label{hamins}
\no \mathbb{I}_1&=&1-g^2(\ham_{L^{(12)}}^{(1)}+\ham_{L^{(1)}}^{(1)})+\ldots\;, \\
\mathbb{I}_2&=&1-g^2(\ham_{L^{(12)}}^{(2)}+\ham_{L^{(2)}}^{(2)})+\ldots\;,\\
\mathbb{I}_3&=&1-g^2(\ham_{L^{(31)}}^{(3)}+\ham_{L^{(3)}}^{(3)})+\ldots\;.  \no
\end{eqnarray}
\def\sla{\slash\!\!  \!} \def\IR{{\mathbb{R}}} \def\IC{{\mathbb{C}}}
\def\IZ{{\mathbb{Z}}}
The knowledge of the Hamiltonian insertions at higher order is one of the
main obstructions in computing the three-point function at two loop
and higher.  The other obstructions are to take into account the
dressing factor and the wrapping contributions.  Let us show now how
to compute the three-point function at one loop.  First, we are going
to choose carefully the inhomogeneities corresponding to the three
operators.  Since we are splitting and joining the chains, it is
convenient to have the same values of the inhomogeneities for the
pieces that we are matching, {\it i.e.}
  \begin{eqnarray}
\thu=\thuv\cup\thuw, \quad \thv=\thuv\cup\thvw,\quad \thw=\thuw\cup
\thvw.  \nonumber
\end{eqnarray}
Moreover, we are going to choose the values of the three groups of
inhomogeneities as follows
 \begin{eqnarray}
 \la{thetaBDS} \theta^{(ab)}_l=2g \sin \frac{2\pi l}{L^{(ab)}}\;,
 \quad l=1,\ldots ,L^{(ab)}\;.
\end{eqnarray}
This choice is compatible with the following values of the
coefficients defined in \eqref{eq:sone} and \eqref{eq:rhosol}
 \begin{eqnarray}
&\sone^{(ab)}_0=\sone^{(ab)}_{L^{(ab)}}=0\;, \quad
\stwo^{(ab)}_0=\stwo^{(ab)}_{L^{(ab)}}=0\; &\no \\
&\stwo^{(ab)}_1=\stwo^{(ab)}_{L^{(ab)}+1}=2g^2\;.&
\end{eqnarray}

To compute the cubic vertex, we shall split each of the
$\Sop^{(a)}$-operators into pieces which commute with the insertions and
among themselves, $\Sop^{(ab)}$, and pieces which do not commute with
the insertions and the rest, $\delta \Sop^{(a)}$,
 \begin{eqnarray}
\Sop^{(1)}=\Sop^{(12)}\,\Sop^{(13)}\,\delta \Sop^{(1)}\;, \quad
\Sop^{(2)}=\Sop^{(12)}\,\Sop^{(23)}\,\delta \Sop^{(2)}\;, \quad
\Sop^{(3)}=\Sop^{(13)}\,\Sop^{(23)}\,\delta \Sop^{(3)}\;.
\end{eqnarray}
For the one-loop three-point function, this splitting is done as
follows
 \begin{eqnarray}
\Sop^{(ab)}=\exp\(i\sum_{k=1}^{L^{(ab)}-1}\sone_k^{(ab)}
\ham_k^{(ab)}-\frac{1}{2}\sum_{k=2}^{L^{(ab)}-1}\stwo_k^{(ab)}\hamone_{k-1}^{(ab)}+\ldots\)
\;,
\end{eqnarray}
\vskip-11pt
 \begin{eqnarray}
\delta
\Sop^{(1)}&=&1-g^2(\hamone_{L^{(1)}}^{(1)}+\hamone_{L^{(12)}}^{(1)})+\ldots\;,
\no \\
\delta
\Sop^{(2)}&=&1-g^2(\hamone_{L^{(2)}}^{(2)}+\hamone_{L^{(12)}}^{(2)})+\ldots\;,
\no \\
\delta
\Sop^{(3)}&=&1-g^2(\hamone_{L^{(3)}}^{(3)}+\hamone_{L^{(13)}}^{(3)})+\ldots\;.
 \label{deltaS}
\end{eqnarray}
Inserting the split expressions into the equation (\ref{tpfho}), one
can see that the $\Sop^{(ab)}$ parts cancel, $\Sop^{(12)}$ because it
commutes with $\CO_{12}$, and $\Sop^{(13)}$ and $\Sop^{(23)}$ because
they act on totally symmetric pieces.  One is left with
\begin{eqnarray}
\label{tpfhoin}
 \langle  {\bf u}^{(1)},  {\bf u}^{(2)},  {\bf u}^{(3)} \rangle 
 &=& {\bf involved}\times {\bf simple}, 
 \\
 {\bf involved} &=& \vbrain \delta \Sop_{2}^{-1} \;\mathbb{I}_{2}\;
 \CO_{12}\;\mathbb{I}_{1}\; \delta \Sop_{1}\; \uketin\ \no \\
{\bf simple} &=& \langle \uparrow\ldots \uparrow\downarrow\ldots \downarrow|
\;\mathbb{I}_{3}\;  \delta \Sop_{3} \; \wketin 
\;, \no
\end{eqnarray}
where we used the notations from \cite{GV} to facilitate the comparison.
The leading (tree-level) contribution to the correlator can be
computed by fixing the inhomogeneities and using the freezing method
of \cite{Foda:2011rr}.  The basic idea of the freezing trick is to get a
sequence of down spins by synchronizing the rapidities of a number of
magnons with the same number of inhomogeneities.  The simplest example
is
 \begin{eqnarray}
 |\uparrow\ldots \uparrow\downarrow\ldots
 \downarrow\rangle=\B(z_1^{(23)})\ldots
 \B(z_{L^{(23)}}^{(23)})|\Omega\rangle= |\raps{z^{(23)};{\bm
 \theta}^{(3)}}\rangle\quad {\rm with} \quad
 z_k^{(ab)}\equiv\theta_k^{(ab)}-\ihalf\;,
\end{eqnarray}
so that we obtain  
\begin{eqnarray}
\label{simplesp}
\langle \uparrow\ldots \uparrow\downarrow\ldots \downarrow\; \wketin
=\langle\raps{z^{(23)};{\bm \theta}^{(3)}}\wketin\; ,
\end{eqnarray}
and, similarly, 
\begin{eqnarray}
\label{involvedsp}
\vbrain \CO_{12}\; \uketin =\langle\raps{u^{(2)}\cup {\bf
z}^{(13)};{\bm \theta}^{(1)}}\uketin\;.
\end{eqnarray}
Both of these expressions are scalar products of a Bethe eigenstate with
an off-shell vector, and as such they can be expressed in terms of
Slavnov determinants.  Let us point out that the two scalar products
\eqref{simplesp, involvedsp} are no longer symmetric under the
permutation of inhomogeneities ${\bm \theta}^{(3)}$ and ${\bm
\theta}^{(1)}$, respectively.  Instead, as we show in Appendix
\ref{reduction} (see also \cite{SL}) the expression \eqref{simplesp}
does not depend at all on the group of inhomogeneities ${\bm
\theta}^{(23)}$ but only on ${\bm \theta}^{(13)}$ and, similarly,
\eqref{involvedsp} does not depend on ${\bm \theta}^{(13)}$ but only
on ${\bm \theta}^{(12)}$.

Now let us proceed with the calculation of the one-loop corrections. 
The norms are computed as
\begin{equation}
\label{normestheta}
  \langle {\bf u}^{(a)}  |
 {\bf u}^{(a)}  
 \rangle=
  \langle {\bf u}^{(a)} , {\bm \theta ^{(a)}}|
 {\bf u}^{(a)} , {\bm \theta ^{(a)} 
 }\rangle ,
\end{equation}
where the inhomogeneities on the r.h.s.\ are given by the values
(\ref{thetaBDS}). 
The corrections to the cubic vertex coming from the Hamiltonian
insertions and the insertions of the $\Sop$-operators can be easiest
evaluated by transforming them into derivatives.  For this purpose, we
use the relations
 \begin{align}
 \ham_{L^{(a)}}^{(a)} \uthi
 &=(\QQ_2^{(a)}-\D_{L^{(a)}}^{(a)})\uthi\;,\no\\ 
  \ham_{L^{(ab)}}^{(a)} \uthi
  &= -\D_{L^{(ba)}}^{(a)} \uthi\;,\no\\
   \hamone_{L^{(ba)}}^{(a)}\uthi
   &=\(\half(\D_{L^{(ba)}}^{(a)2}-\D_{L^{(ba)}+1}^{(a)2})
   +(\D_{L^{(ba)}}^{(a)}-\D_{L^{(ba)}+1}^{(a)})\) \uthi\;, \no \\ 
     \hamone_{L^{(a)}}^{(a)} \uthi
     &=\(\half(\D_{L^{(a)}}^{(a)2}
     -\D_{1}^{(a)2})+(\D_{L^{(a)}}^{(a)}-\D_{1}^{(a)})
     -\delta _{{\rm bound}}^{(a)}\) \uthi\;, \label{boundaryterms}
\end{align}
with
\begin{align}
\delta _{{\rm bound}}^{(a)}
\uthi
&=\(\half{\QQ^{(a)2}_2}+i\QQ_3^{(a)}+\p_{1L^{(a)}}\QQ_2^{(a)}\)
\uthi \no\\ \no %
&=
\(i\EE_3^{(a)}-\half{\EE^{(a)2}_2}+\EE_2^{(a)}(1+\D_{L^{(a)}}^{(a)})\)\uthi\;.
\end{align}
Above, it is understood that the inhomogeneities are set to zero after
acting with the derivatives.  After performing the algebra, see
appendix \ref{app:threepcal}, we obtain for the factor \mbox{\bf
simple}:
\begin{eqnarray}
\label{simplefin}
{\bf simple} &=& \langle\raps{z^{(23)};{\bm \theta}^{(3)}}|
\;\mathbb{I}_{3}\; \delta \Sop_{3} \; \wketin 
\\
&=& \langle\raps{z^{(23)};{\bm \theta}^{(3)}} \wketin \no \\
  &&+\,
  g^2\(\partial_1^{(3)}\partial_2^{(3)}-i\EE_2^{(3)}\partial_1^{(3)}
  +i\EE_3^{(3)}-\frac{1}{2}\EE_2^{(3)2}\)\,
  \langle\raps{z^{(23)};{\bm \theta}^{(3)}} \wketin
  \Big|_{\theta=0}\; , \no
\end{eqnarray}
and for the factor \mbox{{\bf involved}}:
\begin{eqnarray}
\label{involvedfin}
 {\bf involved} &=& \vbrain\, \delta \Sop_{2}^{-1}\,\mathbb{I}_{2}\;
 \CO_{12}\;\mathbb{I}_{1}\; \delta \Sop_{1}\; \uketin \no \\
 &=& \langle\raps{u^{(2)}\cup {\bf z}^{(13)};{\bm
 \theta}^{(1)}}\uketin \\
 &&+\,
 {g^2}\(\partial_1^{(1)}\partial_2^{(1)}-i\delta\EE_2\partial_1^{(1)}
 +i\delta\EE_3-\frac{1}{2}\delta\EE_2^2\)
 \langle\raps{u^{(2)}\cup {\bf z}^{(13)};{\bm \theta}^{(1)}}\uketin
 \Big|_{\theta=0}\;, \no
\end{eqnarray}
where we have used the notation $\delta\EE_r=\EE_r^{(1)}-\EE_r^{(2)}$.
In the main terms in the equations \eqref{simplefin} and
\eqref{involvedfin} the inhomogeneities are set to their BDS values
\eqref{thetaBDS}, while in the last term they are set to zero.  Since
here we are interested only in the one-loop order, it is not important
whether we set the inhomogeneities to zero or not, after taking the
derivatives.  The two expressions \eqref{simplefin} and
\eqref{involvedfin} look similar, the first being a particular limit
of the second.  When computing the three-point function, only the
modulus square of the overlaps is relevant, since the phase can be
always changed by a redefinition of the states.  By this argument, we
should drop the imaginary part in the above expressions, {\it e.g.}
the terms containing $\EE_3^{(a)}$ with $a=1,2,3$.  Gromov and Vieira
argued in \cite{GV} that the term containing
$i\delta\EE_2\partial_1^{(1)}+\frac{1}{2}\delta\EE_2^2$ is also
imaginary.  By the same argument, the term containing
$i\EE_2^{(3)}\partial_1^{(3)}+\frac{1}{2}\EE_2^{(3)2}$ should be
imaginary, too.

Let us check now that our results are compatible with those of
\cite{Gromov:2012vu, GV}.  Written in our notations, their result is
given by\footnote{We neglected the factors $d(u_j)
d(v_j)$ because they cancel with the norms in the denominator, and we
dropped the superscript from $\caA^-_{{\bf u},{\bm \theta}}$ to use just
$\caA_{{\bf u},{\bm \theta}}$.} (see  \eqref{scalarpr} and Appendix \ref{reduction})
\begin{align}
{\bf simple}
&=\(1-\frac{g^2}{2}\sum_{k=1}^{L^{(3)}} \D_k^{(3)2}\)\;\caA_{{\bf
u}^{(3)},\,{\bm \theta}^{(13)}}+\ldots,\label{eq:simp1}
\\
{\bf involved}
&=\(1-\frac{g^2}{2}\sum_{k=1}^{L^{(1)}} \D_k^{(1)2}\)\;\caA_{{\bf
u}^{(1)}\cup {\bf u}^{(2)},\,{\bm
\theta}^{(12)}}+\ldots,
\label{eq:inv1}
\end{align}
where the dots on the r.h.s.\ stand for terms which are supposed to be imaginary or
of higher order in $g$. The functionals $\caA _{{\bf u}^{(3)},\,{\bm
\theta}^{(13)}}$ and $\caA _{{\bf u}^{(1)}\cup {\bf u}^{(2)},\,{\bm
\theta}^{(12)}}$ are not symmetric in all the variables ${\bm
\theta}^{(3)}$ and ${\bm \theta}^{(1)}$, respectively, since they do
not depend at all on ${\bm \theta}^{(23)}$ and ${\bm \theta}^{(13)}$,
respectively.  This means that the action of the derivatives does not
simply amount to the substitution of the inhomogeneities by the BDS
values \eqref{thetaBDS} (as it would be the case for symmetric
functionals).  The symmetry default can be cured by rewriting
\eqref{eq:simp1,eq:inv1} as symmetric differential operators in the
variables ${\bm \theta}^{(13)}$ and ${\bm \theta}^{(12)}$ acting on
the functionals $\caA$:
\begin{align}
{\bf simple} &=\(1-\frac{g^2}{2}\sum_{k=1}^{L^{(13)}}
\D_k^{(13)2}+\frac{g^2}{2}\(\D_{L^{(13)}}^{(13)2}-D_{L^{(3)}}^{(3)2}
-D_{L^{(13)}}^{(3)2}\)\)\;\caA
_{{\bf u}^{(3)},\,{\bm \theta}^{(13)}}+\ldots\no\\
&=\caA_{{\bf u}^{(3)},\,{\bm
\theta}^{(13)}}+{g^2}\partial_1\partial_{L^{(13)}}\;\caA _{{\bf
u}^{(3)},\,{\bm \theta}^{(13)}}+\ldots
\\ \no {\bf involved} &=\(1-\frac{g^2}{2}\sum_{k=1}^{L^{(12)}}
\D_k^{(12)2}+\frac{g^2}{2}\(\D_{L^{(12)}}^{(12)2}-D_{L^{(1)}}^{(1)2}
-D_{L^{(12)}}^{(1)2}\)\)\;\caA
_{{\bf u}^{(1)}\cup {\bf u}^{(2)},\,{\bm \theta}^{(12)}}+\ldots\no\\
&=\;\caA  _{{\bf u}^{(1)}\cup
{\bf u}^{(2)},\,{\bm
\theta}^{(12)}}+g^2\partial_1\partial_{L^{(12)}}\;\caA  _{{\bf
u}^{(1)}\cup {\bf u}^{(2)},\,{\bm
\theta}^{(12)}}+\ldots
\end{align}
These are exactly the results in \eqref{simplefin} and
\eqref{involvedfin}, up to the terms supposed to be imaginary.  We
conclude that our results agree with those of \cite{Gromov:2012vu,
GV}.  The advantage of our formulation is that we can
straightforwardly take the classical limit, while in the formulation
of \cite{GV} this limit is hardly possible to take, due to the
complexity of the answer.  Knowing the insertions at two loops would
allow to compute the two-loop correlation function in the same manner
as above.  If the two-loop insertions are restricted to a few sites
around the splitting points, then the corresponding correction would
be given by terms containing four derivatives with respect to
inhomogeneities around the splitting points.  This kind of
contribution is subdominant in the Frolov--Tseytlin limit, as we show
in Section \ref{subsec:compKK}.  We hope to be able to report on this
point in a separate work.
  
\section{Three-Point Functions in the Semi-Classical Limit  }
 \label{sec:classical limit}

In this section we evaluate the one-loop three-point function obtained in the previous section, in the limit of three heavy operators,
also called semi-classical or thermodynamical limit. 
We send $N, L\to\infty$  but the mode numbers are kept finite.
In this limit, which is interesting from the point of view of
comparison with string theory, the Bethe roots arrange themselves into
a small number of macroscopic strings \cite{Sutherland,Kazakov:2004qf}.  The solution of the Bethe
equations in this limit is described by a Riemann surface with a
finite number of cuts.%
 \footnote{In \cite{Kazakov:2004qf}, the spectral parameter was
rescaled as $u = L x$ with $x\sim 1$.  We will not introduce a new
rescaled variable, but will keep in mind that $u\sim L$.} 
The general
finite zone solution in the $\sutwo$ sector is described by a
hyperelliptic complex curve \cite{Kazakov:2004qf}.


\subsection{Scalar Products and Norms in the Semi-Classical Limit}

An $N$-magnon Bethe state with magnon rapidities ${\bf u} = \{u_1,
\dots, u_N\}$
%
is characterized by its quasi-momentum $p(u)$, which is defined modulo
$\pi $ by (\ref{defquasimom}).
The quasi-momentum of an on-shell Bethe state satisfies $N$ conditions
\begin{eqnarray}
e^{2i p(u)}\Big|_{u=u_j}= -1 \quad (j=1,2,,\dots, N),
\end{eqnarray}
which are equivalent to the Bethe equations for the roots $ {\bf u}$.
 In the thermodynamical limit the quasi-momentum is given by
\begin{eqnarray}
\la{inhomquasim} p(u) \simeq G_{\bf u} (u) - \hf G_ { \bm {\theta }
}(u) + \pi n,
\end{eqnarray}
 with $ G_{\bf u} $ and $G_{ \bm {\theta } }$ being the resolvents for
 the magnon rapidities and the inhomogeneities,
\begin{eqnarray}
\label{defresu}
 G_{\bf u} = \partial \log Q_{\bf u} , \qquad G_ { \bm {\theta } }=
 \partial \log Q_{\bm \theta} .
\end{eqnarray}
The resolvent corresponding to the distribution of the rapidities
(\ref{eq:valimp}) is
%
\begin{eqnarray}
\la{resthth} G_ { \bm {\theta } } (u) = {L \over \sqrt{u^2-4g^2 }}.
\end{eqnarray}

The semi-classical limit of the scalar product and the norm follow from
that of the functional (\ref{defcaA}) \cite{Kostov:2012jr,SL}\footnote{
The  two expressions differ by a phase factor.
This also follows from the functional relation for the dilogarithm
$ \Li({1\over  \omega } )= - \Li(\omega) - {\pi^2\over 6} -
{1\over 2} \log^2(-\omega).$
}
\begin{eqnarray}
\label{classicalA}
\log \caA^\pm  _{{\bf u}, {\bm \theta}} =\pm  \oint \limits _ {\CC}
\frac{du}{2\pi } \ \text{Li}_2\big( e^{\pm  i G_{\bf u}(u) \mp  i G_{\bm
\theta} (u) } \big)
     + \CO(1) , \qquad L\to\infty , \ \ N/L \sim 1,
\end{eqnarray}
where the contour $\CC$ surrounds the rapidities ${\bf u}$ and leaves outside ${\bm
\theta}$.
As a consequence, the scalar product (\ref{scalarpr}) is
expressed through the sum of the two quasi-momenta:
   \begin{eqnarray}
   \label{scalarprcl}
\log\,   \langle {\bf u} ^{(1)} ; {\bm \theta} |{\bf u} ^{(2)}; {\bm \theta}
  \rangle = \oint \limits _ {\CC^{(1)} \cup\ \CC^{(2)} }
  \frac{du}{2\pi } \ \text{Li}_2\big( e^{ i p ^{(1)}(u) + i p
  ^{(2)}(u) } \big) ,
\end{eqnarray}
where the contour $\CC^{(a)}$ surrounds the set of rapidities ${\bf
u}^{(a)}$ and leaves outside the set of the inhomogeneities ${\bm
\theta}$.  In the classical limit the derivative of the quasimomentum
$p^{(a)} $ is defined on a four-sheeted Riemann surface and the
discrete set of points ${\bf u}^{(a)}$ condenses into a set of cuts on
the main sheet (similarly for the set ${\bm \theta} ^{(a)}$).

The norm of a Bethe eigenstate is obtained in the
classical limit by taking ${\bf u} ^{(1)} = {\bf u} ^{(2)}={\bf u} $ in
(\ref{scalarprcl}):
   \begin{eqnarray}
   \la{continorm}
\log\,   \langle {\bf u} ; {\bm \theta} |{\bf u} ; {\bm \theta} \rangle =
  \oint \limits _ {\CC } \frac{du}{2\pi } \ \text{Li}_2\big( e^{ 2i p
  (u) } \big) ,
\end{eqnarray}
where the contour of integration  surrounds the rapidities ${\bf u}$ and leaves outside ${\bm
\theta}$.
The determination of the contour is a subtle issue because of the
logarithmic branch cuts starting at the points where the argument of
the dilogarithm equals 1.  The contour must avoid these cuts and its
choice depends on the analytic properties of the quasimomenta.

 \def\DA{A}

\subsection{One-Loop Three-Point Function in the Semi-Classical Limit}
   \def\hf{ {\textstyle{1\over 2}} } By the computation of the
   previous subsection, the structure constant up to two-loop
   corrections is given by
\begin{eqnarray}
\la{C123-1} \langle {\bf u}^{(1)},{\bf u}^{(2)},{\bf u}^{(3)}\rangle
\equiv e^{ F_{123}} &=& \( 1+g^2 \hat \Delta\) e^{ F_{123}({\bm
\theta})}+ \CO(g^{4}) , \\
  \label{PhaseF123} F_{123}({\bm \theta})&\equiv &\log \caA  _{ {\bf u}
  ^{(2)} \cup {\bf u} ^{(1)}, { { \bm {\theta } } ^{(12)}}} \, \,
  +\log \caA_{{\bf u} ^{(3)} , { \bm {\theta } }^{(13)} }.
\end{eqnarray}
where the differential operator $\hat\Delta$ is defined as
($\delta\EE_r=\EE_r^{(1)}-\EE_r^{(2)}$)
\begin{equation}
\hat\Delta=
\(\partial_1^{(3)}\partial_2^{(3)}-i\EE_2^{(3)}\partial_1^{(3)}+i\EE_3^{(3)}-
\hf\EE_2^{(3)2}\)+
\(\partial_1^{(1)}\partial_2^{(1)}-i\delta\EE_2\partial_1^{(1)}+i\delta\EE_3-\hf\delta\EE_2^2\).
\end{equation}
Thus the one-loop result for the structure constant is expressed in
terms of the tree-level quasiclassical expression with the
inhomogeneities entering as free parameters (\ref{scalarprcl}).  Using
the quasiclassical formula (\ref{classicalA}), one obtains in the
thermodynamical limit
\begin{align}
\label{clsF123}
F_{123}({\bm \theta}) 
\simeq& \oint \limits _ {\CC ^{(1)} \cup\
\CC^{(2)} } \frac{du}{2\pi } \ \text{Li}_2\big( e^{ i p ^{(1)}(u) + i
p ^{(2)}(u) - i q ^{(3)}(u) } \big)\no\\ 
&+
  \oint \limits _ { \CC^{(3)}} \frac{du}{2\pi } \ \text{Li}_2\big(
 e^{ i p ^{(3)}(u) + i q^{(1)}(u) - i q^{(2)}(u) } \big)
\, .
\end{align}
Here $p^{(a)}$ are the three quasimomenta and $q^{(a)}$ are their
singular parts:
\begin{eqnarray}
\label{quasimomresolv}
\qquad p^{(a) } &= G_{{\bf u} ^{(a)}} + q^{(a) } \qquad q^{(a) } = -
\hf G_{ { \bm {\theta } }^{(a)}} \qquad (a=1,2,3).
\end{eqnarray}
For the complete phase in (\ref{C123-1}) we obtain
\begin{equation}
\la{F123} F_{123}= F_{123}({\bm \theta}) + g^2 \, \delta F_{123} +
\CO(g^4) ,
\end{equation}
%
  where the inhomogeneities in the first term on the r.h.s.\ are fixed to
  their BDS values, and the second term
 \begin{eqnarray}
 \label{deltaF123}
  \delta F_{123}= e^{- F_{123}({\bm \theta})} \hat \Delta\,
  e^{F_{123}({\bm \theta})} \Big|_{{\bm \theta}=0}
    \end{eqnarray}
will be computed below.  The first term $F_{123} ({\bm \theta })$ is
an infinite series in $g^2$ from which only the $\CO(g^0)$ and the
$\CO(g^2)$ terms should be retained.
     
       In order to evaluate $\delta F_{123}$ one should compute the
       derivatives in $\theta _{1,2}$ of the phase $F_{123} ({\bm
       \theta })$.  The computation of the derivatives in $\theta
       _{1,2}$ is done
 using the representation (\ref{classicalA}) of the $\caA$-functional:
\begin{eqnarray}
   \la{ddth}
\begin{aligned}
 {\d \over \d \theta _1}\log \caA_{{\bf u} , { \bm {\theta } }}
& =-i \oint\limits_{\CC} \frac{du}{2\pi i}\, {1\over u^2 }\, \log\( 1- e^{ i G_{\bf
u} - iG_ { \bm {\theta } }}\),
\\
 {\d \over \d \theta _1} {\d \over \d \theta _2} \log \caA_{{\bf u} ,
 { \bm {\theta } }} &= - \oint \limits_{\CC}\frac{du}{2\pi i}\, {1\over u^4 }\, \
 {1\over 1- e^{ i G_{\bf u} -i G_ { \bm {\theta } } }} .
\end{aligned}
\end{eqnarray}
 Below we will neglect the term with the second derivative, which is
 of order $1/L$ compared to the other terms.  Then we have
%
\begin{eqnarray}
\label{oneloop}
 \delta F_{123} &=& i\EE_3^{(3)}-\hf\EE_2^{(3)2}
 +i\EE_3^{(1)}-i\EE_3^{(2)} -\hf ( \EE_3^{(1)}-\EE_3^{(2)} ) ^2 \no
\\
&&- ( \EE_2^{(1)}- \EE_2^{(2)}) \oint_{\CC^{(1)}\cup \CC^{(2)}}
\frac{du/u^2}{2\pi i}\, \log\( 1- e^{ i p ^{(1)} + i p ^{(2)}- i q
^{(3)} } \) \no
\\
&&- \EE_2^{(3)} \oint_{ \CC^{(3)}} \frac{du/u^2}{2\pi i}\, \log\( 1-
e^{ i p ^{(3)} + i q ^{(1)}- i q ^{(2)} } \) \no
\\
  &&- \[ \oint_{\CC^{(3)} }\frac{du/u^2 }{2\pi i}\, \log\(1- e^{ i p
  ^{(3)} + i q ^{(1)}- i q ^{(2)} }\) \]^2 \no
\\
&&-
  \[ \oint _{\CC^{(1)}\cup \CC^{(2)}}\frac{du/u^2 }{2\pi i}\,
   \log\( 1- e^{ ip^{(1)} +ip^{(2)} - i q^{(3)} }\)\]^2 .
 \end{eqnarray}
The complete result for $\log C_{123}$ is obtained by subtracting from
$F_{123}+\delta F_{123}$ the logarithms of the norms of the three
states, given by the contour integrals (\ref{continorm}).

 As we mentioned earlier, the choice of the integration contours is a
 non-trivial problem.  The heuristic derivations of the quasiclassical
 limit in \cite{GSV, SL} require that the contour of integration
 $\CC^{(a)}$ encircles the cuts ${\bf u}^{(a)}$ and leaves outside the
 the $\bm \theta$-cut.  However this prescription does not determine
 the contours completely because it says nothing about the logarithmic
 singularities of the integrand at the points where the argument of
 the dilogarithm takes value 1.
A necessary condition on the integration contours is that they should
not cross any of the cuts produced by these singularities.  In the
contour integral along $\CC^{(a)}\cup \CC^{(b)}$ the positions of the
singularities depend on the analytic properties of both $p^{(a)}$ and
$p^{(b)}$.  Let us denote by $\CC^{(ab|c)}$ the contour which
encircles the cuts ${\bf u}^{(a)}$ and $ {\bf u}^{(b)} $, leaves
outside the $\bm \theta$-cut and does not cross any of the logarithmic
cuts ending at the other singularities of the integrand:
%
  \begin{eqnarray}
  \CC^{(a)}\cup \CC^{(b)} \to \CC^{(ab|c)}\, .
  \no
  \end{eqnarray} 
 In order to determine the contour of integration $\CC^{(ab|c)} $, one
 can consider a family of solutions characterised by their global
 filling fractions $\a^{(a)} = N^{(a)}/L^{(a)}$, solve for the
 singular points in the limit $\a^{(a)} \ll 1$ $(a=1,2,3)$ and place
 the contours $\CC^{(ab|c)}$ so that they return to the same sheet.
 When $\a^{(a)} $ increases, the contour deforms in a continuous way.

The above rule works only if the logarithmic singularities at the
points where the argument of the dilogarithm equals 1 are
macroscopically far from the cuts formed by condensation of Bethe
roots.  If a singular point gets close or crosses such a cut, the
integration contour should be closed on the second sheet, possibly
through infinity, as in the example considered in \cite{SL}.

\subsection{Comparison with the String Theory Results}
\label{subsec:compKK}

The semiclassical limit of the
one-loop result  in the SYM theory is expected to match the strong coupling
result in
 the Frolov--Tseytlin \cite{Frolov-Tseytlin}  limit, where the gauge
coupling $g$ is large, but the typical length $L$ is even larger, so that
the effectve coupling $g'=g/L$ is small.  
  This is however not obvious 
   because of the order-of-limits problem \cite{Callan:2003xr,Serban:2004jf}. 
   
The hope that such a comparison is meaningful is based on the
observation that the first two orders of the expansion in
$g'^2=g^2/L^2$ of the anomalous dimension of a heavy operator in the
weakly coupled gauge theory, and of the energy of the corresponding
classical string state, coincide.  Since the computation of the
correlation function requires the knowledge of the wave functions one
order beyond, it is reasonable to expect that for the three-point
functions the match is to the linear order in $g'^2$.
 
A string theory computation of the three-point function at strong
coupling was carried out very recently by Kazama and Komatsu
\cite{KKnew}.  Kazama and Komatsu expressed the three-point function
in terms of the quasimomenta $p^{(1)}, p^{(2)}, p^{(3)}$ obtained from
the monodromy matrix for a solution of the $\sofour$ sigma model at
strong coupling.  They obtained for the logarithm of the structure
constant an expression in terms of contour integrals, very similar to
(\ref{clsF123}).  The arguments of the dilogarithm functions are
$p^{(a)}+p^{(b)}-p^{(c)}$ for $a,b,c\in\{1,2,3 \}$, as well as
$p^{(1)}+p^{(2)}+ p^{(3)}$, and the expression is symmetric in the
permutations of the three operators.

Here we will compare the Frolov--Tseytlin limit of the strong-coupling
answer of \cite{KKnew} with the quasiclassical limit of our solution
(\ref{cuv}) to the linear order in $g'^2$, assuming that the
integration contours coincide, which is very likely to be the case.
The main obstacle in going to two loops is that the Hamiltonian
insertions have not yet been computed, although the computation seems
doable and we hope to be able to report on it separately.

Let $L$ be the length scale such that $L^{(a)}/L \sim 1$ for
$a=1,2,3$.  The operators $\CO^{(a)}$ correspond to solutions of the
Bethe equations consisting of a few macroscopic Bethe strings.  Since
the typical distance between the roots forming such a string is $\sim
1$, the spectral parameter scales as $u\sim L$, which implies for the
conserved charges $E_r\sim L^{1-r}$ ($r$=1,2, \dots).  As a
consequence, the correction $\delta F_{123}$ to the phase (\ref{F123})
scales as\footnote{The fact that $\delta F_{123}$ does not contain a factor of  $L$ 
in the Frolov-Tseytlin limit is not trivial in our computation presented in Section \ref{oneloop3pf}, 
because Hamiltonian insertions scale as $L g'^2$
  and  $\delta S$ also contains terms that scale as $L g'^2$.  
 These two contributions nicely cancel each other leaving us with a net result 
 scaling as $g'^2$.  We suspect that  similar cancellations will happen also at higher loop orders.
}
\begin{eqnarray}
  g^2 \, \delta F _{123}\sim g'^2 .
   \end{eqnarray}
On the other hand, the one-loop correction in $F_{123}({\bm \theta}) $
due to the inhomogeneities, which comes from replacing $L/u\to G_{\bm
\theta} = L/u + 2 L g^2/u^3+\dots$, scales as $ L \times L  g^2/ L^3=
 L g'^2$.%
 \footnote{The additional factor of $L$ comes from the differential $du$ in \eqref{clsF123}.}
Therefore the correction  $g^2  \delta F
_{123}/ F_{123}\sim 1/L$ can be neglected in the Frolov-Tseytlin limit 
 and our one-loop result reads simply
\begin{eqnarray}
\la{C123-FT} \langle {\bf u}^{(1)},{\bf u}^{(2)},{\bf u}^{(3)}\rangle
\simeq \exp F_{123}({\bm \theta}) .
 \end{eqnarray}
  The fact that $ \delta F _{123}$ disappears in the Frolov--Tseytlin
  limit is easy to explain: unlike the ``bulk'' corrections, the
  insertions are localised at the splitting points, and are suppressed
  by a factor of $1/L$.

In the Frolov--Tseytlin limit the result of Kazama and Komatsu for
$F_{123}$ (section 7.5 of \cite{KKnew}) consists of four terms,
 \begin{eqnarray}
\begin{aligned}
\label{clsF123KK}
F_{123}^{\text{KK}}
\simeq& \oint
 \text{Li}_2\big( e^{ i p ^{(1)} + i
p ^{(2)}- i p ^{(3)}} \big)
+
  \oint 
   \
   \text{Li}_2\big(
 e^{ i p ^{(3)} + i p^{(1)}- i p^{(2)}} \big)
\\ +&
  \oint
  \ \text{Li}_2\big(
 e^{ i p ^{(2)} + i p^{(3)}- i p^{(1)} } \big)
+
  \oint 
   \ \text{Li}_2\big(
 e^{ i p ^{(2)} + i p^{(3)} + i p^{(1)}} \big)
\, .  
\end{aligned}
 \end{eqnarray}
Comparing this with (\ref{clsF123}), we see that the first two terms
resemble the two terms of (\ref{clsF123}), while the last two terms do
not have counterparts in the weak coupling result.  The correspondence
with the gauge theory requires that the last two terms vanish, but it
is not clear if this is the case.  In this paper we will discuss only
the first two terms.

We will compare the first two terms (\ref{clsF123KK}) with the
one-loop result (\ref{clsF123}).  We will give an interpretation of
the exponent in (\ref{C123-FT}) in terms of the complex curves of the
three heavy fields.  Obviously the asymmetric form of the tree-level
expression (\ref{clsF123}) is a consequence of the specific choice of
the $\sutwo$ sectors for the three operators ($\CO_1, \CO_2\in
\sutwo_{R}$ and $\CO_2\in \sutwo_L$).  Since the left and the right
$\sutwo$ sectors do not talk to each other perturbatively, the
dependence on the third operator factors out.  This factorisation is
accidental and is a consequence of the choice of the three $\sutwo$
sectors and the weak coupling limit.  At strong coupling, there is no
reason to expect that the three-point function factorises.

\begin{figure}
\centering
\includegraphicsbox{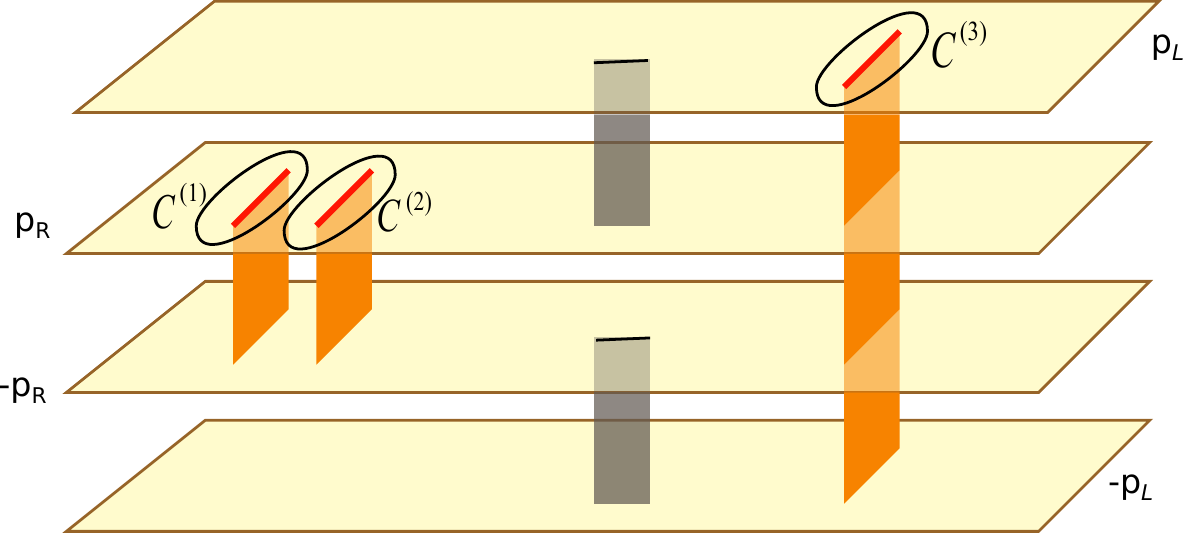} $\dots$ \caption{The Riemann surface
for the three quasimomenta in the $u$-parametrization.  For simplicity
we assumed one-cut solutions.  The left (sheets 1,4) and the right
(sheets 2,3) sectors are connected by Zhukowsky cuts.  In the limit
$g\to 0$ the Zhukovsky cuts shrink to points and the $\sofour$ Riemann
surface decomposes into two disconnected two-sheet Riemann surfaces
describing the $\sutwo_{_R}$ and the $\sutwo_{_L}$ sectors.  }
\label{fig:RiemannSO4}
\end{figure}

Below we are going to show that the arguments of the dilogarithm
function in (\ref{clsF123}) are the $g/L\to 0$ limit of symmetric
combinations of the three quasimomenta, {\it e.g.} $p ^{(3)} + q^{(1)}
- q^{(2)}$ is obtained as a limit of $p^{(3)}+p^{(1)}-p^{(2)}$.  For
that we assume that the three operators are on-shell Bethe states from
the $\sofour$ sector.  This makes sense at strong coupling when the
$\sofour$ sector is closed.\footnote{The $\sofour$ sectors at weak and
at strong coupling have different nature and the comparison should be
taken with caution, see the discussion in \cite{Minahan:2005jq}.  In
the XXX spin chain (with or without inhomogeneities) the length of the
chain $L= \Delta|_{g=0}$ is expressed in terms of the two conserved
$R$-charges.  At perturbative level the length of an operator is
conserved, since the dimension $\Delta$ of the states that contain $n$
pairs $X \bar X$ and have the same $R$-charges is separated by a gap
$2 n$ from the states belonging to the $\sutwo$ sector and are
unreachable perturbatively.  On the contrary, in the sigma model there
is no such gap and to the states of given charge one can add $X$ and
$\bar X$ as constituent fields, since this combination has zero total
charge.  The length of a state is not a conserved charge and it is not
defined at strong coupling.  The $\sofour$ sector is therefore not
stable for finite $g$, but in the limit $g\to\infty$ it becomes stable
again, as the $\sofour$ sigma model is classically stable.  } Then the
linear combination of the three quasimomenta is a meromorphic function
with a four-sheeted Riemann surface as the one depicted in
\figref{fig:RiemannSO4}.

The natural parametrization of the momenta in the strong coupling
limit is by the Zhukovsky variable $x$ defined by (\ref{eq:Zhuk}).
The $a$-th quasimomentum is determined by the set of $N^{(a)}$
rapidities ${\bf x}^{(a)} = \{ {x}^{(a)}_1, \dots, {x}^{(a)}_{N^{(a)}}
\}$, which are related to the rapidities ${\bf u}^{(a)}$ by the
Zhukovsky map (\ref{eq:Zhuk}).  Instead of (\ref{quasimomresolv}), we
have
\begin{eqnarray}
\la{pofx}
p(x) = \CG(x)- {\Delta/2\over x-g^2/x},
\end{eqnarray}
where $\Delta = L+ \delta$ is the conformal dimension and  the resolvent
\begin{eqnarray}
\la{deftGx} \CG(x) = \sum_j {x'_j\over x- x_j}, \qquad x'_j\equiv
{1\over 1-g^2/x_j^2}\; ,\
\end{eqnarray}
is related to the resolvent in the $u$-plane by
\begin{eqnarray}
\la{deftG} G(u) = \CG(x) + \CG(g^2/x) - \CG(0) .
\end{eqnarray}

The left and the right $\sutwo$ sectors in $\sofour$ are related by
the inversion symmetry $x\leftrightarrow g^2 /x$, which exchanges
right and left quasimomenta, $p_{_R}$ and $p_{_L}$
\cite{Kazakov:2004qf, Beisert:2004ag}:
 \begin{eqnarray}
 \la{LRsymmetry} p_{_R} (x) = - p_{_L}(g^2 /x) - 2\pi m, \quad m\in
 \IZ.
 \end{eqnarray}
This allows to go from the four-sheeted Riemann surface in the
$u$-parametrization to a two-sheet Riemann surface in the
$x$-parametrization.  We will use the convention
 \begin{eqnarray}
 \la{pLpRp} p_{_R} (u) = p(x) \Big|_{|x|>g}, \qquad p_{_L}( x) = p(x)
 \Big|_{|x|<g}.
 \end{eqnarray}
With this convention the left and right quasimomenta are assembled
into a single quasimomentum $p(x)$ without inversion symmetry, defined
on the whole $x$-plane \cite{Kazakov:2004qf}.  The quasimomentum
$p(x)$ is thus an analytic function defined on a hyper-elliptic
Riemann surface, with poles at $x=0, x=\infty$ and at the fixed points
of the inversion symmetry $x=\pm g$.  The behavior of the
quasimomentum near these poles is \cite{Kazakov:2004qf}\footnote{In
our convention the quasimomentum has negative sign compared to
\cite{Kazakov:2004qf}.}

\begin{eqnarray} 
\la{pnearpoles}
p (x) \simeq \begin{cases} {\ \ (N - \hf L)/ x}&
\quad (x\to\infty); \\ - \hf \Delta /(x- g^2/x)& \quad (x\to\pm g);
\\ 2\pi m + \hf L\, x/g^2 & \quad (x\to 0).  \end{cases} \la{condp3}
\end{eqnarray}

For the problem we are interested in, $p^{(1)}$ and $p^{(2)}$ belong
to the $\sutwo_R$ sector, while $p^{(3)}$ belongs to the $\sutwo_L$
sector.  Therefore the linear combinations of the type
$p^{(1)}+p^{(2)}-p^{(3)}$ should be understood as
\begin{eqnarray}
\begin{aligned}\la{ppp}
p^{(1)}+p^{(2)}-p^{(3)} \ \ &\to  \ \ \ p^{(1)}(x) +p^{(2)}(x) +
p^{(3)}(g^2/x); \\
p^{(3)}+p^{(1)}-p^{(2)} \ \ &\to  \ \ \ p^{(3)}(x) -p^{(1)}(g^2/x) +
p^{(2)}(g^2/x).
\end{aligned}
\end{eqnarray}
In the limit $g^2 \to 0$, as it is clear from the asymptotics
(\ref{pnearpoles}) of the quasimomentum at the origin, we obtain
exactly the combination that appeared in the arguments of the
dilogarithm in (\ref{clsF123})!  Since the quasimomentum appears only
in the exponent, the term $2\pi m$ can be neglected.

Now let us see if the the r.h.s.\ of (\ref{ppp}) and the arguments of
the dilogarithm in (\ref{clsF123}) match at linear order in
$g'^2=g^2/L^2$.  This will be the case if the function $p(g^2/x)+
q(x)$ vanishes up to $g'^4$.  We have from (\ref{pofx})
\begin{eqnarray}
\la{pexpan}
p (g^2/ x) + q(x) & =& \sum_{j=1}^N {x'_j\over g^2 /x -x_j} + {\Delta/2\over
x- g^2/x}-  {L/2\over
x- g^2/x} \no
\\
&=& 2\pi m 
+ g^4\( {E_2\over x^3} - {2 E_3\over x^2}\)
+  \CO(g^6)  .
\end{eqnarray}
Therefore, if the second two terms in (\ref{clsF123KK}) can be
ignored, the Frolov--Tseytlin limit of the strong coupling result from
the string theory side matches, up to the subtleties related to the
choice of the contour, with the one-loop result from the SYM
side at order $g^2/L^2$.  In any case, if the results
match at tree level, they will match also at one loop.  Note that if
the Hamiltonian insertions at two loops are located only at the
splitting points, there will be disagreement at two-loop order in the
Frolov--Tseytlin limit.

We also see that the factorisation of the structure constant into two
pieces, the first depending on ${\bf u}^{(1)}$ and ${\bf u}^{(2)}$ and
the second depending on ${\bf u}^{(3)}$, takes place only in the weak
coupling limit and it is a consequence of the fact that at $g\to 0$
the spectral curve for the $\sofour$ sector splits into two components
connected by a vanishing cycle (the Zhukowsky circle $|x|=g$).
Returning to the $u$-parametrization, the three operators are defined
on the Riemann surface for the $\sofour$ sector sketched in
\figref{fig:RiemannSO4}.  The Riemann surface splits into two
disjoined hyperelliptic surfaces in the limit $g\to 0$, when the two
Zhukovsky cuts disappear.

\section{Conclusions \& Outlook}

In this work we have considered the relation between inhomogeneous and
boost-induced long-range spin chains
which share the same spectrum.%
\footnote{Up to wrapping interactions.} We followed the philosophy
that both models can be generated from a homogeneous XXX spin chain
using different kinds of transformations.  In one case the generators
of the transformation are the boost operators studied in
\cite{Bargheer:2008jt,Bargheer:2009xy} and the transformation can be
written as a singular unitary operator $\Sop_\boo$.  In the other
case, the transformation from a homogeneous to an inhomogeneous chain
is generated by $\Sop_\inhom$ and agrees at least up to terms of order
$g^2$ with an inhomogeneous version of Baxter's corner transfer
matrix.  Since both deformations have the same spectrum, they should
be related by a unitary non-singular operator $\Sop=\Sop_\boo
\Sop_\inhom^{-1}$.  Using the map between the two models, we have
determined the scalar products of the
long-range model.%
 \footnote{The explicit expression of the operator $\Sop$ is not
 important for computing the scalar product.} We have determined the
 unitary operator $\Sop$ up to terms of order $g^3$, the highest order
 being obtained from the comparison with the corner transfer matrix.
 The method works for a large class of long-range deformations of the
 spin 1/2 XXX spin chain and can be straightforwardly extended to
 similar deformations of higher-rank or higher-spin models.
 
The map that we have discussed here is also a morphism of Yangian
algebras.  In particular, in the case of $\mathcal{N} = 4$ super
Yang--Mills theory, this morphism allows to relate the Yangian algebra
of the higher loop dilatation operator to the Yangian of an
inhomogeneous spin chain.  A similar Yangian algebra was found for
scattering amplitudes in this gauge theory (see {\it e.g.}\
\cite{Bargheer:2011mm} and references therein).  It would be
interesting to investigate whether this morphism can also be used to
exploit the integrability of amplitudes at higher loops.

We have used the map between long-range and inhomogeneous spin chains
in order to compute the three-point function of three operators in
different $\alg{su}(2)$ sectors of $\mathcal{N}=4$ super Yang--Mills
theory.  The necessary ingredients are the wave functions of the
dilatation operator at higher loop order, plus the diagrammatic
field-theoretical corrections.  These corrections have not yet been
computed at two-loop order, and thus we have not performed the
computation of the three-point function at two loops.

We have re-derived the results of Gromov and Viera at one loop
\cite{Gromov:2012vu,GV} at finite length, in a form that allows to
straightforwardly take the semi-classical limit.  In the so-called
Frolov--Tseytlin limit the results of the classical limit agree with
the conjecture in \cite{Serban:2012dr}.  We have compared the one-loop
computation with the strong coupling result obtained recently by
Kazama and Komatsu \cite{KKnew} and we have found that if the results
match at tree order, they match also at one loop.  If
the Hamiltonian insertions at two loops are located only at the
splitting points, there will possibly be disagreement at two-loop order.

In order to go to three loops and beyond, one has to take into
account the \emph{dressing phase} as well.  To include the dressing
phase into this framework, we note that the generator of the
corresponding long-range model $\Sop_{\biloc{\charge}{\charge}}$ is
known from the investigations in
\cite{Bargheer:2008jt,Bargheer:2009xy}.  For the dressing phase
contributions, this operator furnishes the analogue of the boost
generator $\Sop_\boo$ discussed above.  However, extending the
correspondence to an asymptotically dual model, like the inhomogeneous
spin chain, should be more involved because the values of the
inhomogeneities will be state-dependent.

\pdfbookmark[2]{Acknowledgements}{acknowledgements}
\subsection*{Acknowledgements}

We would like to thank O.\ Foda, S.\ Komatsu, N.\ Gromov and M.\ Wheeler
for stimulating discussions. 
We thank N.\ Gromov and P.\ Vieira for useful remarks on the initial versions of the preprint.
FL would like to thank Till Bargheer for
initial collaboration on studying the relation between boost operators
and theta derivatives as well as for helpful discussions.  FL would
also like to thank Peter Orland for a discussion on corner transfer
matrices.  IK and DS would like to thank Melbourne University, and YJ,
IK and DS would like to thank SCGP, YITP and IPMU for their warm
hospitality.  This work received funding from the PHC Sakura 27588UA,
the European Programme IRSES UNIFY No 269217 and People Programme
(Marie Curie Actions) of the European Union's Seventh Framework
Programme FP7/2007-2013/ under REA Grant Agreement No 317089.  The
work of FL was supported by a fellowship within the Postdoc-Program of
the German Academic Exchange Service (DAAD).



\appendix
\addcontentsline{toc}{section}{Appendix}
\addtocontents{toc}{\protect\setcounter{tocdepth}{-1}}


\section{Inhomogeneous CTM at Order $\theta^3$}
\label{app:CTMthree}
 It is instructive to expand the inhomogeneous CTM to order
 $\theta^3$.  In the bulk we find up to terms proportional to the
 identity
\begin{align}
\ctm_\inhom(0)=\exp\bigg[\dots+\frac{1}{6}
\sum_{k=1}^L\Big(
-\sthree_k^{[a]}\,\ham_{k}
+\sthree_k^{[b]}\, \big(\ham_{k}[\ham]_{k-1}-[\ham]_{k-1}\big)
+\sthree_k^{[c]}\,[[\ham]]_{k-1}\Big)
+\order{\theta^4}\bigg],
\end{align}
where the individual coefficients are functions of the inhomogeneities given by
\begin{align}
\sthree_k^{[a]}&=-2\sum_{x=1}^k\theta_x^3-2(\theta_{k+1}
-\theta_k)\sum_{x=1}^k\sone_{x-1}\theta_x -\hat\stwo_k(2\theta_k
+\theta_{k+1})-\sone_k(\theta_k^2+\theta_k\theta_{k+1}),\nonumber\\
\sthree_k^{[b]}&=-2\sum_{x=1}^k\theta_x^3-3\hat\stwo_k\theta_k
-2\sone_k\theta_k^2,\nonumber\\
\sthree_k^{[c]}
&=-2\sum_{x=1}^k \theta_x^3+(\theta_{k+1}-\theta_k)\sum_{x=1}^k\sone_{x-1}\theta_x
-\hat\stwo_k(2\theta_k+\theta_{k+1})-2\sone_k\theta_k^2,
\end{align}
The coefficients $\sthree$ vanish for $\theta_k=u$ as expected.


\section{The BDS Charges from Boost Deformations}
\label{app:BDScharges}
Here we give explicit solutions for the BDS charges up to four loop
order.  We may restrict the construction introduced in
\cite{Bargheer:2008jt,Bargheer:2009xy} to the BDS model using the
above expression \eqref{eq:BDSPi} for $\tcon_k$ to find the
deformation equation \eqref{eq:remarkdefeqtwo} for the BDS Hamiltonians:
\begin{equation}\label{eq:defeqBDS}
\frac{d}{dg}\chargeBDS_r(g)= \sum_{k=1}^\infty
2g^{2k-1}\Big(2ki\comm{\boost{\chargeBDS_{2k+1}(g)}}{\chargeBDS_r(g)}
+(r+2k-1)\chargeBDS_{r+2k}(g)\Big).
\end{equation}
Solving the above equation perturbatively one finds the following
contributions at order $g^2$, $g^4$ and $g^6$:
\begin{align}
\chargeBDS_r^\ups{2}=&2i\comm{\boost{\chargeBDS_3^\ups{0}}}{\chargeBDS_r^\ups{0}}
+(r+1)\chargeBDS_{r+2}^\ups{0},\nonumber\\
\chargeBDS_r^\ups{4}
=&\frac{1}{2}\Big[2i\comm{\boost{\chargeBDS_3^\ups{2}}}{\chargeBDS_r^\ups{0}}
+2i\comm{\boost{\chargeBDS_3^\ups{0}}}{\chargeBDS_r^\ups{2}}
+(r+1)\chargeBDS_{r+2}^\ups{2}\nonumber\\
&+4i\comm{\boost{\chargeBDS_5^\ups{0}}}{\chargeBDS_r^\ups{0}}
+(r+3)\chargeBDS_{r+4}^\ups{0}\Big],\nonumber\\
\chargeBDS_r^\ups{6}
=&\frac{1}{3}\Big[2i\comm{\boost{\chargeBDS_3^\ups{4}}}{\chargeBDS_r^\ups{0}}
+2i\comm{\boost{\chargeBDS_3^\ups{2}}}{\chargeBDS_r^\ups{2}}
+2i\comm{\boost{\chargeBDS_3^\ups{0}}}{\chargeBDS_r^\ups{4}}
+(r+1)\chargeBDS_{r+2}^\ups{4}\nonumber\\
&+4i\comm{\boost{\chargeBDS_5^\ups{2}}}{\chargeBDS_r^\ups{0}}
+4i\comm{\boost{\chargeBDS_5^\ups{0}}}{\chargeBDS_r^\ups{2}}
+(r+3)\chargeBDS_{r+4}^\ups{2}\nonumber\\
&+6i\comm{\boost{\chargeBDS_7^\ups{0}}}{\chargeBDS_r^\ups{0}}
+(r+5)\chargeBDS_{r+6}^\ups{0}\Big].
\label{eq:BDSsol}
\end{align}
When the $\chargeBDS_r^\ups{0}$ are chosen to be the XXX charges, these
expressions give the BDS Hamiltonians at two, three and four gauge
theory loops.
\section{Derivation of the S-operator at order $g^2$}
\label{app:Satgsquare}

In this appendix we compute the S-operator to higher order.  We
explicitly evaluate the right hand side of \eqref{eq:shiftS} to find
\begin{align}
\Sop \shift_\inhom&\Sop^{-1}=\nonumber\\
\shift_0&\bigg[1+i\sum_{k=1}^L\sone_{k-1}\ham_{k}
-\frac{1}{2}\sum_{k,l=1}^L\sone_{k-1}\sone_{l-1}\ham_{k}\ham_{l}
-\frac{1}{2}\sum_{k=1}^L\stwo_{k-1}\hamone_{k-1}\bigg]\nonumber\\
\times&\bigg[1-i\sum_{k=1}^L\theta_k\ham_{k}
-\frac{1}{2}\sum_{k,l=1}^L\theta_k\theta_l\ham_{k}\ham_{l}
-\frac{1}{2}\sum_{k=1}^L\theta_{k-1}\theta_k\hamone_{k-1}\bigg]\nonumber\\
\times&\bigg[1-i\sum_{k=1}^L\sone_{k}\ham_{k}
-\frac{1}{2}\sum_{k,l=1}^L\sone_{k}\sone_{l}\ham_{k}\ham_{l}
+\frac{1}{2}\sum_{k=1}^L\stwo_{k}\hamone_{k-1}\bigg]+\order{g^3}.
\end{align}
Here we assumed periodicity of $\sone_k$ and $\stwo_k$ to commute the
shift operator with the first bracket.  Making use of the above
constraints on the paramter $\sone_k$ that guarantee a vanishing
contribution at $g^1$, this immediately evaluates to
\begin{align}
\dots
=\shift_0\bigg[1&+\frac{1}{2}\sum_{k,l=1}^L\big(2\sone_{k-1}\theta_l
-2\theta_k\sone_l+2\sone_{k-1}\sone_l-\sone_{k-1}\sone_{l-1}
-\sone_k\sone_l\big)\ham_{k}\ham_{l}\nonumber\\
&-\frac{1}{2}\sum_{k,l=1}^L\theta_k\theta_l\ham_{k}\ham_{l}
-\frac{1}{2}\sum_{k=1}^L\theta_{k-1}\theta_k\hamone_{k-1}
+\frac{1}{2}\sum_{k=1}^L(\stwo_k-\stwo_{k-1})\hamone_{k-1}\bigg]
+\order{g^3}.
\end{align}
and using again \eqref{eq:inhomone} the first line can be simplified
according to
\begin{equation}
\big(2\sone_{k-1}\theta_l-2\theta_k\sone_l+2\sone_{k-1}\sone_l
-\sone_{k-1}\sone_{l-1}-\sone_k\sone_l)\ham_{k}\ham_{l}
=\theta_k\theta_l \ham_{k}\ham_{l}+\sone_k\theta_l
\comm{\ham_{k}}{\ham_{l}}.
\end{equation}
Combining the terms finally results in \eqref{eq:finally}.


\section{From Permutations to Derivatives}
\label{sec:pdrel}
In this appendix, we explain how to convert the action of any
permutation operators on the monodromy matrix of Bethe states into
derivatives with respect to impurities.  We shall call this kind of
relations PD relations.  We derive the PD relations both in the bulk
and at the boundary.\par
\subsection{PD Relations in the Bulk}
We start with algebraic Bethe ansatz.  For simplicity, we choose a
different normalization from the main text.  The R-matrix at each
site is given by
\begin{align}
\mathrm{R}'_{\a n}(u)=\mathrm{I}_{\a n}+\frac{i}{u}\mathrm{P}_{\a n},\quad
n=1,\ldots,L
\end{align}
and it is related to the one in the main text by
\begin{align}
\mathrm{R}'_{\a\b}(u)=\frac{u+i}{u}\mathrm{R}_{\a\b}(u)\;.
\end{align}
Here $\a$ denotes the auxiliary space and $n$ is the quantum space.
$\mathrm{I}$ and $\mathrm{P}$ are identity and permutation operators,
respectively.  The monodromy matrix is defined as in the main text
\begin{align}
\mathrm{M}_\a(u,\bm{\theta})\equiv\prod_{n=1}^L\mathrm{R}'_{\a n}(u-\theta_n-i/2)
\end{align}
which becomes, in the homogeneous limit where $\theta_k\to 0$,
\begin{align}
\mathrm{M}_\a(u)=\prod_{n=1}^L\mathrm{R}'_{\a n}(u-i/2)\;.
\end{align}
The authors in \cite{Gromov:2012vu} found the following relation
\begin{align}\label{PD1}
[\mathrm{P}_{k,k+1},\mathrm{M}_\a(u)]=\left.i(\partial_k
-\partial_{k+1})\mathrm{M}_\a(u,\bm{\theta})\right|_{\bm{\theta}=0}\;,
\quad \partial_k\equiv\frac{\partial}{\partial\theta_k}\;,
\end{align}
where in the r.h.s.\ one first takes the derivatives with respect to
the impurities and then sends all impurities to zero.  For simplicity,
we will denote the r.h.s.\ of (\ref{PD1}) by
$i(\partial_k-\partial_{k+1})\mathrm{M}_{\a}(u)$ and adopt the same
convention for all PD relations.  As in the main text, we introduce
the following notation
\begin{align}
\mathrm{D}_k\equiv i(\partial_k-\partial_{k+1}),\quad
\mathrm{D}_L=i(\partial_L-\partial_1)\;.
\end{align}
We will generalize (\ref{PD1}) to the case when several permutations
act on the monodromy matrix.  To this end, we first notice that if the
action of permutation and derivatives do not overlap, they will act
independently.  This means, for example
\begin{align}
\partial_j[\mathrm{P}_{k,k+1},\mathrm{M}_\a(u)]=\partial_j\mathrm{D}_k\mathrm{M}_\a(u),
\quad\text{if
}j\ne k,k+1\;.
\end{align}
The case where permutations and derivatives overlap needs to be
considered more carefully.  From the definition of monodromy matrix,
one can derive the following relations
\begin{align}\label{interaction1}
\partial_{k}^n[\mathrm{P}_{k,k+1},\mathrm{M}_\a(u)]&=-(\partial_k^n
-\partial_{k+1}^n)\mathrm{M}_\a(u)\mathrm{P}_{k,k+1}+\frac{1}{n+1}(i\partial_k^{n+1}
-i\partial_{k+1}^{n+1})\mathrm{M}_\a(u)
\\\nonumber
\partial_{k+1}^n[\mathrm{P}_{k,k+1},\mathrm{M}_\a(u)]&=(\partial_k^n
-\partial_{k+1}^n)\mathrm{M}_\a(u)\mathrm{P}_{k,k+1}+\frac{1}{n+1}(i\partial_k^{n+1}-i\partial_{k+1}^{n+1})\mathrm{M}_\a(u)\\\nonumber
\partial_k^m\partial_{k+1}^n[\mathrm{P}_{k,k+1},\mathrm{M}_\a(u)]&
=\frac{m!n!}{(m+n+1)!}(i\partial_k^{m+n+1}-i\partial_{k+1}^{m+n+1})\mathrm{M}_\a(u)
\end{align}
for any $m,n\in\mathbb{N}$.  Relations (\ref{interaction1}) can also
be written as
\begin{align}\label{p1}
\mathrm{P}_{k,k+1}\partial_{k+1}^n\mathrm{M}_\a(u)&=\partial_k^n\mathrm{M}_{\a}(u)\mathrm{P}_{k,k+1}
+\frac{1}{n+1}(i\partial_{k}^{n+1}-i\partial_{k+1}^{n+1})\mathrm{M}_\a(u)\\\nonumber
\mathrm{P}_{k,k+1}\partial_{k}^n\mathrm{M}_\a(u)&=\partial_{k+1}^n\mathrm{M}_{\a}(u)\mathrm{P}_{k,k+1}
+\frac{1}{n+1}(i\partial_{k}^{n+1}-i\partial_{k+1}^{n+1})\mathrm{M}_\a(u)\\\nonumber
\mathrm{P}_{k,k+1}\partial_k^m\partial_{k+1}^n\mathrm{M}_\a(u)&=\frac{m!n!}{(m+n+1)!}(i\partial_k^{m+n+1}
-i\partial_{k+1}^{m+n+1})\mathrm{M}_\a(u)+\partial_k^m\partial_{k+1}^n\mathrm{M}_\a(u)\mathrm{P}_{k,k+1}\;.
\end{align}
By the help of (\ref{p1}), we can derive the general PD relation.  To
see how this works, let us consider the following example
\begin{align}\label{pp1}
[\mathrm{P}_{k,k-1}\mathrm{P}_{k,k+1},\mathrm{M}_\a(u)]&=[\mathrm{P}_{k,k-1},
\mathrm{M}_\a(u)]\mathrm{P}_{k,k+1}+\mathrm{P}_{k,k-1}[\mathrm{P}_{k,k+1},
\mathrm{M}_\a(u)]
\end{align}
\vskip-25pt
\begin{align}
\nonumber
&=\mathrm{D}_{k-1}\mathrm{M}_\a(u)\mathrm{P}_{k,k+1}+i\mathrm{P}_{k,k-1}(\partial_k
-\partial_{k+1})\mathrm{M}_\a(u)
\\\nonumber
&=\mathrm{D}_{k-1}\mathrm{M}_\a(u)\mathrm{P}_{k,k+1}+\frac{i^2}{2}(\partial_{k-1}^2
-\partial_k^2)\mathrm{M}_\a(u)+i\partial_{k-1}\mathrm{M}_\a(u)\mathrm{P}_{k,k-1}\\\nonumber
&-i\partial_{k+1}\mathrm{M}_\a(u)\mathrm{P}_{k,k-1}-i^2\partial_{k+1}(\partial_{k-1}
-\partial_k)\mathrm{M}_\a(u)\\\nonumber
&=\frac{1}{2}(\mathrm{D}_{k-1}^2+2\mathrm{D}_{k-1}\mathrm{D}_k)\mathrm{M}_\a(u)
+\mathrm{D}_{k-1}\mathrm{M}_\a(u)\mathrm{P}_{k,k+1}+(\mathrm{D}_{k-1}
+\mathrm{D}_k)\mathrm{M}_\a(u)\mathrm{P}_{k,k-1}\;.
\end{align}
Similarly, we can derive
\begin{align}\label{pp2}
[\mathrm{P}_{k,k+1}\mathrm{P}_{k,k-1},&\mathrm{M}_\a(u)]=\\ \nonumber
&=\frac{1}{2}(\mathrm{D}_k^2+2\mathrm{D}_{k-1}\mathrm{D}_k)\mathrm{M}_\a(u)
+\mathrm{D}_k\mathrm{M}_\a(u)\mathrm{P}_{k,k-1}
+(\mathrm{D}_{k-1}+\mathrm{D}_k)\mathrm{M}_\a(u)\mathrm{P}_{k,k+1}\;.
\end{align}
It is straightforward to generalize this calculation to
$[\mathcal{P},\mathrm{M}_\a(u)]$ where $\mathcal{P}$ is a product of
$\mathrm{P}_{k,k+1}$.
 In order to apply PD relation
on a Bethe state instead of monodromy matrix, one has also need to
show the PD relation has morphism property.  This means, given two
functions of the monodromy matrix $\mathrm{X}(u)$ and $\mathrm{Y}(u)$,
we have
\begin{align}
[\mathrm{P}_{k,k-1}\mathrm{P}_{k,k+1},\mathrm{XY}]=\frac{1}{2}(\mathrm{D}_k^2
+2\mathrm{D}_{k-1}\mathrm{D}_k)(\mathrm{XY})+\mathrm{D}_k(\mathrm{XY})\mathrm{P}_{k,k-1}
+(\mathrm{D}_{k-1}+\mathrm{D}_k)(\mathrm{XY})\mathrm{P}_{k,k+1}\;.
\end{align}
One can show this is true by explicit calculation.  Using PD relation
and morphism property we can derive the following relations, which
will be useful in later discussion
\begin{align}\label{H1}
\mathrm{H}_{k-1}\mathrm{H}_{k}|\mathbf{u}\rangle&=[\mathrm{P}_{k,k-1},
[\mathrm{P}_{k,k+1},\mathrm{B}(\mathbf{u})]]|\Omega\rangle\\\nonumber
&=[\mathrm{P}_{k,k-1}\mathrm{P}_{k,k+1},\mathrm{B}(\mathbf{u})]|\Omega\rangle
-[\mathrm{P}_{k,k+1},\mathrm{B}(\mathbf{u})]|\Omega\rangle-[\mathrm{P}_{k,k-1},
\mathrm{B}(\mathbf{u})]|\Omega\rangle\\\nonumber
&=\frac{1}{2}(\mathrm{D}_{k-1}^2+2\mathrm{D}_k\mathrm{D}_{k-1})
|\mathbf{u}\rangle+\mathrm{D}_{k-1}|\mathbf{u}\rangle\;,
\end{align}
where we use the shorthand notation
$\mathrm{B}(\mathbf{u})\equiv\mathrm{B}(u_1)\cdots\mathrm{B}(u_N)$.
Similarly, we have
\begin{align}\label{H2}
\mathrm{H}_{k}\mathrm{H}_{k-1}|\mathbf{u}\rangle=\textstyle{\frac{1}{2}}(\mathrm{D}_{k}^2
+2\mathrm{D}_k\mathrm{D}_{k-1})|\mathbf{u}\rangle+\mathrm{D}_{k}|\mathbf{u}\rangle\;.
\end{align}
Taking the sum and difference of (\ref{H1}) and (\ref{H2}), we obtain
\begin{align}\label{2}
[\mathrm{H}_{k-1},\mathrm{H}_{k}]|\mathbf{u}\rangle&=[\mathrm{H}]_{k-1}|\mathbf{u}\rangle
=\left(\textstyle{\frac{1}{2}}(\mathrm{D}_{k-1}^2-\mathrm{D}^2_k)+\mathrm{D}_{k-1}-\mathrm{D}_k
\right)|\mathbf{u}\rangle\\ \label{eq:anticom}
\{\mathrm{H}_{k-1},\mathrm{H}_{k}\}|\mathbf{u}\rangle&
=\left(\textstyle{\frac{1}{2}}(\mathrm{D}_{k-1}^2+\mathrm{D}_k^2)
+\mathrm{D}_{k-1}+\mathrm{D}_k+2\mathrm{D}_k\mathrm{D}_{k-1}
\right)|\mathbf{u}\rangle\;.
\end{align}
Higher order PD relations can be determined along the same lines.

\subsection{PD Relations at the Boundary}
The PD relations at the boundary are more subtle.  In this section, we
will derive the boundary PD relations for one and two overlapping
permutations, at least one of them involving the bond $1L$.  The key
observation is to notice that $\mathrm{D}_k$ should satisfy the
following trivial constraint
\begin{align}
\sum_{k=1}^L\mathrm{D}_k=0\;.
\end{align}
At first order, we have
\begin{align}
\mathrm{D}_L|\mathbf{u}\rangle=-\sum_{k=1}^{L-1}\mathrm{D}_k|\mathbf{u}\rangle
=\sum_{k=1}^{L-1}\mathrm{H}_{k}|\mathbf{u}\rangle=\EE_2|\mathbf{u}\rangle-\mathrm{H}_{L}|\mathbf{u}\rangle\;,
\end{align}
hence we find the boundary term at first order,
\begin{align}\label{e3}
\mathrm{H}_{L}|\mathbf{u}\rangle=-\mathrm{D}_L|\mathbf{u}\rangle+E_2|\mathbf{u}\rangle\;.
\end{align}
We consider now  the square,
\begin{align}
\mathrm{D}_L^2=(\mathrm{D}_1+\cdots\mathrm{D}_{L-1})^2
\end{align}
such that
\begin{align}
\textstyle{\frac{1}{2}}(\mathrm{D}_L^2&-\mathrm{D}_1^2)|\mathbf{u}\rangle
&=\textstyle{\frac{1}{2}}
(\mathrm{D}_2^2+2\mathrm{D}_1\mathrm{D}_2)|\mathbf{u}\rangle+\dots
+\textstyle{\frac{1}{2}}
(\mathrm{D}_{L-1}^2+2\mathrm{D}_{L-2}\mathrm{D}_{L-1})|\mathbf{u}\rangle+\text{non-connected}
\end{align}
where ``non-connected'' are the terms
$2\mathrm{D}_j\mathrm{D}_k|\mathbf{u}\rangle$ with $|j-k|\ge2$.  Using
(\ref{H2}),
\begin{align}
\frac{1}{2}(\mathrm{D}_k^2+2\mathrm{D}_{k-1}\mathrm{D}_k)|\mathbf{u}\rangle
=\mathrm{H}_{k}\mathrm{H}_{k-1}|\mathbf{u}\rangle-\mathrm{D}_k|\mathbf{u}\rangle
\end{align}
we have
\begin{align}
\frac{1}{2}(\mathrm{D}_L^2-\mathrm{D}_1^2)|\mathbf{u}\rangle=&(\mathrm{H}_{2}\mathrm{H}_{1}
+\cdots+\mathrm{H}_{L-1}\mathrm{H}_{L-2})|\mathbf{u}\rangle-(\mathrm{D}_2
+\cdots\mathrm{D}_{L-1})|\mathbf{u}\rangle+\text{non-connected}\;,
\end{align}
which is the same as
\begin{align}\label{boundary1}
\left(\frac{1}{2}(\mathrm{D}_L^2-\mathrm{D}_1^2)+(\mathrm{D}_L-\mathrm{D}_1)\right)|\mathbf{u}\rangle
=&\sum_{k=2}^{L-1}\mathrm{H}_{k}\mathrm{H}_{k-1}|\mathbf{u}\rangle+2\mathrm{D}_L|\mathbf{u}\rangle
+\text{non-connected}\;.
\end{align}
Similarly, using \eqref{H1} we have
\begin{align}\label{boundary2}
\left(\frac{1}{2}(\mathrm{D}_L^2-\mathrm{D}_{L-1}^2)+(\mathrm{D}_L
-\mathrm{D}_{L-1})\right)|\mathbf{u}\rangle=&\sum_{k=2}^{L-1}\mathrm{H}_{k-1}\mathrm{H}_{k}|\mathbf{u}\rangle
+2\mathrm{D}_L|\mathbf{u}\rangle+\text{non-connected}\;.
\end{align}
Taking the difference of (\ref{boundary1}) and (\ref{boundary2}), we
have
\begin{align}\label{Q3}
&\left(\frac{1}{2}(\mathrm{D}_L^2-\mathrm{D}_1^2)+(\mathrm{D}_L-\mathrm{D}_1)\right)|\mathbf{u}\rangle
-\left(\frac{1}{2}(\mathrm{D}_L^2-\mathrm{D}_{L-1}^2)+(\mathrm{D}_L
-\mathrm{D}_{L-1})\right)|\mathbf{u}\rangle\\\nonumber
=&-\sum_{k=2}^{L-1}[\mathrm{H}_{k-1},\mathrm{H}_{k}]|\mathbf{u}\rangle
=(2i\mathrm{Q}_3+[\mathrm{H}_{L-1},\mathrm{H}_{L}]+[\mathrm{H}_{L},
\mathrm{H}_{1}])|\mathbf{u}\rangle=(2iE_3+[\mathrm{H}]_{L-1}+[\mathrm{H}]_{L})|\mathbf{u}\rangle\;.
\end{align}
If we take instead the sum of (\ref{boundary1}) and (\ref{boundary2}), we have
\begin{align}\label{boundary3}
&\left(\frac{1}{2}(\mathrm{D}_L^2-\mathrm{D}_1^2)+(\mathrm{D}_L
-\mathrm{D}_1)\right)|\mathbf{u}\rangle+\left(\frac{1}{2}(\mathrm{D}_L^2
-\mathrm{D}_{L-1}^2)+(\mathrm{D}_L-\mathrm{D}_{L-1})\right)|\mathbf{u}\rangle\\\nonumber
=&\sum_{k=2}^{L-1}\{\mathrm{H}_{k-1},\mathrm{H}_{k}\}|\mathbf{u}\rangle
+4\mathrm{D}_L|\mathbf{u}\rangle+\text{cross
terms}=
\left(\sum_{k=1}^{L-1}\mathrm{H}_{k}\right)^2|\mathbf{u}\rangle
+2\mathrm{D}_L|\mathbf{u}\rangle\;,
\end{align}
where we have used the fact that
\begin{align}
\sum_{k=1}^{L-1}\mathrm{H}_{k}^2|\mathbf{u}\rangle=2\sum_{k=1}^{L-1}\mathrm{H}_{k}|\mathbf{u}\rangle
=-2\sum_{k=1}^{L-1}\mathrm{D}_k|\mathbf{u}\rangle=2\mathrm{D}_L|\mathbf{u}\rangle\;.
\end{align}
Now we plug
$
\sum_{k=1}^{L-1}\mathrm{H}_{k}=\mathrm{Q}_2-\mathrm{H}_{L}
$
into (\ref{boundary3}),
\begin{align}\label{Q2}
\(\frac{1}{2}(\mathrm{D}_L^2-\mathrm{D}_1^2\)&+(\mathrm{D}_L-\mathrm{D}_1))|\mathbf{u}\rangle
+\left(\frac{1}{2}(\mathrm{D}_L^2-\mathrm{D}_{L-1}^2)+(\mathrm{D}_L-\mathrm{D}_{L-1})\right)|\mathbf{u}\rangle=\\\nonumber
=&(\mathrm{Q}_2-\mathrm{H}_{L})^2|\mathbf{u}\rangle+2\mathrm{D}_L|\mathbf{u}\rangle
=(\EE_2^2+2\EE_2-2\EE_2\mathrm{H}_{L})|\mathbf{u}\rangle-[\mathrm{Q}_2,\mathrm{H}_{L}]|\mathbf{u}\rangle\\\nonumber
=&(\EE_2^2+2\EE_2-2\EE_2\mathrm{H}_{L})|\mathbf{u}\rangle-[\mathrm{H}]_{L-1}|\mathbf{u}\rangle+[\mathrm{H}]_L|\mathbf{u}\rangle\;,
\end{align}
where we have used (\ref{e3}).  Taking the sum of (\ref{Q3}) and
(\ref{Q2}), we find that
\begin{align}\label{boundary}
[\mathrm{H}]_L|\mathbf{u}\rangle&=\left(\frac{1}{2}(\mathrm{D}_L^2-\mathrm{D}_{1}^2)+(\mathrm{D}_L-\mathrm{D}_{1})\right)|\mathbf{u}\rangle+\EE_2\mathrm{H}_{L}|\mathbf{u}\rangle-C(\mathbf{u})|\mathbf{u}\rangle\;,\\\no
[\mathrm{H}]_{L-1}|\mathbf{u}\rangle&=\left(\frac{1}{2}(\mathrm{D}_{L-1}^2-\mathrm{D}_{L}^2)+(\mathrm{D}_L-\mathrm{D}_{1})\right)|\mathbf{u}\rangle-\EE_2\mathrm{H}_{L}|\mathbf{u}\rangle+C(\mathbf{u})|\mathbf{u}\rangle-2i\EE_3|\mathbf{u}\rangle\;.
\end{align}
where $C(\mathbf{u})$ is a function of rapidities defined by
\begin{align}
C(\mathbf{u})=\frac{1}{2}[\EE_2^2(\mathbf{u})+2\EE_2(\mathbf{u})+2i\EE_3(\mathbf{u})]\;.
\end{align}
In the derivation above, we use the fact that the second and third
conserved charge read
\begin{align}
\mathrm{Q}_2=\sum_{k=1}^L\mathrm{H}_{k},\quad
\mathrm{Q}_3=\frac{i}{2}\sum_{k=1}^L[\mathrm{H}]_k
\end{align}
and
$
\mathrm{Q}_r|\mathbf{u}\rangle=E_r|\mathbf{u}\rangle
$
when $|\mathbf{u}\rangle$ is on-shell.

\section{From $\Sop$-Transformation to Theta-Morphism}

In this section, we will show how the theta morphism can be derived
from the S-operator.  Up to the order $g^2$, the $S$-operator reads
\begin{align}
\mathrm{S}=\exp\left(\sum_{k=1}^L
i\nu_k\mathrm{H}_{k}-\frac{1}{2}\rho_k[\mathrm{H}]_{k-1} \right)\;.
\end{align}
Let us first recall the main result of this section
\begin{align}\label{main}
\mathrm{S}|\mathbf{u};\theta\rangle=\left(1-g^2
E_2\mathrm{H}_{L}-\frac{g^2}{2}\sum_{k=1}^L
(\mathrm{D}_k^2+2\mathrm{D}_k)
\right)|\mathbf{u}\rangle-g^2C|\mathbf{u}\rangle\;.
\end{align}
For simplicity, we compute the action of $\Sop$ on an eigenstate
$|\mathbf{u}\rangle$, but the action on a product of elements of the
monodromy matrix can be computed along the same lines.  The derivation
of (\ref{main}) makes use of the PD relations.  At first order, we use
\begin{align}\label{1}
\mathrm{H}_{k}|\mathbf{u}\rangle&=-\mathrm{D}_k|\mathbf{u}\rangle,\qquad
k=1,\cdots L-1\\\nonumber
\mathrm{H}_{L}|\mathbf{u}\rangle&=-\mathrm{D}_L|\mathbf{u}\rangle+\EE_2|\mathbf{u}\rangle
\end{align}
At the second order, we use (\ref{2}) in the bulk and (\ref{boundary})
at the boundary and
\begin{align}\label{HD}
\mathrm{H}_{j}\mathrm{D}_k|\mathbf{u}\rangle=-\mathrm{H}_{j}\mathrm{H}_{k}|\mathbf{u}\rangle
\end{align}
Now we start the derivation.  The BDS eigenstate can be obtain from
the homogeneous XXX state as follows
\begin{align}
|\mathbf{u}\rangle_{\text{BDS}}=\mathrm{S}|\mathbf{u};\theta\rangle=\mathrm{S}
\exp\(\sum_{k=1}^L\theta_k\partial_k\)|\mathbf{u}\rangle
=\mathrm{S}\exp\(i\sum_{k=1}^L\nu_k\mathrm{D}_k\)|\mathbf{u}\rangle
=\mathscr{D}_\theta|\mathbf{u}\rangle
\end{align}
where $\theta_k$ is related to $\sone_k$ by the relation
\eqref{eq:inhomone} with $\sone_L=0$, and we define the operator
\begin{align}
\mathscr{D}_\theta\equiv \mathrm{S}
\exp\(i\sum_{k=1}^L\nu_k\mathrm{D}_k\)=1+\sum_{k=1}^\infty g^k
\mathscr{D}_\theta^{(k)}
\end{align}
We shall show this operator reproduce theta-morphism up to order
$g^2$.  At first order,
\begin{align}
g\mathscr{D}_\theta^{(1)}|\mathbf{u}\rangle&=\(i\sum_{k=1}^L\nu_k\mathrm{H}_{k}
+i\sum_{k=1}^L\nu_k\mathrm{D}_k\)|\mathbf{u}\rangle=\(-i\sum_{k=1}^L
\nu_k\mathrm{D}_k+i\sum_{k=1}^L\nu_k\mathrm{D}_k\)|\mathbf{u}\rangle=0\;,
\end{align}
where we have used (\ref{1}).  Hence the first order contribution
vanishes.  Note that by our choice of parameter $\mu_L=\mu_0=0$ hence
we do not need to consider the boundary operator.  As for the second
order, we consider separately the non-local and local contributions,
\begin{align}
\mathscr{D}_\theta^{(2)}=\mathscr{D}_{\text{NL}}^{(2)}+\mathscr{D}_{\text{L}}^{(2)}\;.
\end{align}
By non-local contribution we mean the case where two operators act
independently
\begin{align}
g^2\mathscr{D}_{\text{NL}}^{(2)}|\mathbf{u}\rangle=\sum_{|j-k|\ge
2}\nu_j\nu_k\(-\frac{1}{2}\mathrm{H}_{j}\mathrm{H}_{k}-\frac{1}{2}\mathrm{D}_j\mathrm{D}_k
+\mathrm{H}_{j}\mathrm{D}_k\)|\mathbf{u}\rangle=0\;,
\end{align}
hence the non-local terms do not contribute.  Note that again since
$\sone_L=0$, we do not need consider the boundary terms for non-local
terms.  For the local terms, we have
\begin{align}
g^2\mathscr{D}_{\text{L}}^{(2)}|\mathbf{u}\rangle&=-\frac{1}{2}\sum_{k=2}^L\rho_k[\mathrm{H}]_{k-1}|\mathbf{u}\rangle
-\frac{1}{2}\sum_{k=1}^L\nu_k\nu_{k-1}\{\mathrm{H}_{k-1},\mathrm{H}_{k}\}|\mathbf{u}\rangle\\\nonumber
&-\sum_{k=1}^L\nu_k\nu_{k-1}\mathrm{D}_k\mathrm{D}_{k-1}|\mathbf{u}\rangle-\sum_{k=1}^L\nu_k\nu_{k-1}(\mathrm{H}_{k}\mathrm{D}_{k-1}+\mathrm{H}_{k-1}\mathrm{D}_k)|\mathbf{u}\rangle\\\nonumber
&-\frac{1}{2}\sum_{k=1}^L\nu_k^2(\mathrm{H}^2_{k}+\mathrm{D}_k^2+2\mathrm{H}_{k}\mathrm{D}_k)|\mathbf{u}\rangle-\frac{\rho_1}{2}[\mathrm{H}]_L|\mathbf{u}\rangle\;.
\end{align}
Using (\ref{HD}),
\begin{align}\label{DL2}
g^2\mathscr{D}_{\text{L}}^{(2)}|\mathbf{u}\rangle&=-\frac{1}{2}\sum_{k=2}^L\rho_k[\mathrm{H}]_{k-1}|\mathbf{u}\rangle
+\frac{1}{2}\sum_{k=1}^L\nu_k\nu_{k-1}\{\mathrm{H}_{k-1},\mathrm{H}_{k}\}|\mathbf{u}\rangle\\\nonumber
&-\sum_{k=1}^L\nu_k\nu_{k-1}\mathrm{D}_k\mathrm{D}_{k-1}|\mathbf{u}\rangle-\frac{1}{2}\sum_{k=1}^L\nu_k^2(\mathrm{D}_k^2+2\mathrm{D}_k)|\mathbf{u}\rangle-g^2[\mathrm{H}]_L|\mathbf{u}\rangle\;.
\end{align}
Now we use (\ref{2}) to simplify the first line of (\ref{DL2}),
\begin{align}\label{DL22}
\frac{1}{2}\sum_{k=2}^L\rho_k[\mathrm{H}]_{k-1}|\mathbf{u}\rangle&=\frac{1}{4}\sum_{k=2}^L\rho_k[(\mathrm{D}_{k-1}^2+2\mathrm{D}_{k-1})-(\mathrm{D}_k^2+2\mathrm{D}_k)]|\mathbf{u}\rangle\\\nonumber
&=\frac{1}{4}\sum_{k=1}^L(\rho_{k+1}-\rho_k)(\mathrm{D}_k^2+2\mathrm{D}_k)|\mathbf{u}\rangle
+\frac{g^2}{2}[(\mathrm{D}_1^2+2\mathrm{D}_1)-(\mathrm{D}_L^2+2\mathrm{D}_L)]|\mathbf{u}\rangle\;.
\end{align}
We now use the equation \eqref{eq:sonestwo} with $\tcon_3=2g^2$, 
\begin{align}
\rho_{k+1}-\rho_k=2g^2+(\nu_{k+1}-2\nu_k+\nu_{k-1})\nu_k\;, 
\end{align}
that we substitute into \eqref{DL22}
\begin{align}\label{rterm}
-\frac{1}{2}\sum_{k=2}^L\rho_k[\mathrm{H}]_{k-1}|\mathbf{u}\rangle
&=-\frac{g^2}{2}\sum_{k=1}^L(\mathrm{D}_k^2+2\mathrm{D}_k)|\mathbf{u}\rangle-\frac{1}{4}\sum_{k=1}^L(\nu_{k-1}\nu_k+\nu_{k+1}\nu_k)(\mathrm{D}_k^2+2\mathrm{D}_k)|\mathbf{u}\rangle\\\nonumber
&+\frac{1}{2}\sum_{k=1}^L\nu_k^2(\mathrm{D}_k^2+2\mathrm{D}_k)-\frac{g^2}{2}[(\mathrm{D}_1^2+2\mathrm{D}_1)-(\mathrm{D}_L^2+2\mathrm{D}_L)]|\mathbf{u}\rangle\;.
\end{align}
We can also express the action of the anticommutators, using equation \eqref{eq:anticom},
\begin{align}\label{sterm}
&\frac{1}{2}\sum_{k=1}^L\nu_k\nu_{k-1}\{\mathrm{H}_{k-1},\mathrm{H}_{k}\}|\mathbf{u}\rangle\\\nonumber
&=\frac{1}{4}\sum_{k=1}^L\nu_{k-1}\nu_k[(\mathrm{D}_{k-1}^2+2\mathrm{D}_{k-1})
+(\mathrm{D}_k^2+2\mathrm{D}_k)]|\mathbf{u}\rangle+\sum_{k=1}^L\nu_{k-1}\nu_k\mathrm{D}_k\mathrm{D}_{k-1}|\mathbf{u}\rangle\\\nonumber
&=\frac{1}{4}\sum_{k=1}^L(\nu_k\nu_{k-1}+\nu_k\nu_{k+1})(\mathrm{D}_k^2+2\mathrm{D}_k)|\mathbf{u}\rangle+\sum_{k=1}^L\nu_{k-1}\nu_k\mathrm{D}_k\mathrm{D}_{k-1}|\mathbf{u}\rangle\;.
\end{align}
Last, we consider the boundary term, from (\ref{boundary}),
\begin{align}\label{bterm}
-g^2[\mathrm{H}]_{L}|\mathbf{u}\rangle&=-\frac{g^2}{2}[(\mathrm{D}_L^2+2\mathrm{D}_L)-(\mathrm{D}_1^2+2\mathrm{D}_1)]|\mathbf{u}\rangle-g^2\EE_2\mathrm{H}_{1,L}|\mathbf{u}\rangle-g^2C|\mathbf{u}\rangle\;.
\end{align}
Plugging (\ref{rterm}), (\ref{sterm}) and (\ref{bterm}) into
(\ref{DL2}), we obtain
\begin{align}
g^2\mathscr{D}_{\text{L}}^{(2)}|\mathbf{u}\rangle&=-\frac{g^2}{2}\sum_{k=1}^L(\mathrm{D}_k^2+2\mathrm{D}_k)|\mathbf{u}\rangle-g^2\EE_2\mathrm{H}_{1,L}|\mathbf{u}\rangle-g^2C|\mathbf{u}\rangle=g^2\mathscr{D}_\theta^{(2)}|\mathbf{u}\rangle\;,
\end{align}
since $\mathscr{D}_{\text{NL}}^{(2)}|\mathbf{u}\rangle=0$.  Therefore,
we have derived our main result (\ref{main}).

 \section{ Reduction Formula }
  \label{reduction}
  
In this section we prove a reduction formula for the functional $\caA$
that we use together with the freezing method.

\bigskip 

\noindent {\it Reduction formula:}
%
Let $\tilde {\bm \theta} =\{ \tilde
\theta _l\}_{l=1}^{\tilde L}$ and $ 
\tilde {\bm \theta}^\pm 
= \{ \tilde
\theta _l \pm \frac{i}{2} \}_{l=1}^{\tilde L}$.    Then
\begin{align}
\la{AuzA} \caA ^\pm _{ {\bf u} \cup \tilde {\bm \theta } ^\pm  , {\bm \theta}
\cup{\tilde {\bm \theta} }} = \caA^\pm _{{\bf u} ,{\bm \theta } } \, .
 \end{align}
The proof is based on the representation of the scalar product
(\ref{scalarpr}) and a reduction formula for the functional $\caA$
defined by (\ref{defcaA}).

  \bigskip
  \noindent {\it Proof:} By the definition (\ref{defcaA}),
\begin{align}
\la{defCA1biss} \caA ^\pm _{{\bf u} , {\bm \theta} } ={1\over \Delta_{\bf
u} } \prod_j\( 1- {Q_{\bm \theta } (u_j\mp \frac{i}{2})\over Q_{\bm \theta }
(u_j\pm \ihalf)} \, e^{\pm  i \, \d/\d u_j } \) \Delta_{\bf u} \, .
 \end{align}
Now compute the l.h.s., replacing in the last expression ${\bf u} \to
 {\bf u}    \cup\tilde {\bm \theta } ^\pm $ and ${\bm \theta } \to { \bm
\theta } \cup\tilde {\bm \theta } $:
\begin{align}
\la{defCA1sttt}\caA ^\pm _{ {\bf u} \cup \tilde {\bm \theta } ^\pm ,
{\bm \theta} \cup{\tilde {\bm \theta} }} &={1\over \Delta_{\bf u}
\Delta_{ \tilde {\bm \theta} ^\pm } \prod_{j,l} (u_j-\tilde \theta _l
\mp \ihalf)} \prod_{j} \(1 - {Q_{ \tilde {\bm \theta} \cup {\bm
\theta} } (u_j\mp \ihalf)\over Q_{ \tilde {\bm \theta} \cup {\bm
\theta} }(u_j\pm \ihalf )} \, e^{ \pm i \, \d/\d u_j } \) \\
 &\times \prod_{l=1}^{\tilde L} \( 1- {Q_{ \tilde {\bm \theta} \cup
 {\bm \theta} } (\tilde  \theta_l )\over Q_{ \tilde {\bm \theta} \cup
 {\bm \theta} }(\tilde \theta _l\pm i )} \ e^{\pm  \d/\d \tilde \theta _l} \)
 \
  \Delta_{\bf u} \Delta_{  \tilde
		  {\bm \theta}   ^\pm  }
		  \prod_{j,l} (u_j-\tilde \theta _l \mp \ihalf)
.
    \end{align}
   Since $Q_{ \tilde {\bm \theta} \cup {\bm \theta}}(\tilde \theta _l
   )=0$, the factors containing shift operators in $\tilde \theta_l$ are
   equal to 1.  But then we can also remove the Vandermonds
   $\Delta_{  \tilde
		  {\bm \theta}   ^\pm  } $ from both sides and write, using that
   $Q_{{\bm \theta} }(u-i/2 )= Q_{{\bm \theta} ^\pm }(u)$,
  \begin{align}
 \caA ^\pm _{{\bf u} \cup \tilde {\bm \theta} ^\pm , {\bm \theta} \cup
 {\tilde {\bm \theta} }} &= {1\over \Delta_{\bf u} \prod_{j,l}
 (u_j-\tilde \theta _l \mp \sfrac{i}{2})} \ \prod_{j} \(1 - {Q_{
 \tilde {\bm \theta} \cup {\bm \theta} } (u_j\mp \sfrac{i}{2})\over
 Q_{ \tilde {\bm \theta} \cup {\bm \theta} }(u_j\pm \sfrac{i}{2} )} \,
 e^{ \pm i \, \d/\d u_j } \) \Delta_{\bf u} \prod_{j,l} (u_j-\tilde
 \theta _l \mp \sfrac{i}{2}) \no \\
  \no &= {1\over \Delta_{\bf u} } \prod_{j} \(1 - {Q_{ \tilde {\bm
  \theta} } (u_j\mp \ihalf )\over Q_{ \tilde {\bm \theta} } (u_j \pm
  \ihalf )} {Q_{ \tilde {\bm \theta} \cup {\bm \theta} } (u_j\mp
  \ihalf)\over Q_{ \tilde {\bm \theta} \cup {\bm \theta} }(u_j\pm
  \ihalf )} \, e^{\pm i \, \d/\d u_j } \) \Delta_{\bf u} \\
   \no
  &= {1\over \Delta_{\bf u} } \prod_{j} \(1 - {Q_{{\bm \theta} }
  (u_j\mp \ihalf )\over Q_{{\bm \theta} }(u_j\pm \ihalf )} \, e^{\pm i \, \d/\d u_j }
  \) \Delta_{\bf u} = \caA^\pm  _{{\bf u} , {\bm \theta} } \;.\qquad \qquad\qquad \qquad \qquad\square
    \end{align}
%
 As a consequence of the reduction formula,
 denoting ${\bf z} =  {\bm \theta}^-$,
\begin{eqnarray}
\label{simple}
 \langle\raps{z^{(23)};{\bm \theta}^{(3)}}\wketin =
 \caA_{\raps{z^{(23)}} \cup {\bf u}^{(3)}, {\bm \theta}^{(3)} } =
 \caA_{\raps{z^{(23)}} \cup {\bf u}^{(3)}, {\bm \theta}^{(13)}\cup
 {\bm \theta}^{(23)} }= \caA_{ {\bf u}^{(3)}, {\bm \theta}^{(13)} } \;
 ;
\end{eqnarray}
\begin{eqnarray}
\label{involved}
 \langle\raps{u^{(2)}\cup {\bf z}^{(13)};{\bm \theta}^{(1)}}\uketin =
 \caA_{\raps{ {\bf u}^{(2)}\cup z^{(13)}} , {\bm \theta}^{(1)} } =
 \caA_{\raps{ {\bf u}^{(2)}\cup z^{(13)}} , {\bm \theta}^{(12)} \cup
 {\bm \theta}^{(13)} } = \caA_{\raps{ {\bf u}^{(2)} } , {\bm
 \theta}^{(12)} }\;.
\end{eqnarray}

\section{Calculation of Three-Point Function}
\label{app:threepcal}

In this section, we give the details of the computation of the three-
point function.  We have to compute the two factors, denoted {\bf
simple}, respectively {\bf involved} in \cite{GV},
\begin{eqnarray}
\label{simplesp1}
{\bf simple}&=&\langle \uparrow\ldots \uparrow\downarrow\ldots \downarrow|
\;\mathbb{I}_{3}\;  \delta \Sop_{3} \; \wketin,
\\
\no
{\bf involved}&=&\vbrain\delta \Sop_{2}^{-1}\,\mathbb{I}_{2}\; \CO_{12}\;\mathbb{I}_{1}\; \delta \Sop_{1}\;  \uketin.
\end{eqnarray}
The Hamiltonian insertions $\mathbb{I}_j$ and the operators $\delta
\Sop_j$ are given in equations \eqref{hamins} and \eqref{deltaS},
respectively.  As explained in the main text, we are going to use the
freezing trick, which allows to express
\begin{eqnarray}
\label{simplespfr}
\langle \uparrow\ldots \uparrow\downarrow\ldots \downarrow\; \wketin
&=&\langle\raps{z^{(23)};{\bm \theta}^{(3)}} \; \wketin\; ,\\
\no
\vbrain\;\CO_{12}\;  \uketin 
&=&\langle\raps{u^{(2)}\cup {\bf
z}^{(13)};{\bm \theta}^{(1)}}|\; \uketin\;.
\end{eqnarray}
 We have shown in the previous appendix that
 $\langle\raps{z^{(23)};{\bm \theta}^{(3)}}\wketin$ does not depend on
 the inhomogeneities ${\bm \theta}^{(23)}$ and moreover it is a
 symmetric function of the remaining inhomogeneities ${\bm
 \theta}^{(13)}$.  Using the equations \eqref{boundaryterms} to
 transform the permutations into derivatives, we obtain
\begin{align}
&\langle\raps{z^{(23)};{\bm \theta}^{(3)}}|
\;\mathbb{I}_{3}\;  \delta \Sop_{3} \; \wketin=\langle\raps{z^{(23)};{\bm \theta}^{(3)}} \wketin
+\\ 
\no &g^2\(\!  \D_{1}^{(3)}\!  +\!  \D_{L^{(13)}
+1}^{(3)}+\EE_2^{(3)}\D_{L^{(3)}}^{(3)}+i\EE_3^{(3)}+\textstyle{\frac{1}{2}}
( \D_{1}^{(3)2}\!  \!\!  + \D_{L^{(13)} \!  +1}^{(3)2}
-\D_{L^{(3)}}^{(3)2}-\D_{L^{(13)} }^{(3)2}\!  -\!  \EE_2^{(3)2})\)\,
\langle \raps{z}|\raps{u}^{(3)}\rangle\\ \no
&=\langle\raps{z^{(23)};{\bm \theta}^{(3)}}
\wketin+g^2\(-i\EE_2^{(3)}\partial_1^{(3)}+i\EE_3^{(3)}+\partial_1^{(3)}\partial_2^{(3)}-\textstyle{\frac{1}{2}}\EE_2^{(3)2}\)\,
\langle \raps{z}|\raps{u}^{(3)}\rangle
\end{align}
The  {\bf involved}  factor in (\ref{tpfho}) can be evaluated similarly; let us first consider
\begin{align}
 \CO_{12}\;\mathbb{I}_{1}\; \delta \Sop_{1}&\; \uketin=\CO_{12}\; \uketin 
\no\\ 
&+g^2\Big(\D_{1}^{(1)}+\D_{L^{(12)} +1}^{(1)}+\EE_2^{(1)}\D_{L^{(1)}}^{(1)}\no\\
&\qquad\qquad+i\EE_3^{(1)}+\frac{1}{2}(\D_{1}^{(1)2}+\D_{L^{(12)}
+1}^{(1)2}-\D_{L^{(1)}}^{(1)2}-\D_{L^{(12)}
}^{(1)2}-\EE_2^{(1)2})\Big) \CO^{(12)} |\raps{u}^{(1)}\rangle\no\\
& =\CO_{12}\; \uketin 
\no\\ 
&+g^2\Big(\D_{1}^{(2)}+\D_{L^{(12)} +1}^{(2)}+\EE_2^{(1)}\D_{L^{(2)}}^{(2)}\no\\
&\qquad\qquad+i\EE_3^{(1)}+\textstyle{\frac{1}{2}}(\D_{1}^{(2)2}+\D_{L^{(12)}
+1}^{(2)2}-\D_{L^{(2)}}^{(2)2}-\D_{L^{(12)}
}^{(2)2}-\EE_2^{(1)2})\Big) \CO_{12} |\raps{u}^{(1)}\rangle.
\end{align}
In the last line we have used that $\CO_{12}|\raps{v}\rangle$ does not depend neither on ${\bm \theta}^{(23)}$ nor on ${\bm \theta}^{(13)}$, so we can replace
the derivatives $\D_k^{(1)}$  by $\D_k^{(2)}$. 
Similarly, we obtain for the action on the bra vector
\begin{align}
& \vbrain\, \delta \Sop_{2}^{-1}\,\mathbb{I}_{2}\;= \vbrain +\\ \no
&g^2\big[\!  -\D_{1}^{(2)}\!  -\!  \D_{L^{(12)} \!  +\!  1}^{(2)}\!
-\!  \EE_2^{(2)}\D_{L^{(2)}}^{(2)}\!  -i \EE_3^{(2)}\!
+\textstyle{\frac{1}{2}}(\D_{1}^{(2)2}\!  +\D_{L^{(12)}
+1}^{(2)2}-\D_{L^{(2)}}^{(2)2}\!  -\D_{L^{(12)} }^{(2)2}\!
-\EE_2^{(2)2})\big] \langle \raps{u}^{(1)}|,
\end{align}
where we used that the action of the operators $\D_k$ on the left
vectors is $(\D_k|\raps{u}\rangle)^\dagger=-\D_k\langle\raps{u}|$.
Using the Leibniz rule, we have
$\D_k(\langle\raps{u}|\raps{v}\rangle)=\langle\raps{u}|\D_k|\raps{v}\rangle+\D_k(\langle\raps{u}|)|\raps{v}\rangle$.
This quantity is zero unless $k=L^{(a)}$ or $L^{(ab)}$, for the type
of vectors we use in this section.  We also have
\begin{equation}
D_k^2(\langle\raps{u}|\raps{v}\rangle)=D_k^2(\langle\raps{u}|)|\raps{v}\rangle+\langle\raps{u}|D_k^2(|\raps{v}\rangle)+2D_k(\langle\raps{u}|)D_k(|\raps{v}\rangle),
\end{equation}
and
$2D_k(\langle\raps{u}|)D_k(|\raps{v}\rangle)=4\langle\raps{u}|D_k|\raps{v}\rangle$.
For $k=L^{(12)} $ we use that
\begin{equation}
D_{L^{(12)} }^2(\langle\raps{u}|\CO_{12}|\raps{v}\rangle)=D_{L^{(12)} }^{(2)2}(\langle\raps{u}|)\CO_{12}|\raps{v}\rangle+\langle\raps{u}|D_{L^{(12)} }^{(1)2}(|\raps{v}\rangle)-2\langle\raps{u}|\ham_{L^{(12)} }^{(2)}\CO_{12} \ham_{L^{(12)} }^{(1)}|\raps{v}\rangle,
\end{equation}
with the last term being zero because $\ham_{L^{(12)} }^{(2)}\CO_{12}
\ham_{L^{(12)} }^{(1)}=0$, as noticed already in \cite{GV}.  For
$k=L^{(2)}$ one has
\begin{equation}
D_{L^{(2)}}^2(\langle\raps{u}|\CO_{12}|\raps{v}\rangle)=D_{L^{(2)}}^{(2)2}(\langle\raps{u}|)\CO_{12}|\raps{v}\rangle+\langle\raps{u}|D_{L^{(1)}}^{(1)2}(|\raps{v}\rangle)-2\langle
\raps{u}|\D_{L^{(2)}}^{(2)}\CO_{12} \D_{L^{(1)}}^{(1)}|\raps{v}\rangle\;.
\end{equation}
Proceeding as previously, we get
\begin{equation}
0=\langle\raps{u}|\ham_{L^{(2)}}^{(2)}\CO_{12}
\ham_{L_{1}}^{(1)}|\raps{v}\rangle=\langle\raps{u}|(E_2^{(2)}-\D_{L^{(2)}}^{(2)})\CO_{12}(E_2^{(1)}-\D_{L_{1}}^{(1)})
|\raps{v}\rangle\;,
\end{equation} 
so that 
\begin{equation}
\langle\raps{u}|\D_{L_{2}}^{(2)}\CO_{12}\D_{L_{1}}^{(1)}|\raps{v}\rangle=\langle\raps{u}|(E_2^{(2)}+E_2^{(1)})\D_{L_{2}}^{(2)}\CO_{12}
|\raps{v}\rangle-E_2^{(2)}E_2^{(1)}\langle\raps{u}|\CO_{12}
|\raps{v}\rangle\;.
\end{equation} 
Putting together the various identities above, we obtain for {\bf involved}
\begin{align}
\vbrain\, \delta \Sop_{2}^{-1}\,&\mathbb{I}_{2}\; \CO_{12}\;\mathbb{I}_{1}\; \delta \Sop_{1}\; \uketin
= \vbrain\, \CO_{12}\, \uketin\\
&+\frac{g^2}{2}\big(\D_{1}^{(2)2}+\D_{L^{(12)}
+1}^{(2)2}-\D_{L^{(2)}}^{(2)2}-\D_{L^{(12)} }^{(2)2}\no\\
&\qquad\qquad\qquad-(\EE_2^{(1)}-\EE_2^{(2)})^2+i(\EE_3^{(1)}-\EE_3^{(2)})\big)\vbrain \, \CO_{12}\,  \uketin.
\no
\end{align}
Since the scalar product $\vbrain\, \CO_{12}\, \uketin$ does not
depend on the inhomogeneities ${\bm \theta}^{(13)}$ or ${\bm
\theta}^{(23)}$ and is a symmetric function of the inhomogeneities
${\bm \theta}^{(12)}$, one can write
\begin{align}
 &\vbrain\, \delta \Sop_{2}^{-1}\,\mathbb{I}_{2}\; \CO_{12}\;\mathbb{I}_{1}\; \delta \Sop_{1}\; \uketin= \vbrain\, \CO_{12}\, \uketin\\
&\qquad\qquad+\frac{g^2}{2}(2\partial_1\partial_2-(\EE_2^{(1)}-\EE_2^{(2)})^2+i(\EE_3^{(1)}-\EE_3^{(2)}))\vbrain
\, \CO_{12}\, \uketin \no\;.
\end{align}
This finishes our derivation of the three-point function at one loop.

\begin{bibtex}{\jobname}

@article{Drummond:2009fd,
     author         = "Drummond, James M. and Henn, Johannes M. and Plefka, Jan",
     title          = "{Yangian symmetry of scattering amplitudes in N=4 super
                       Yang-Mills theory}",
     journal        = "JHEP",
     volume         = "0905",
     pages          = "046",
     doi            = "10.1088/1126-6708/2009/05/046",
     year           = "2009",
     eprint         = "0902.2987",
     archivePrefix  = "arXiv",
     primaryClass   = "hep-th",
     reportNumber   = "HU-EP-09-06, LAPTH-1308-09",
     SLACcitation   = "
}
@article{Bargheer:2009qu,
     author         = "Bargheer, Till and Beisert, Niklas and Galleas,
                       Wellington and Loebbert, Florian and McLoughlin, Tristan",
     title          = "{Exacting N=4 Superconformal Symmetry}",
     journal        = "JHEP",
     volume         = "0911",
     pages          = "056",
     doi            = "10.1088/1126-6708/2009/11/056",
     year           = "2009",
     eprint         = "0905.3738",
     archivePrefix  = "arXiv",
     primaryClass   = "hep-th",
     reportNumber   = "AEI-2009-048",
    SLACcitation   = "
}

@article{Korchemsky:2010ut,
      author         = "Korchemsky, G.P. and Sokatchev, E.",
      title          = "{Superconformal invariants for scattering amplitudes in
                        N=4 SYM theory}",
      journal        = "Nucl.Phys.",
      volume         = "B839",
      pages          = "377-419",
      doi            = "10.1016/j.nuclphysb.2010.05.022",
      year           = "2010",
      eprint         = "1002.4625",
      archivePrefix  = "arXiv",
      primaryClass   = "hep-th",
      SLACcitation   = "
}

@article{Drummond:2010zv,
      author         = "Drummond, J.M. and Ferro, L. and Ragoucy, E.",
      title          = "{Yangian symmetry of light-like Wilson loops}",
      journal        = "JHEP",
      volume         = "1111",
      pages          = "049",
      doi            = "10.1007/JHEP11(2011)049",
      year           = "2011",
      eprint         = "1011.4264",
      archivePrefix  = "arXiv",
      primaryClass   = "hep-th",
      reportNumber   = "CERN-PH-TH-2010-275, HU-EP-10-81",
      SLACcitation   = "
}

@article{Beisert:2010gn,
     author         = "Beisert, Niklas and Henn, Johannes and McLoughlin,
                       Tristan and Plefka, Jan",
     title          = "{One-Loop Superconformal and Yangian Symmetries of
                       Scattering Amplitudes in N=4 Super Yang-Mills}",
     journal        = "JHEP",
     volume         = "1004",
     pages          = "085",
     doi            = "10.1007/JHEP04(2010)085",
     year           = "2010",
     eprint         = "1002.1733",
     archivePrefix  = "arXiv",
     primaryClass   = "hep-th",
     reportNumber   = "AEI-2010-019, HU-EP-10-06",
     SLACcitation   = "
}

@article{Alday:2010vh,
     author         = "Alday, Luis F. and Maldacena, Juan and Sever, Amit and
                       Vieira, Pedro",
     title          = "{Y-system for Scattering Amplitudes}",
     journal        = "J.Phys.",
     volume         = "A43",
     pages          = "485401",
     doi            = "10.1088/1751-8113/43/48/485401",
     year           = "2010",
     eprint         = "1002.2459",
     archivePrefix  = "arXiv",
     primaryClass   = "hep-th",
     SLACcitation   = "
}

@article{CaronHuot:2011kk,
     author         = "Caron-Huot, Simon and He, Song",
     title          = "{Jumpstarting the All-Loop S-Matrix of Planar N=4 Super
                       Yang-Mills}",
     journal        = "JHEP",
     volume         = "1207",
     pages          = "174",
     doi            = "10.1007/JHEP07(2012)174",
     year           = "2012",
     eprint         = "1112.1060",
     archivePrefix  = "arXiv",
     primaryClass   = "hep-th",
     SLACcitation   = "
}

@article{Correa:2012hh,
      author         = "Correa, Diego and Maldacena, Juan and Sever, Amit",
      title          = "{The quark anti-quark potential and the cusp anomalous
                        dimension from a TBA equation}",
      journal        = "JHEP",
      volume         = "1208",
      pages          = "134",
      doi            = "10.1007/JHEP08(2012)134",
      year           = "2012",
      eprint         = "1203.1913",
      archivePrefix  = "arXiv",
      primaryClass   = "hep-th",
      SLACcitation   = "
}

@article{Drukker:2012de,
      author         = "Drukker, Nadav",
      title          = "{Integrable Wilson loops}",
      journal        = "JHEP",
      volume         = "1310",
      pages          = "135",
      doi            = "10.1007/JHEP10(2013)135",
      year           = "2013",
      eprint         = "1203.1617",
      archivePrefix  = "arXiv",
      primaryClass   = "hep-th",
      SLACcitation   = "
}

@article{Sever:2012qp,
     author         = "Sever, Amit and Vieira, Pedro and Wang, Tianheng",
     title          = "{From Polygon Wilson Loops to Spin Chains and Back}",
     journal        = "JHEP",
     volume         = "1212",
     pages          = "065",
     doi            = "10.1007/JHEP12(2012)065",
     year           = "2012",
     eprint         = "1208.0841",
     archivePrefix  = "arXiv",
     primaryClass   = "hep-th",
     SLACcitation   = "
}

@article{Ferro:2013dga,
     author         = "Ferro, Livia and Lukowski, Tomasz and Meneghelli, Carlo
                       and Plefka, Jan and Staudacher, Matthias",
     title          = "{Spectral Parameters for Scattering Amplitudes in N=4
                       Super Yang-Mills Theory}",
     year           = "2013",
     eprint         = "1308.3494",
     archivePrefix  = "arXiv",
     primaryClass   = "hep-th",
     reportNumber   = "HU-MATHEMATIK-2013-12, HU-EP-13-33, AEI-2013-235,
                       DESY-13-488,  --ZMP-HH-13-15",
     SLACcitation   = "
}

@article{Chicherin:2013ora,
     author         = "Chicherin, D. and Derkachov, S. and Kirschner, R.",
     title          = "{Yang-Baxter operators and scattering amplitudes in
                       $\mathcal{N} = 4$ super-Yang-Mills theory}",
     year           = "2013",
     eprint         = "1309.5748",
     archivePrefix  = "arXiv",
     primaryClass   = "hep-th",
     SLACcitation   = "
}

@article{Muller:2013rta,
     author         = "MŸller, Dennis and MŸnkler, Hagen and Plefka, Jan and
                       Pollok, Jonas and Zarembo, Konstantin",
     title          = "{Yangian Symmetry of smooth Wilson Loops in $\mathcal{N}
                       = $ 4 super Yang-Mills Theory}",
     journal        = "JHEP",
     volume         = "1311",
     pages          = "081",
     doi            = "10.1007/JHEP11(2013)081",
     year           = "2013",
     eprint         = "1309.1676",
     archivePrefix  = "arXiv",
     primaryClass   = "hep-th",
     reportNumber   = "HU-EP-13-42, NORDITA-2013-64, UUITP-10-13",
     SLACcitation   = "
}

@article{Basso:2013vsa,
     author         = "Basso, Benjamin and Sever, Amit and Vieira, Pedro",
     title          = "{Space-time S-matrix and Flux-tube S-matrix at Finite
                       Coupling}",
     journal        = "Phys.Rev.Lett.",
     volume         = "111",
     pages          = "091602",
     doi            = "10.1103/PhysRevLett.111.091602",
     year           = "2013",
     eprint         = "1303.1396",
     archivePrefix  = "arXiv",
     primaryClass   = "hep-th",
     SLACcitation   = "
}

@article{Basso:2013aha,
     author         = "Basso, Benjamin and Sever, Amit and Vieira, Pedro",
     title          = "{Space-time S-matrix and Flux tube S-matrix II.
                       Extracting and Matching Data}",
     year           = "2013",
     eprint         = "1306.2058",
     archivePrefix  = "arXiv",
     primaryClass   = "hep-th",
     SLACcitation   = "
}

@article{Bissi:2011dc,
      author         = "Bissi, A. and Kristjansen, C. and Young, D. and Zoubos,
                        K.",
      title          = "{Holographic three-point functions of giant gravitons}",
      journal        = "JHEP",
      volume         = "1106",
      pages          = "085",
      doi            = "10.1007/JHEP06(2011)085",
      year           = "2011",
      eprint         = "1103.4079",
      archivePrefix  = "arXiv",
      primaryClass   = "hep-th",
      SLACcitation   = "
}

@article{Bargheer:2013faa,
      author         = "Bargheer, Till and Minahan, Joseph A. and Pereira, Raul",
      title          = "{Computing Three-Point Functions for Short Operators}",
      year           = "2013",
      eprint         = "1311.7461",
      archivePrefix  = "arXiv",
      primaryClass   = "hep-th",
      reportNumber   = "DESY-13-201, UUITP-17-13",
      SLACcitation   = "
}

@article{Minahan:2012fh,
      author         = "Minahan, Joseph A.",
      title          = "{Holographic three-point functions for short operators}",
      journal        = "JHEP",
      volume         = "1207",
      pages          = "187",
      doi            = "10.1007/JHEP07(2012)187",
      year           = "2012",
      eprint         = "1206.3129",
      archivePrefix  = "arXiv",
      primaryClass   = "hep-th",
      reportNumber   = "UUITP-02-12",
      SLACcitation   = "
}

@article{Klose:2011rm,
      author         = "Klose, Thomas and McLoughlin, Tristan",
      title          = "{A light-cone approach to three-point functions in $\mathrm{AdS}_5 \times \mathrm{S}^5$}",
      journal        = "JHEP",
      volume         = "1204",
      pages          = "080",
      doi            = "10.1007/JHEP04(2012)080",
      year           = "2012",
      eprint         = "1106.0495",
      archivePrefix  = "arXiv",
      primaryClass   = "hep-th",
      reportNumber   = "UUITP-16-11, AEI-2011-031",
      SLACcitation   = "
}

@article{Costa:2011mg,
      author         = "Costa, Miguel S. and Penedones, Joao and Poland, David
                        and Rychkov, Slava",
      title          = "{Spinning Conformal Correlators}",
      journal        = "JHEP",
      volume         = "1111",
      pages          = "071",
      doi            = "10.1007/JHEP11(2011)071",
      year           = "2011",
      eprint         = "1107.3554",
      archivePrefix  = "arXiv",
      primaryClass   = "hep-th",
      reportNumber   = "LPTENS-11-22, NSF-KITP-11-128",
      SLACcitation   = "
}

@article{Wheeler:2013zja,
      author         = "Wheeler, M.",
      title          = "{Multiple integral formulae for the scalar product of
                        on-shell and off-shell Bethe vectors in SU(3)-invariant
                        models}",
      journal        = "Nucl.Phys.",
      volume         = "B875",
      pages          = "186-212",
      doi            = "10.1016/j.nuclphysb.2013.06.015",
      year           = "2013",
      eprint         = "1306.0552",
      archivePrefix  = "arXiv",
      primaryClass   = "math-ph",
      SLACcitation   = "
}

@article{Elvang:2013cua,
      author         = "Elvang, Henriette and Huang, Yu-tin",
      title          = "{Scattering Amplitudes}",
      year           = "2013",
      eprint         = "1308.1697",
      archivePrefix  = "arXiv",
      primaryClass   = "hep-th",
      SLACcitation   = "
}

@article{Maldacena:1997re,
      author         = "Maldacena, Juan Martin",
      title          = "{The Large N limit of superconformal field theories and
                        supergravity}",
      journal        = "Adv.Theor.Math.Phys.",
      volume         = "2",
      pages          = "231-252",
      year           = "1998",
      eprint         = "hep-th/9711200",
      archivePrefix  = "arXiv",
      primaryClass   = "hep-th",
      reportNumber   = "HUTP-97-A097",
      SLACcitation   = "
}

@article{Beisert:2010jr,
      author         = "Beisert, Niklas and Ahn, Changrim and Alday, Luis F. and
                        Bajnok, Zoltan and Drummond, James M. and others",
      title          = "{Review of AdS/CFT Integrability: An Overview}",
      journal        = "Lett.Math.Phys.",
      volume         = "99",
      pages          = "3-32",
      doi            = "10.1007/s11005-011-0529-2",
      year           = "2012",
      eprint         = "1012.3982",
      archivePrefix  = "arXiv",
      primaryClass   = "hep-th",
      reportNumber   = "AEI-2010-175, CERN-PH-TH-2010-306, HU-EP-10-87,
                        HU-MATH-2010-22, KCL-MTH-10-10, UMTG-270, UUITP-41-10",
      SLACcitation   = "
}

@article{KKnew,
      author         = "Kazama, Yoichi and Komatsu, Shota",
      title          = "{Three-point functions in the SU(2) sector at strong
                        coupling}",
      year           = "2013",
      eprint         = "1312.3727",
      archivePrefix  = "arXiv",
      primaryClass   = "hep-th",
      reportNumber   = "UT-KOMABA-13-16",
      SLACcitation   = "
}

@article{Sklyanin:1991ss,
      author         = "Sklyanin, E.K.",
      title          = "{Quantum inverse scattering method. Selected topics}",
      year           = "1991",
      eprint         = "hep-th/9211111",
      archivePrefix  = "arXiv",
      primaryClass   = "hep-th",
      reportNumber   = "HU-TFT-91-51",
      SLACcitation   = "
}

@article{Bargheer:2011mm,
      author         = "Bargheer, Till and Beisert, Niklas and Loebbert, Florian",
      title          = "{Exact Superconformal and Yangian Symmetry of Scattering
                        Amplitudes}",
      journal        = "J.Phys.",
      volume         = "A44",
      pages          = "454012",
      doi            = "10.1088/1751-8113/44/45/454012",
      year           = "2011",
      eprint         = "1104.0700",
      archivePrefix  = "arXiv",
      primaryClass   = "hep-th",
      reportNumber   = "AEI-2011-016, LPT-ENS-11-12, UUITP-11-11",
      SLACcitation   = "
}

@article{Beisert:2003tq,
      author         = "Beisert, N. and Kristjansen, C. and Staudacher, M.",
      title          = "{The Dilatation operator of conformal N=4 superYang-Mills
                        theory}",
      journal        = "Nucl.Phys.",
      volume         = "B664",
      pages          = "131-184",
      doi            = "10.1016/S0550-3213(03)00406-1",
      year           = "2003",
      eprint         = "hep-th/0303060",
      archivePrefix  = "arXiv",
      primaryClass   = "hep-th",
      reportNumber   = "AEI-2003-028",
      SLACcitation   = "
}

@article{Rej:2005qt,
      author         = "Rej, Adam and Serban, Didina and Staudacher, Matthias",
      title          = "{Planar N=4 gauge theory and the Hubbard model}",
      journal        = "JHEP",
      volume         = "0603",
      pages          = "018",
      doi            = "10.1088/1126-6708/2006/03/018",
      year           = "2006",
      eprint         = "hep-th/0512077",
      archivePrefix  = "arXiv",
      primaryClass   = "hep-th",
      reportNumber   = "AEI-2005-164, SPHT-T05-190, NSF-KITP-05-84",
      SLACcitation   = "
}

@article{Bernard:1993va,
      author         = "Bernard, Denis and Gaudin, M. and Haldane, F.D.M. and
                        Pasquier, V.",
      title          = "{Yang-Baxter equation in long range interacting system}",
      journal        = "J.Phys.",
      volume         = "A26",
      pages          = "5219-5236",
      doi            = "10.1088/0305-4470/26/20/010",
      year           = "1993",
      SLACcitation   = "
}

@article{GV,
      author         = "Gromov, Nikolay and Vieira, Pedro",
      title          = "{Tailoring Three-Point Functions and Integrability IV.
                        Theta-morphism}",
      year           = "2012",
      eprint         = "1205.5288",
      archivePrefix  = "arXiv",
      primaryClass   = "hep-th",
      SLACcitation   = "
}

@article{Didina-Dunkl-2,
      author         = "Serban, Didina",
      title          = "{Eigenvectors and scalar products for long range
                        interacting spin chains II: the finite size effects}",
      journal        = "JHEP",
      volume         = "1308",
      pages          = "128",
      doi            = "10.1007/JHEP08(2013)128",
      year           = "2013",
      eprint         = "1302.3350",
      archivePrefix  = "arXiv",
      primaryClass   = "hep-th",
      SLACcitation   = "
}

@article{Serban:2013jua,
      author         = "Serban, Didina",
      title          = "{Eigenvectors and scalar products for long range
                        interacting spin chains II: the finite size effects}",
      journal        = "JHEP",
      volume         = "1308",
      pages          = "128",
      doi            = "10.1007/JHEP08(2013)128",
      year           = "2013",
      eprint         = "1302.3350",
      archivePrefix  = "arXiv",
      primaryClass   = "hep-th",
      SLACcitation   = "
}

@article{Serban:2012dr,
      author         = "Serban, Didina",
      title          = "{A note on the eigenvectors of long-range spin chains and
                        their scalar products}",
      journal        = "JHEP",
      volume         = "1301",
      pages          = "012",
      year           = "2013",
      eprint         = "1203.5842",
      archivePrefix  = "arXiv",
      primaryClass   = "hep-th",
      SLACcitation   = "
}

@article{Beisert:2004hm,
      author         = "Beisert, N. and Dippel, V. and Staudacher, M.",
      title          = "{A Novel long range spin chain and planar N=4 super
                        Yang-Mills}",
      journal        = "JHEP",
      volume         = "0407",
      pages          = "075",
      doi            = "10.1088/1126-6708/2004/07/075",
      year           = "2004",
      eprint         = "hep-th/0405001",
      archivePrefix  = "arXiv",
      primaryClass   = "hep-th",
      reportNumber   = "AEI-2004-036",
      SLACcitation   = "
}

@article{Serban:2004jf,
      author         = "Serban, Didina and Staudacher, Matthias",
      title          = "{Planar N=4 gauge theory and the Inozemtsev long range
                        spin chain}",
      journal        = "JHEP",
      volume         = "0406",
      pages          = "001",
      doi            = "10.1088/1126-6708/2004/06/001",
      year           = "2004",
      eprint         = "hep-th/0401057",
      archivePrefix  = "arXiv",
      primaryClass   = "hep-th",
      reportNumber   = "AEI-2004-001, SACLAY-SPHT-T04-002",
      SLACcitation   = "
}

@article{Inozemtsev:2002vb,
      author         = "Inozemtsev, V.I.",
      title          = "{Integrable Heisenberg-van Vleck chains with variable
                        range exchange}",
      journal        = "Phys.Part.Nucl.",
      volume         = "34",
      pages          = "166-193",
      year           = "2003",
      eprint         = "hep-th/0201001",
      archivePrefix  = "arXiv",
      primaryClass   = "hep-th",
      SLACcitation   = "
}

@Article{Tetelman:1981xx,
     author    = "M.G. Tetelman",
     title     = "{Lorentz group for two-dimensional integrable lattice systems}",
     journal   = "Sov. Phys. JETP.",
     volume    = "55",
     year      = "1982",
     pages     = "306",
     SLACcitation  = "
}

@article{Bargheer:2008jt,
      author         = "Bargheer, Till and Beisert, Niklas and Loebbert, Florian",
      title          = "{Boosting Nearest-Neighbour to Long-Range Integrable Spin
                        Chains}",
      journal        = "J.Stat.Mech.",
      volume         = "0811",
      pages          = "L11001",
      doi            = "10.1088/1742-5468/2008/11/L11001",
      year           = "2008",
      eprint         = "0807.5081",
      archivePrefix  = "arXiv",
      primaryClass   = "hep-th",
      reportNumber   = "AEI-2008-052",
      SLACcitation   = "
}

@article{Bargheer:2009xy,
      author         = "Bargheer, Till and Beisert, Niklas and Loebbert, Florian",
      title          = "{Long-Range Deformations for Integrable Spin Chains}",
      journal        = "J.Phys.",
      volume         = "A42",
      pages          = "285205",
      doi            = "10.1088/1751-8113/42/28/285205",
      year           = "2009",
      eprint         = "0902.0956",
      archivePrefix  = "arXiv",
      primaryClass   = "hep-th",
      reportNumber   = "AEI-2009-009",
      SLACcitation   = "
}

@article{Loebbert:2012yd,
      author         = "Loebbert, Florian",
      title          = "{Recursion Relations for Long-Range Integrable Spin
                        Chains with Open Boundary Conditions}",
      journal        = "Phys.Rev.",
      volume         = "D85",
      pages          = "086008",
      doi            = "10.1103/PhysRevD.85.086008",
      year           = "2012",
      eprint         = "1201.0888",
      archivePrefix  = "arXiv",
      primaryClass   = "hep-th",
      reportNumber   = "LPT-ENS-12-01, AEI-2012-000",
      SLACcitation   = "
}

@Article{Thacker:1985gz,
     author    = "Thacker, H. B.",
     title     = "{Corner Transfer Matrices and Lorentz Invariance on a
                  Lattice}",
     journal   = "Physica",
     volume    = "18D",
     year      = "1986",
     pages     = "348-359",
     SLACcitation  = "
}

@article{Itoyama:1986ad,
      author         = "Itoyama, H. and Thacker, H.B.",
      title          = "{Lattice Virasoro Algebra and Corner Transfer Matrices in
                        the Baxter Eight Vertex Model}",
      journal        = "Phys.Rev.Lett.",
      volume         = "58",
      pages          = "1395",
      doi            = "10.1103/PhysRevLett.58.1395",
      year           = "1987",
      reportNumber   = "FERMILAB-PUB-86-152-T",
      SLACcitation   = "
}

@article{Baxter:1976uh,
      author         = "Baxter, R.J.",
      title          = "{Corner transfer matrices of the eight-vertex model. 1.
                        Low-temperature expansions and conjectured properties}",
      journal        = "J.Statist.Phys.",
      volume         = "15",
      pages          = "485-503",
      doi            = "10.1007/BF01020802",
      year           = "1976",
      SLACcitation   = "
}

@article{Loebbert:2010thesis,
      author         = "Loebbert, Florian",
      title          = "{Integrable Spin Chains in $\mathcal{N}=4$ super Yang--Mills Theory}",
       note= "PhD thesis",
      year= "2010",
}

@article{Beisert:2013voa,
      author         = "Beisert, Niklas and FiŽvet, Lucas and de Leeuw, Marius
                        and Loebbert, Florian",
      title          = "{Integrable Deformations of the XXZ Spin Chain}",
      journal        = "J.Stat.Mech.",
      volume         = "2013",
      pages          = "P09028",
      doi            = "10.1088/1742-5468/2013/09/P09028",
      year           = "2013",
      eprint         = "1308.1584",
      archivePrefix  = "arXiv",
      primaryClass   = "math-ph",
      SLACcitation   = "
}

@article{sz,
      author         = "Kostov, Ivan and Matsuo, Yutaka",
      title          = "{Inner products of Bethe states as partial domain wall
                        partition functions}",
      journal        = "JHEP",
      volume         = "1210",
      pages          = "168",
      doi            = "10.1007/JHEP10(2012)168",
      year           = "2012",
      eprint         = "1207.2562",
      archivePrefix  = "arXiv",
      primaryClass   = "hep-th",
      reportNumber   = "IPHT-T12-049, UT12-16",
      SLACcitation   = "
}

@article{Kazakov:2004qf,
      author         = "Kazakov, V.A. and Marshakov, A. and Minahan, J.A. and
                        Zarembo, K.",
      title          = "{Classical/quantum integrability in AdS/CFT}",
      journal        = "JHEP",
      volume         = "0405",
      pages          = "024",
      doi            = "10.1088/1126-6708/2004/05/024",
      year           = "2004",
      eprint         = "hep-th/0402207",
      archivePrefix  = "arXiv",
      primaryClass   = "hep-th",
      reportNumber   = "LPTENS-04-09, FIAN-TD-04-04, MPG-ITEP-11-04,
                        IHES-P-04-05, UUITP-05-04, CTP-MIT-3472",
      SLACcitation   = "
}

@article{EGSV,
      author         = "Escobedo, Jorge and Gromov, Nikolay and Sever, Amit and
                        Vieira, Pedro",
      title          = "{Tailoring Three-Point Functions and Integrability}",
      journal        = "JHEP",
      volume         = "1109",
      pages          = "028",
      doi            = "10.1007/JHEP09(2011)028",
      year           = "2011",
      eprint         = "1012.2475",
      archivePrefix  = "arXiv",
      primaryClass   = "hep-th",
      SLACcitation   = "
}

@article{Alday:2005nd,
      author         = "Alday, Luis F. and David, Justin R. and Gava, Edi and
                        Narain, K.S.",
      title          = "{Structure constants of planar N = 4 Yang Mills at one
                        loop}",
      journal        = "JHEP",
      volume         = "0509",
      pages          = "070",
      doi            = "10.1088/1126-6708/2005/09/070",
      year           = "2005",
      eprint         = "hep-th/0502186",
      archivePrefix  = "arXiv",
      primaryClass   = "hep-th",
      reportNumber   = "SPIN-05-06, ITP-05-08",
      SLACcitation   = "
}

@article{Kostov:2012jr,
      author         = "Kostov, Ivan",
      title          = "{Classical Limit of the Three-Point Function of N=4
                        Supersymmetric Yang-Mills Theory from Integrability}",
      journal        = "Phys.Rev.Lett.",
      volume         = "108", 
      pages          = "261604",
      doi            = "10.1103/PhysRevLett.108.261604",
      year           = "2012",
      eprint         = "1203.6180",
      archivePrefix  = "arXiv",
      primaryClass   = "hep-th",
      reportNumber   = "IPHT-T12-023",
      SLACcitation   = "
}

@article{Foda:2011rr,
      author         = "Foda, Omar",
      title          = "{N=4 SYM structure constants as determinants}",
      journal        = "JHEP",
      volume         = "1203",
      pages          = "096",
      doi            = "10.1007/JHEP03(2012)096",
      year           = "2012",
      eprint         = "1111.4663",
      archivePrefix  = "arXiv",
      primaryClass   = "math-ph",
      SLACcitation   = "
}

@article{NSlavnov1,
	Author = {Slavnov, N. A.},
	Journal = {Theoretical and Mathematical Physics},
	Pages = {502-508},
	Title = {Calculation of scalar products of wave functions and form factors in the framework of the algebraic Bethe ansatz},
	Volume = {79},
	Year = {1989}}

@article{Gromov:2012vu,
      author         = "Gromov, Nikolay and Vieira, Pedro",
      title          = "{Quantum Integrability for Three-Point Functions}",
      journal        = "Phys.Rev.Lett.",
      volume         = "111",
      pages          = "211601",
      doi            = "10.1103/PhysRevLett.111.211601",
      year           = "2013",
      eprint         = "1202.4103",
      archivePrefix  = "arXiv",
      primaryClass   = "hep-th",
      SLACcitation   = "
}

@article{Haldane:1987gg,
      author         = "Haldane, F.D.M.",
      title          = "{Exact Jastrow-Gutzwiller resonating valence bond ground
                        state of the spin 1/2 antiferromagnetic Heisenberg chain
                        with $1/r^2$ exchange}",
      journal        = "Phys.Rev.Lett.",
      volume         = "60",
      pages          = "635",
      doi            = "10.1103/PhysRevLett.60.635",
      year           = "1988",
      SLACcitation   = "
}
	
@article{Shastry:1987gh,
      author         = "Sriram Shastry, B.",
      title          = "{Exact solution of an S = 1/2 Heisenberg
                        antiferromagnetic chain with long ranged interactions}",
      journal        = "Phys.Rev.Lett.",
      volume         = "60",
      pages          = "639",
      doi            = "10.1103/PhysRevLett.60.639",
      year           = "1988",
      SLACcitation   = "
}

@article{Arutyunov:2004vx,
      author         = "Arutyunov, Gleb and Frolov, Sergey and Staudacher,
                        Matthias",
      title          = "{Bethe ansatz for quantum strings}",
      journal        = "JHEP",
      volume         = "0410",
      pages          = "016",
      doi            = "10.1088/1126-6708/2004/10/016",
      year           = "2004",
      eprint         = "hep-th/0406256",
      archivePrefix  = "arXiv",
      primaryClass   = "hep-th",
      reportNumber   = "AEI-2004-046",
      SLACcitation   = "
}

@article{Beisert:2006ib,
      author         = "Beisert, Niklas and Hernandez, Rafael and Lopez,
                        Esperanza",
      title          = "{A Crossing-symmetric phase for $\mathrm{AdS}_5 \times \mathrm{S}^5$ strings}",
      journal        = "JHEP",
      volume         = "0611",
      pages          = "070",
      doi            = "10.1088/1126-6708/2006/11/070",
      year           = "2006",
      eprint         = "hep-th/0609044",
      archivePrefix  = "arXiv",
      primaryClass   = "hep-th",
      reportNumber   = "AEI-2006-068, CERN-PH-TH-2006-176, IFT-UAM-CSIC-06-44,
                        PUTP-2208",
      SLACcitation   = "
}

@article{BES,
      author         = "Beisert, Niklas and Eden, Burkhard and Staudacher,
                        Matthias",
      title          = "{Transcendentality and Crossing}",
      journal        = "J.Stat.Mech.",
      volume         = "0701",
      pages          = "P01021",
      doi            = "10.1088/1742-5468/2007/01/P01021",
      year           = "2007",
      eprint         = "hep-th/0610251",
      archivePrefix  = "arXiv",
      primaryClass   = "hep-th",
      reportNumber   = "AEI-2006-079, ITP-UU-06-44, SPIN-06-34",
      SLACcitation   = "
}

@article{SL,
      author         = "Kostov, Ivan",
      title          = "{Three-point function of semiclassical states at weak
                        coupling}",
      journal        = "J.Phys.",
      volume         = "A45",
      pages          = "494018",
      doi            = "10.1088/1751-8113/45/49/494018",
      year           = "2012",
      eprint         = "1205.4412",
      archivePrefix  = "arXiv",
      primaryClass   = "hep-th",
      reportNumber   = "IPHT-T12-035",
      SLACcitation   = "
}

@article{EldadI,
	Author = {Eldad Bettelheim, Ivan Kostov},
	Title = {in preparation}
}

@article{Sutherland,
      author         = "Sutherland, Bill",
      title          = "{Low-Lying Eigenstates of the One-Dimensional Heisenberg
                        Ferromagnet for any Magnetization and Momentum}",
      journal        = "Phys.Rev.Lett.",
      volume         = "74",
      pages          = "816-819",
      doi            = "10.1103/PhysRevLett.74.816",
      year           = "1995",
      SLACcitation   = "
}

@article{Beisert:2003xu,
	Archiveprefix = {arXiv},
	Author = {Beisert, N. and Minahan, J. A. and Staudacher, M. and Zarembo, K.},
	Eprint = {hep-th/0306139},
	Journal = {JHEP},
	Pages = {010},
	Slaccitation = {
	Title = {{Stringing spins and spinning strings}},
	Volume = {09},
	Year = {2003},
	}

@article{Minahan:2005jq,
	Archiveprefix = {arXiv},
	Author = {Minahan, Joseph A.},
	Date-Added = {2009-05-23 15:19:29 +0200},
	Date-Modified = {2009-05-23 15:19:29 +0200},
	Doi = {10.1002/prop.200410204},
	Eprint = {hep-th/0503143},
	Journal = {Fortsch. Phys.},
	Pages = {828-838},
	Slaccitation = {
	Title = {{The SU(2) sector in AdS/CFT}},
	Volume = {53},
	Year = {2005},
}

@article{Beisert:2004ag,
	Archiveprefix = {arXiv},
	Author = {Beisert, N. and V. Kazakov and Sakai, K.},
	Date-Modified = {2009-10-15 08:45:06 +0200},
	Eprint = {hep-th/0410253},
	Journal = {Commun. Math. Phys.},
	Pages = {611-657},
	Slaccitation = {
	Title = {{Algebraic curve for the SO(6) sector of AdS/CFT}},
	Volume = {263},
	Year = {2006},
	}

@article{Frolov-Tseytlin,
      author         = "Frolov, S. and Tseytlin, Arkady A.",
      title          = "{Semiclassical quantization of rotating superstring in
                        $\mathrm{AdS}_5 \times \mathrm{S}^5$}",
      journal        = "JHEP",
      volume         = "0206",
      pages          = "007",
      doi            = "10.1088/1126-6708/2002/06/007",
      year           = "2002",
      eprint         = "hep-th/0204226",
      archivePrefix  = "arXiv",
      primaryClass   = "hep-th",
      SLACcitation   = "
}

@article{Didina-globalangles,
	Author = {Didina Serban},
	Date-Added = {2013-11-15 11:10:53 +0000},
	Date-Modified = {2013-11-15 11:11:26 +0000},
	Title = {Unpublished}
}

@article{GSV,
      author         = "Gromov, Nikolay and Sever, Amit and Vieira, Pedro",
      title          = "{Tailoring Three-Point Functions and Integrability III.
                        Classical Tunneling}",
      journal        = "JHEP",
      volume         = "1207",
      pages          = "044",
      doi            = "10.1007/JHEP07(2012)044",
      year           = "2012",
      eprint         = "1111.2349",
      archivePrefix  = "arXiv",
      primaryClass   = "hep-th",
      SLACcitation   = "
}

@article{Callan:2003xr,
      author         = "Callan, Curtis G., Jr. and Lee, Hok Kong and McLoughlin,
                        Tristan and Schwarz, John H. and Swanson, Ian and others",
      title          = "{Quantizing string theory in $\mathrm{AdS}_5\times \mathrm{S}^5$: Beyond the pp
                        wave}",
      journal        = "Nucl.Phys.",
      volume         = "B673",
      pages          = "3-40",
      doi            = "10.1016/j.nuclphysb.2003.09.008",
      year           = "2003",
      eprint         = "hep-th/0307032",
      archivePrefix  = "arXiv",
      primaryClass   = "hep-th",
      reportNumber   = "PUPT-2089, CALT-68-2426",
      SLACcitation   = "
}

@article{Sobko:3ptsl2,
      author         = "Sobko, Evgeny",
      title          = "{A new representation for two- and three-point
                        correlators of operators from sl(2) sector}",
      year           = "2013",
      eprint         = "1311.6957",
      archivePrefix  = "arXiv",
      primaryClass   = "hep-th",
      reportNumber   = "LPT-ENS-13-22",
      SLACcitation   = "
}

 @article{Pedro:sl23pt,
      author         = "Vieira, Pedro and Wang, Tianheng",
      title          = "{Tailoring Non-Compact Spin Chains}",
      year           = "2013",
      eprint         = "1311.6404",
      archivePrefix  = "arXiv",
      primaryClass   = "hep-th",
      SLACcitation   = "
}

@article{Foda:su3,
      author         = "Foda, Omar and Jiang, Yunfeng and Kostov, Ivan and
                        Serban, Didina",
      title          = "{A tree-level 3-point function in the su(3)-sector of
                        planar N=4 SYM}",
      journal        = "JHEP",
      volume         = "1310",
      pages          = "138",
      doi            = "10.1007/JHEP10(2013)138",
      year           = "2013",
      eprint         = "1302.3539",
      archivePrefix  = "arXiv",
      primaryClass   = "hep-th",
      reportNumber   = "IPHT-T13-034",
      SLACcitation   = "
}

@article{Alday:higherspin3pt,
      author         = "Alday, Luis F. and Bissi, Agnese",
      title          = "{Higher-spin correlators}",
      journal        = "JHEP",
      volume         = "1310",
      pages          = "202",
      doi            = "10.1007/JHEP10(2013)202",
      year           = "2013",
      eprint         = "1305.4604",
      archivePrefix  = "arXiv",
      primaryClass   = "hep-th",
      SLACcitation   = "
}

@article{Kazakov:3pttwist2,
      author         = "Kazakov, Vladimir and Sobko, Evgeny",
      title          = "{Three-point correlators of twist-2 operators in N=4 SYM
                        at Born approximation}",
      journal        = "JHEP",
      volume         = "1306",
      pages          = "061",
      doi            = "10.1007/JHEP06(2013)061",
      year           = "2013",
      eprint         = "1212.6563",
      archivePrefix  = "arXiv",
      primaryClass   = "hep-th",
      reportNumber   = "LPT-ENS-12-47",
      SLACcitation   = "
}

@article{Plefka:3pttwist2,
      author         = "Plefka, Jan and Wiegandt, Konstantin",
      title          = "{Three-Point Functions of Twist-Two Operators in N=4 SYM
                        at One Loop}",
      journal        = "JHEP",
      volume         = "1210",
      pages          = "177",
      doi            = "10.1007/JHEP10(2012)177",
      year           = "2012",
      eprint         = "1207.4784",
      archivePrefix  = "arXiv",
      primaryClass   = "hep-th",
      reportNumber   = "HU-EP-12-23",
      SLACcitation   = "
}

@article{Komatsu:3pt2,
      author         = "Kazama, Yoichi and Komatsu, Shota",
      title          = "{Wave functions and correlation functions for GKP strings
                        from integrability}",
      journal        = "JHEP",
      volume         = "1209",
      pages          = "022",
      doi            = "10.1007/JHEP09(2012)022",
      year           = "2012",
      eprint         = "1205.6060",
      archivePrefix  = "arXiv",
      primaryClass   = "hep-th",
      reportNumber   = "UT-KOMABA-12-4",
      SLACcitation   = "
}

@article{Caetano:3ptstrong,
      author         = "Caetano, J. and Toledo, J.",
      title          = "{$\chi$-Systems for Correlation Functions}",
      year           = "2012",
      eprint         = "1208.4548",
      archivePrefix  = "arXiv",
      primaryClass   = "hep-th",
      SLACcitation   = "
}

@article{Plefka:3ptlength5,
      author         = "Georgiou, George and Gili, Valeria and Grossardt, Andre
                        and Plefka, Jan",
      title          = "{Three-point functions in planar $\mathcal{N}=4$ super Yang-Mills
                        Theory for scalar operators up to length five at the
                        one-loop order}",
      journal        = "JHEP",
      volume         = "1204",
      pages          = "038",
      doi            = "10.1007/JHEP04(2012)038",
      year           = "2012",
      eprint         = "1201.0992",
      archivePrefix  = "arXiv",
      primaryClass   = "hep-th",
      reportNumber   = "HU-EP-12-02, QMUL-PH-11-23",
      SLACcitation   = "
}

@article{Bissi:holographic3pt,
      author         = "Bissi, Agnese and Harmark, Troels and Orselli, Marta",
      title          = "{Holographic 3-Point Function at One Loop}",
      journal        = "JHEP",
      volume         = "1202",
      pages          = "133",
      doi            = "10.1007/JHEP02(2012)133",
      year           = "2012",
      eprint         = "1112.5075",
      archivePrefix  = "arXiv",
      primaryClass   = "hep-th",
      SLACcitation   = "
}
@article{Komatsu:3pt1,
      author         = "Kazama, Yoichi and Komatsu, Shota",
      title          = "{On holographic three point functions for GKP strings
                        from integrability}",
      journal        = "JHEP",
      volume         = "1201",
      pages          = "110",
      doi            = "10.1007/JHEP06(2012)150, 10.1007/JHEP01(2012)110",
      year           = "2012",
      eprint         = "1110.3949",
      archivePrefix  = "arXiv",
      primaryClass   = "hep-th",
      reportNumber   = "UT-KOMABA-11-9",
      SLACcitation   = "
}

@article{Tseytlin:3pt2,
      author         = "Buchbinder, E.I. and Tseytlin, A.A.",
      title          = "{Semiclassical correlators of three states with large $\mathrm{S}^{5}$ charges in string theory in $\mathrm{AdS}_5\times \mathrm{S}^{5}$}",
      journal        = "Phys.Rev.",
      volume         = "D85",
      pages          = "026001",
      doi            = "10.1103/PhysRevD.85.026001",
      year           = "2012",
      eprint         = "1110.5621",
      archivePrefix  = "arXiv",
      primaryClass   = "hep-th",
      SLACcitation   = "
}

@article{Janik:3pt2,
      author         = "Janik, Romuald A. and Wereszczynski, Andrzej",
      title          = "{Correlation functions of three heavy operators: The AdS
                        contribution}",
      journal        = "JHEP",
      volume         = "1112",
      pages          = "095",
      doi            = "10.1007/JHEP12(2011)095",
      year           = "2011",
      eprint         = "1109.6262",
      archivePrefix  = "arXiv",
      primaryClass   = "hep-th",
      SLACcitation   = "
}

@article{Caetano:4ptweak,
      author         = "Caetano, Joao and Escobedo, Jorge",
      title          = "{On four-point functions and integrability in N=4 SYM:
                        from weak to strong coupling}",
      journal        = "JHEP",
      volume         = "1109",
      pages          = "080",
      doi            = "10.1007/JHEP09(2011)080",
      year           = "2011",
      eprint         = "1107.5580",
      archivePrefix  = "arXiv",
      primaryClass   = "hep-th",
      SLACcitation   = "
}

@article{Georgiou:sl2,
      author         = "Georgiou, George",
      title          = "{SL(2) sector: weak/strong coupling agreement of
                        three-point correlators}",
      journal        = "JHEP",
      volume         = "1109",
      pages          = "132",
      doi            = "10.1007/JHEP09(2011)132",
      year           = "2011",
      eprint         = "1107.1850",
      archivePrefix  = "arXiv",
      primaryClass   = "hep-th",
      SLACcitation   = "
}

@article{EGSV:Tailoring2,
      author         = "Escobedo, Jorge and Gromov, Nikolay and Sever, Amit and
                        Vieira, Pedro",
      title          = "{Tailoring Three-Point Functions and Integrability II.
                        Weak/strong coupling match}",
      journal        = "JHEP",
      volume         = "1109",
      pages          = "029",
      doi            = "10.1007/JHEP09(2011)029",
      year           = "2011",
      eprint         = "1104.5501",
      archivePrefix  = "arXiv",
      primaryClass   = "hep-th",
      SLACcitation   = "
}

@article{Costa:3pt,
      author         = "Costa, Miguel S. and Monteiro, Ricardo and Santos, Jorge
                        E. and Zoakos, Dimitrios",
      title          = "{On three-point correlation functions in the
                        gauge/gravity duality}",
      journal        = "JHEP",
      volume         = "1011",
      pages          = "141",
      doi            = "10.1007/JHEP11(2010)141",
      year           = "2010",
      eprint         = "1008.1070",
      archivePrefix  = "arXiv",
      primaryClass   = "hep-th",
      SLACcitation   = "
}

@article{Zarembo:3pt1,
      author         = "Zarembo, K.",
      title          = "{Holographic three-point functions of semiclassical
                        states}",
      journal        = "JHEP",
      volume         = "1009",
      pages          = "030",
      doi            = "10.1007/JHEP09(2010)030",
      year           = "2010",
      eprint         = "1008.1059",
      archivePrefix  = "arXiv",
      primaryClass   = "hep-th",
      reportNumber   = "ITEP-TH-29-10, UUITP-25-10",
      SLACcitation   = "
}

@article{Tseytlin:3ptvertex,
      author         = "Buchbinder, E.I. and Tseytlin, A.A.",
      title          = "{On semiclassical approximation for correlators of closed
                        string vertex operators in AdS/CFT}",
      journal        = "JHEP",
      volume         = "1008",
      pages          = "057",
      doi            = "10.1007/JHEP08(2010)057",
      year           = "2010",
      eprint         = "1005.4516",
      archivePrefix  = "arXiv",
      primaryClass   = "hep-th",
      reportNumber   = "IMPERIAL-TP-AT-2010-2",
      SLACcitation   = "
}

@article{Janik:3pt1,
      author         = "Janik, Romuald A. and Surowka, Piotr and Wereszczynski,
                        Andrzej",
      title          = "{On correlation functions of operators dual to classical
                        spinning string states}",
      journal        = "JHEP",
      volume         = "1005",
      pages          = "030",
      doi            = "10.1007/JHEP05(2010)030",
      year           = "2010",
      eprint         = "1002.4613",
      archivePrefix  = "arXiv",
      primaryClass   = "hep-th",
      SLACcitation   = "
}

@article{Roiban:2004va,
      author         = "Roiban, Radu and Volovich, Anastasia",
      title          = "{Yang-Mills correlation functions from integrable spin
                        chains}",
      journal        = "JHEP",
      volume         = "0409",
      pages          = "032",
      doi            = "10.1088/1126-6708/2004/09/032",
      year           = "2004",
      eprint         = "hep-th/0407140",
      archivePrefix  = "arXiv",
      primaryClass   = "hep-th",
      SLACcitation   = "
}

@article{Okuyama:2004bd,
      author         = "Okuyama, Kazumi and Tseng, Li-Sheng",
      title          = "{Three-point functions in N = 4 SYM theory at one-loop}",
      journal        = "JHEP",
      volume         = "0408",
      pages          = "055",
      doi            = "10.1088/1126-6708/2004/08/055",
      year           = "2004",
      eprint         = "hep-th/0404190",
      archivePrefix  = "arXiv",
      primaryClass   = "hep-th",
      reportNumber   = "EFI-04-14",
      SLACcitation   = "
}

@article{Freedman:1998tz,
      author         = "Freedman, Daniel Z. and Mathur, Samir D. and Matusis,
                        Alec and Rastelli, Leonardo",
      title          = "{Correlation functions in the CFT(d) / AdS(d+1)
                        correspondence}",
      journal        = "Nucl.Phys.",
      volume         = "B546",
      pages          = "96-118",
      doi            = "10.1016/S0550-3213(99)00053-X",
      year           = "1999",
      eprint         = "hep-th/9804058",
      archivePrefix  = "arXiv",
      primaryClass   = "hep-th",
      reportNumber   = "MIT-CTP-2727",
      SLACcitation   = "
}

@article{Lee:1998bxa,
      author         = "Lee, Sangmin and Minwalla, Shiraz and Rangamani, Mukund
                        and Seiberg, Nathan",
      title          = "{Three point functions of chiral operators in $D = 4$, $\mathcal{N}=4$
                        SYM at large $N$}",
      journal        = "Adv.Theor.Math.Phys.",
      volume         = "2",
      pages          = "697-718",
      year           = "1998",
      eprint         = "hep-th/9806074",
      archivePrefix  = "arXiv",
      primaryClass   = "hep-th",
      reportNumber   = "PUPT-1796, IASSNS-HEP-98-51",
      SLACcitation   = "
}

@article{korepin-DWBC,
      author         = "Korepin, V.E.",
      title          = "{Calculation of Norms of Bethe Wave Functions}",
      journal        = "Commun.Math.Phys.",
      volume         = "86",
      pages          = "391-418",
      doi            = "10.1007/BF01212176",
      year           = "1982",
      SLACcitation   = "
}

@article{PhysRevD.23.417,
	Author = {Gaudin, Michel and McCoy, Barry M. and Wu, Tai Tsun},
	Date-Added = {2012-04-21 09:39:34 +0000},
	Date-Modified = {2012-04-21 09:39:34 +0000},
	Doi = {10.1103/PhysRevD.23.417},
	Issue = {2},
	Journal = {Phys. Rev. D},
	Pages = {417--419},
	Publisher = {American Physical Society},
	Title = {Normalization sum for the Bethe's hypothesis wave functions of the Heisenberg-Ising chain},
	Volume = {23},
	Year = {1981},
	Bdsk-Url-1 = {http://link.aps.org/doi/10.1103/PhysRevD.23.417},
	Bdsk-Url-2 = {http://dx.doi.org/10.1103/PhysRevD.23.417}}

\end{bibtex}
\bibliographystyle{nb}
\bibliography{\jobname}

\end{document}